\documentclass[aps, pra,twocolumn]{revtex4-2}
\usepackage{graphicx}
\usepackage{dcolumn}
\usepackage{bm}
\usepackage{amsmath}
\usepackage{physics}
\usepackage{comment}
\usepackage{color}
\usepackage{xr}
\usepackage[dvipsnames]{xcolor}
\usepackage[english]{babel}
\usepackage{amsmath}
\usepackage{amssymb}
\usepackage{amsthm}
\usepackage{amsfonts}
\usepackage{listings}
\usepackage{enumerate}
\usepackage{latexsym}
\usepackage{psfrag}
\usepackage{bm}
\usepackage{blkarray}
\usepackage{array}
\usepackage[normalem]{ulem}
\usepackage[colorlinks]{hyperref}
\usepackage[caption=false]{subfig}

\def\beq{\begin{equation}}
\def\eeq{\end{equation}}
\def\bal{\begin{aligned}}
\def\eal{\end{aligned}}

\begin{document}
\title{Wannier Function Methods for Topological Modes in 1D Photonic Crystals}
\author{Vaibhav Gupta}
\affiliation{Department of Physics and Institute for Condensed Matter Theory, University of Illinois at Urbana-Champaign, Urbana IL, 61801-3080, USA}
\author{Barry Bradlyn}
\affiliation{Department of Physics and Institute for Condensed Matter Theory, University of Illinois at Urbana-Champaign, Urbana IL, 61801-3080, USA}
\email{bbradlyn@illinois.edu}
\date{\today}
\begin{abstract}
    In this work, we use Wannier functions to analyze topological phase transitions in one dimensional photonic crystals. 
    We first review the construction of exponentially localized Wannier functions in one dimension, and show how to numerically construct them for photonic systems. 
    We then apply these tools to study a photonic analog of the Su-Schrieffer-Heeger model. 
    We use photonic Wannier functions to construct a quantitatively accurate approximate model for the topological phase transition, and compute the localization of topological defect states. 
    Finally, we discuss the implications of our work for the study of band representations for photonic crystals.
\end{abstract}
\maketitle

\section{Introduction}

Since their introduction in the seminal work of Haldane and Raghu\cite{haldane2008possible,raghu2008analogs}, topological photonic crystals have attracted much theoretical and experimental attention\cite{Joannopoulos08Book,wang2009observation,wang2008reflection,lu2013weyl,slobozhanyuk2017three,khanikaev2017two,sun2017photonics,rider2019perspective,ozawa2019topological,yang2019realization,kim2020recent}. 
From a practical perspective, boundary states in topological photonic crystals can allow for lossless and unidirectional propagation of light, useful in both communication and quantum information applications. 
From a more theoretical perspective, photonic systems provide the ideal platform to explore the physics of band theory: the lack of interactions between photons in linear dielectrics, combined with the ability to address individual states with precise momentum and frequency resolution allows for the study of topological phenomena in photonic crystals that may be difficult to observe in electronic systems. 
Two timely example are the realization of quantum anomalous Hall ``insulators''
\footnote{Throughout this work, we will refer to gapped photonic systems as insulators in analogy with electronic systems, even though there is no analogous filling of states} 
through modulated coupling of Weyl points in three dimensions\cite{liu2021observation,wieder2020axionic,devescovi2021cubic}, and the observation of higher-order topological corner modes and filling anomalies in two-dimensional photonic crystals\cite{he2020quadrupole,kim2021total,jin2021floquet,proctor2020robustness,zhang2021topological,liu2017novel,xie2018second,ota2019photonic,chen2019direct,xie2019visualization,noh2018topological,mittal2019photonica,el2019corner,li2020higher,cerjan2020observation,zhou2020photonic,wang2021higher,lin2022dirac,xie2021higher}.

One factor enabling the growth of the field of topological photonics is the application of tools for studying electronic band topology to the photonic realm. 
The use of (non-Abelian) Berry phases to diagnose the presence of nontrivial topology in photonic crystals is well-established\cite{blanco2020tutorial,ozawa2019topological,lu2016symmetryprotected,wang2019banda,xiao2014surface}. 
Recently, symmetry-based indicators of band topology\cite{po2017symmetry} have also seen use in diagnosing and engineering topological photonic structures, opening the door to high-throughput screening of potential photonic crystal architectures\cite{de2019engineering,alexandradinata2020crystallographic,christensen2021location}. 
While naturally formulated in reciprocal space, Berry phase and symmetry-indicators are intimately connected to \emph{position} space information\cite{soluyanov2011wannier,z2packvanderbilt,vanderbilt2018berry,bradlyn2017topological}. 
The non-Abelian Berry phase gives the centers of photonic hybrid Wannier functions, localized along one direction of the photonic crystal; the location and flow of these centers is connected to the presence of localized corner and edge states in topologically nontrivial systems. 
Similarly, filling anomalies in one- and two-dimensional insulators can be understood in terms of the centers of maximally localized Wannier functions relative to symmetry centers of the crystal\cite{wieder2020strong,fang2021filling,benalcazar2019quantization}.

In the study of electronic systems, this connection between position-space Wannier functions and momentum space topological invariants has led to new perspectives on topological materials that are only now beginning to be exploited for photonic crystals. 
The study of band representations---symmetry properties of groups of bands that support symmetric, localized Wannier functions---has allowed for deeper understanding of topological crystalline phases and boundary modes in photonic crystals\cite{alexandradinata2020crystallographic,de2019engineering,christensen2021location,longhi2019probing}. 
Although the symmetry properties of photonic band representations have proved useful, little attention has been paid thus far to the localized (hybrid) Wannier functions themselves. 
Photonic Wannier functions have been used to numerically study defect modes in photonic crystals\cite{Romano_2010,leung1993defect,albert2000generalized,albert2002photonic,busch2003wannier,busch2011photonic,romano2018wannier,tanaue2020wannier,garcia2003defect,bruno2003bloch,hermann2008wannier}. 
Here we aim to connect this earlier work on photonic Wannier functions directly to topology, showing how the theory of band representations can be used to study photonic crystals in a new light.

To do so, we will study the properties and applications of photonic hybrid Wannier functions, starting with one-dimensional photonic crystals. 
In Sec.~\ref{sec:construction} we show that the computational methods and localization arguments for electronic Wannier functions carry over to the photonic case -- in particular, we will argue that all bands in a 1D photonic crystal (even the zero-frequency band) admit exponentially localized Wannier functions. 
While this result is not new, we review it here to be self-contained and to show how single- and multi-band hybrid Wannier functions can be computed using the MIT Photonic Bands (MPB) library\cite{johnson2001block}. 
For inversion-symmetric 1D photonic crystals, we show that each isolated band transforms in an elementary band representation, and that the Wannier centers of each band are consistent with the inversion eigenvalues of the occupied bands. 
Using this, in Sec.~\ref{sec:1dmodel} we construct a family of inversion-symmetric 1D photonic crystal structures that exhibit a transition between the two band representations of line group $p\bar{1}1'$. 
We show that this transition is related to the obstructed atomic limit transition in the electronic $1D$ Su-Schrieffer-Heeger (SSH) chain\cite{su1979solitons}. 
We show how our method allows us to design domain walls between these different phases which feature topological boundary modes, which we diagnose via a filling anomaly. 
Lastly, in Sec.~\ref{sec:tb} we show that we can use the photonic Wannier functions to construct tight-binding-like approximations to Maxwell's equations, which captures the essential features of the physics within a given frequency range. 
We show how this tight-binding approach allows us to quantitatively analyze the topological properties of photonic systems, using our SSH-like model as an example.  
We conclude with a discussion of the extension of these ideas to two and three dimensions.

\section{Constructing Photonic Hybrid Wannier Functions}\label{sec:construction}

We shall begin with an overview of Maxwell's equations in a photonic crystal\cite{Joannopoulos08Book}. 
We will review how we can recast the wave equation for a 1D photonic crystal into a Hermitian form analogous to the time-independent Schr\"{o}dinger equation. 
Using this, we will show how, just like in electronic systems, a consideration of the position operator lets us introduce an algorithm for numerically constructing unique Wannier functions in 1D. 
For illustration, we will apply these tools to the simplest 1D photonic crystal, which maps onto a photonic version of the Kronig-Penney model.

\subsection{Properties of Maxwell's Equations in a Photonic Crystal}

In materials which are linear, isotropic, lossless, and dispersionless, the source-free macroscopic Maxwell’s equations can be expressed in the frequency domain as\cite{Joannopoulos08Book}
\beq \label{eq:macroscopic_maxwell}
    \bal    
        \div{[\epsilon(\vb{r}) \vb{E} (\vb{r})]} = 0 \\
        \curl{\vb{E}(\vb{r})} - i \omega \vb{H}(\vb{r}) = 0 \\
        \div{H(\vb{r})} = 0 \\
        \curl{\vb{H}(\vb{r})} + i \omega \varepsilon_0 \epsilon(\vb{r}) \vb{E}(\vb{r}) = 0
    \eal
\eeq
where $\vb{E}(\vb{r})$ and $\vb{H}(\vb{r})$ are the electric and magnetic field vectors, respectively, and the time dependence of the fields is harmonic such that $\vb{E}(\vb{r},t)=\vb{E}(\vb{r})e^{-i\omega t}$ and $\vb{H}(\vb{r},t)=\vb{H}(\vb{r})e^{-i\omega t}$. 
Since we will be working with non-magnetic materials, we set $\mu = 1$ such that $\vb{B} = \vb{H} / \mu_0$. 
The above equations can be decoupled into the second-order wave equations
\beq
    \bal
        \curl{\bqty{\frac{1}{\epsilon({\vb{r}})} \curl{\vb{H}(\vb{r})}}} &= \pqty{\frac{\omega}{c}}^2 \vb{H}(\vb{r}) \\
        \curl{\curl{\vb{E}(\vb{r})}} &= \pqty{\frac{\omega}{c}}^2 \epsilon(\vb{r}) \vb{E}(\vb{r})
    \eal
\eeq

When the medium is inhomogeneous only along a single direction (which we will take to be the $x$-direction for concreteness, $\epsilon(\vb{r})=\epsilon(x)$), then the fields can be taken to be functions of only that direction. 
Furthermore, since we consider a medium in which $\epsilon(x)$ is a scalar, the $\hat{\vb{y}}$ and $\hat{\vb{z}}$ components of the fields are decoupled. 
Focusing on the equations for z-polarized fields $\vb{E}(\vb{r})=E_z(x)\hat{\vb{z}},$ we find
\beq
\bal
    -\dv{}{x} \frac{1}{\epsilon \qty(x)} \dv{}{x} H_y \qty(x) &= \qty(\frac{\omega}{c})^2 H_y \qty(x) \\
    -\dv[2]{}{x} E_z \qty(x) &= \qty(\frac{\omega}{c})^2 \epsilon \qty(x) E_z \qty(x)\label{eq:maxwellE}
\eal
\eeq

An identical set of equations holds for $\vb{E}\parallel \hat{\vb{y}},$ so we omit it for brevity. 
Rescaling the electric fields to $\widetilde{E}_z \qty(x) = \sqrt{\epsilon \qty(x)} E_z \qty(x)$ allows us to rewrite Eq.~(\ref{eq:maxwellE}) in the Hermitian form
\beq \label{eq:rescaled_E_field_eqn}
    -\frac{1}{\sqrt{\epsilon \qty(x)}}\dv[2]{}{x} \frac{1}{\sqrt{\epsilon \qty(x)}} \widetilde{E}_z \qty(x) = \qty(\frac{\omega}{c})^2 \widetilde{E}_z \qty(x)
\eeq
From now on, we will set $c = 1$. 
We label the two Hermitian operators appearing above as
\beq\label{eq:maxwellops}
\bal
    \mathcal{O}_H &\equiv -\dv{}{x} \frac{1}{\epsilon \qty(x)} \dv{}{x}  \\
    \mathcal{O}_E &\equiv -\frac{1}{\sqrt{\epsilon \qty(x)}}\dv[2]{}{x} \frac{1}{\sqrt{\epsilon \qty(x)}}
\eal
\eeq

A special case of the systems considered above is 1D photonic crystals which are made up of a lattice of slabs of a dielectric material (assumed to be arranged along the $x$-axis) with the lattice constant $a$, i.~e the dielectric function has discrete translation symmetry $\epsilon(x+a)=\epsilon(x)$. 
The transverse dimensions of these slabs are very large compared to the spacing between them, allowing us to treat the system as quasi-1D. 
Consequently, Bloch's theorem can be used to write the eigenfunctions of Eq.~\eqref{eq:maxwellops} as $\widetilde{\vb{E}}_{n k} \qty(x) = e^{i k x} \widetilde{\vb{v}}_{n k} \qty(x)$ and $\vb{H}_{n k} \qty(x) = e^{i k x} \vb{u}_{n k} \qty(x)$, where $k$ is the crystal momentum, $n$ is the band index, and $\widetilde{\vb{v}}_{n k} \qty(x)$ and $\vb{u}_{n k} \qty(x)$ are the cell-periodic Bloch functions satisfying
\beq
    \bal
        \widetilde{\vb{v}}_{n k} \qty(x + m a) = \widetilde{\vb{v}}_{n k} \qty(x) \\
        \vb{u}_{n k} \qty(x + m a) = \vb{u}_{n k} \qty(x)
    \eal
\eeq
for any integer $m$. 

In analogy to electronic systems, the eigenvalues $\omega^2(k)$ of $\mathcal{O}_E$ form sets of bands separated by gaps. 
There are, however, a few distinguishing features of photonic band structures as compared with their electronic counterpart. 
First, since the role of energy is played by $\omega^2(k)$ in this second-order formalism, the spectrum is always postive semi-definite. 
Additionally, the dispersion of the lowest band satisfies $\omega^2(k)\rightarrow c^2k^2$ as $k\rightarrow 0$\cite{sipe2000vector,watanabe2018space}. 
Furthermore, as noninteracting bosons, photons do not fill bands and hence we cannot define a notion of Fermi level. 
Nevertheless, since photonic states can be addressed with precise frequency and momentum resolution, we can study (topological) properties of photonic bands.

From here on, we will only focus on the electric fields for simplicity. 
The magnetic fields can be treated similarly without any major changes. 
We can start to study the role of topology in 1D photonic crystals by first looking at the Berry phases for different bands. 
For band $n$, isolated in frequency from all other bands, we define the Berry phase $\phi_n$ as\cite{zak1989berry}
\beq
    \phi_n = \int^{2 \pi / a}_0 \dd k \braket{\widetilde{v}_{n k}}{i \partial_k \widetilde{v}_{n k}}\label{eq:berry}.
\eeq

Like for electronic systems, the Berry phase $\phi_n$ for a 1D photonic crystal gives the eigenvalue of the position operator projected into the band $n$, 
\begin{equation}
PxP = \sum_{kk'}|\widetilde{\vb{E}}_{nk}\rangle\langle \widetilde{\vb{E}}_{nk}|x|\widetilde{\vb{E}}_{nk'}\rangle\langle \widetilde{\vb{E}}_{nk'}|,
\end{equation}
provided we enforce the boundary condition
\begin{equation}
    \widetilde{\vb{E}}_{nk+2\pi/a}(x)=    \widetilde{\vb{E}}_{nk}(x).
\end{equation}
As for electronic systems, we can ask about the eigenstates of $PxP$, which in 1D correspond to the maximally localized Wannier functions for the photonic crystal\cite{marzari1997maximally,kivelson1982wannier,alexandradinata2014wilsonloop}. 
In the subsequent sections, we will show how to numerically construct the Maximally Localized Wannier Functions (MLWFs) for a photonic crystal in 1D. 
Furthermore, we will show how the Wannier functions can be used to derive quantitatively accurate tight-binding models for photonic crystals. 
As a proof-of-principle, we will apply these tools to analyze topologically nontrivial domain walls in a photonic analog of the Su-Schrieffer-Heeger (SSH) model. Before doing so, however, we will focus on a simpler model in order to see how photonic Wannier functions can be numerically computed and analyzed.

\subsection{The Photonic Kronig-Penney (PhKP) Model and Photonic Wannier Functions}\label{sec:PhKP}

We will use the MPB package to design the systems, and then numerically obtain the spectrum and the eigenfunctions of $\mathcal{O}_E$. 
We do this by exploiting the gauge freedom in defining the Wannier functions. 
The result of this procedure is the MLWFs which are smooth and exponentially-localized. 
Next, we will see how to numerically find the eigenstates of the projected position operator, and hence the MLWFs for photonic systems. 
We will review the general strategy here. 
As a concrete example, we will apply the general technique to construct the MLWFs for a photonic analog of the Kronig-Penney model.

To begin, we will first describe the method to numerically compute the Berry phase--and hence the Wannier function centers--for a photonic crystal. 
As mentioned in the previous section, MLWFs in 1D can be obtained by diagonalizing the $PxP$ operator. 
As we know from the study of electronic systems, directly diagonalizing $PxP$ for infinite or periodic systems is numerically untenable. 
Instead we will construct and diagonalize the Wilson loop operator\cite{yu2011equivalent,z2packbernevig}
\begin{equation}
    W^{nm}_{k_f\leftarrow k_i} = \bra{\widetilde{v}_{n{k_f}}}\prod_{k}^{k_f\leftarrow k_i}P(k)\ket{\widetilde{v}_{mk_{i}}},\label{eq:wilsondef}
\end{equation}
where $P=\sum_{n,\text{occ}} \ket{\widetilde{v}_{nk}}\bra{\widetilde{v}_{nk}}$ is the projector onto the bands under consideration. 
When $k_f\equiv k_i$ modulo a reciprocal lattice vector, then $W$ and $PxP$ are isospectral. 
In what follows we will review the expressions for the MLWFs in terms of the Wilson loop. 
The details about the numerical implementation of these results can be found in Appendix \ref{app:numerical_MLWF}.

If there is only a single, isolated band (with index $n$, for concreteness) under consideration, then the Wilson loop reduces to the exponential of the abelian Berry phase $\phi_n$. 
We can compute the Berry phase of isolated bands using the discretization of Eq.~(\ref{eq:berry}) given by\cite{vanderbilt2018berry,blanco2020tutorial}
\beq
    \phi_n \equiv \prod_{k = 0}^{N-1} \braket{\widetilde{v}_{n k}}{\widetilde{v}_{n k+1}}.
\eeq
where we have split the BZ into $N - 1$ segments of length $\Delta k = 2 \pi / \qty(a N)$, $k = 0$ is the $\Gamma$ point and $k = N$ corresponds to the $X$ point at the zone boundary. 
Note that due to the factors of $\sqrt{\epsilon(x)}$ in our definition of $\ket{\tilde{v}_{nk}}$, the overlap here is evaluated using the standard inner product. 
The Berry phase $\phi_n/(2\pi)$ gives the fractional part (in units of the lattice constant) of the eigenvalues of $PxP$. 
We can then construct the Fourier transform
\beq \label{eq:isol_wann_funcs}
    \ket{n R} \equiv \int \frac{\dd k}{2 \pi} \exp \qty[-i k \qty(R + \frac{\phi_n}{2 \pi})] W^n_{k \leftarrow 0} \ket{{E}_{n k}}
\eeq
where $\phi_n$ is the Berry phase and $W^n_{k \leftarrow 0} \equiv \exp \qty(-\int_0^k \dd q \braket{\widetilde{v}_{n q}}{\partial_q \widetilde{v}_{n q}})$ is the Wilson line Eq.~\eqref{eq:wilsondef} from $0$ to $k$ in the Brillouin zone for band $n$. 
The Wilson line in this case can be thought of as a $U(1)$ gauge transformation of the state $\ket{{E}_{nk}}$ chosen to ensure a gauge smooth in $k$. 
Eq.~\eqref{eq:isol_wann_funcs} is then  our Wannier function for an isolated band.
 
We can also find Wannier functions for multiple bands at once by taking advantage of the non-abelian Wilson loop directly. 
If the bands under consideration are band $m$ to $m + l$, we can generalize Eq.~\eqref{eq:isol_wann_funcs} to find the composite band Wannier functions\cite{alexandradinata2014wilsonloop,marzari2012maximally}
\beq \label{eq:comp_wann_funcs}
    \ket{n R} \equiv \sum_{j =m}^{m + l} \int \frac{\dd k}{2 \pi} \exp \qty[-i k \qty(R + \frac{\theta_n}{2 \pi})] Q_{j n} \pqty{k} \ket{{E}_{j k}}.
\eeq
Here $e^{i\theta_n}$ is the $n$\textsuperscript{th} eigenvalue of the Wilson loop operator $W_{k+ 2 \pi / a \leftarrow k}$, and $Q_{jn}\pqty{k}$ is the corresponding eigenvector. 
Note that Eq.~\eqref{eq:comp_wann_funcs} reduces to Eq~\eqref{eq:isol_wann_funcs} when we have only a single band, by virtue of the fact that $Q(k_1+k_2) = W_{k_2\leftarrow k_1}Q(k_1)$.

The Wannier functions $w_{nR}(x)=\bra{x}\ket{nR}$ thus give localized excitations of the electric field $E_z(x)$ formed from superpositions of the eigenmodes $E_{nk}(x)$ for the bands of interest. 
For mathematical convenience, it will also be useful to introduce the rescaled Wannier functions $\ket{\widetilde{nR}}$ defined such that
\begin{equation}
    \bra{x}\ket{\widetilde{nR}} = \sqrt{\epsilon(x)}w_{nR}(x),
\end{equation}
which are linear combinations of the eigenmodes of the Hermitian Maxwell operator Eq.~\eqref{eq:maxwellops}.

By construction, the center of the Wannier function $\ket{nR}$ is given by $r_c^{nR} = R+a\theta_n/(2\phi)$, where $a$ is the lattice constant\cite{zak1989berry,kingsmith1993theory}. 
Since the Berry phases $\theta_n$ are defined modulo $2 \pi$, the Wannier centers are defined up to a translation by a lattice vector (i.~e. up to a choice of the home cell $R=0$). 
In this work, we will focus on systems with inversion symmetry. 
Inversion symmetry forces the eigenvalues of the Wilson loop to be either real, or come in complex conjugate pairs\cite{alexandradinata2014wilsonloop,wieder2018wallpaper,bradlyn2021lecture}. 
In terms of Wannier functions, this implies that for an isolated band, $\phi=0,\pi$ and hence the Wannier functions are centered at either $R$ or $R+a/2$, the two inversion centers in the unit cell.

One subtlety for photonic systems that is not present in electronic systems is the apparent singularity in the dispersion as $k\rightarrow 0$. 
In particular, one may worry that the non-analyticity of $\omega(k\rightarrow 0) \sim |k|$ for the lowest band may introduce power law tails into the Wannier function for this band\cite{de2020equivalence,watanabe2018space}. 
However, for one-dimensional photonic crystals we can circumvent this issue by choosing a global, k-independent polarization for the eigenmodes of the electric field. 
In particular, this lets us derive the scalar equation Eq.~\eqref{eq:rescaled_E_field_eqn}, in which we see that we should identify $\omega^2$ with the eigenvalue of the Hermitian Maxwell operator $\mathcal{O}_E$; $\omega^2$ is an analytic function of $k$ near the $\Gamma$ point, and so we expect no obstruction to forming exponentially localized Wannier functions for one-dimensional photonic crystals.

As an application of these tools, let us construct the MLWFs for the ``Photonic Kronig-Penney'' (PhKP) model of a 1D photonic crystal\cite{Romano_2010}. 
The PhKP model can be realized by a periodic one dimensional array of slabs of width $ab$ and dielectric constant $\varepsilon$ separated by regions of vacuum of width $a(1-b)$, where $0<b<1$. 
Choosing an inversion-symmetric unit cell, we can write the dielectric function in the home ($R=0$) unit cell as
\beq
    \epsilon \pqty{x} = 
        \begin{cases}
            1 & \forall \ x \in [0, a(1 - b)/2[ \ \cup \ ]a(1 + b)/2, a] \\ 
            \varepsilon & \forall \ x \in [a(1 - b)/2, a(1 + b)/2]
        \end{cases}
\eeq
For brevity, we will henceforth set the lattice constant $a=1$.

As a one-dimensional differential equation with periodic potential, Eq.~\eqref{eq:maxwellE} can be solved semianalytically using a transfer matrix approach\cite{romano2018wannier,kohn1959analytic}. 
In particular, we can find solutions in a single unit cell composed of reflected and transmitted waves, and impose the Bloch boundary condition that frequency eigenfunctions be simultaneous eigenfunctions of the lattice translation operator. 
Following an analysis similar to that of the electronic Kronig-Penney model, we find that Bloch wave-like solutions for \eqref{eq:rescaled_E_field_eqn} for this model only exist when the condition $\cos \pqty{k} = f \pqty{\omega^2}$ is satisfied, with
\begin{align}
f \pqty{\omega^2} &\equiv \cos \pqty{\pqty{{1-b}} \alpha} \cos \pqty{b \beta}\nonumber \\
&- \frac{1+\varepsilon}{2 \sqrt{\varepsilon}} \sin \pqty{\pqty{1-b} \alpha} \sin \pqty{b \beta}.
\end{align}
We have introduced $\alpha \equiv \sqrt{\omega^2}$ and $\beta \equiv \sqrt{\varepsilon \omega^2}$. 
The function $f \pqty{\omega^2}$ is analytic with extrema that lie outside the ``physical strip'' $|f(\omega^2)| \leq 1$. 
This means we can find a real, exponentially localized Wannier function with a definite parity under inversion (about one of the inversion centers) for each isolated band. 
In Appendix~\ref{app:energy_function} we discuss in more detail the important analytic properties of $f(\omega^2)$ for the PhKP model, and contrast it with the analogous function for the electronic KP model.

\begin{figure}
    \includegraphics[width=0.48\textwidth]{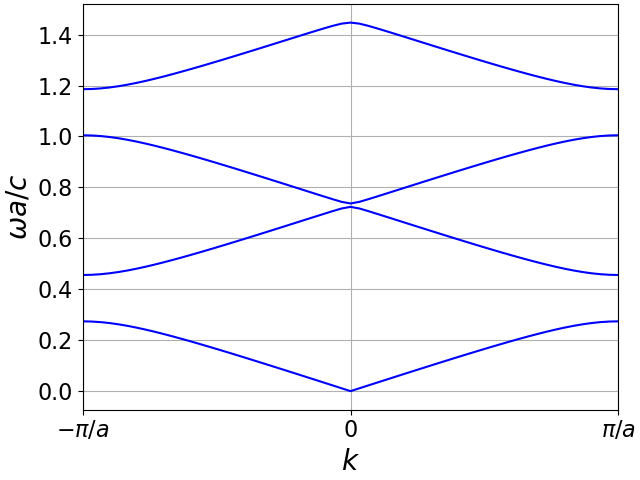}
    \caption{First four bands in the band structure for the PhKP model with $\varepsilon=5$ and $b=0.3$, generated with MPB. 
    Each band is isolated, although the gap between the second and third bands is very small. 
    }\label{fig:kronig_penney_band_diagram}
\end{figure}

\begin{figure}[ht]
    \centering
    \includegraphics[width=0.48\textwidth]{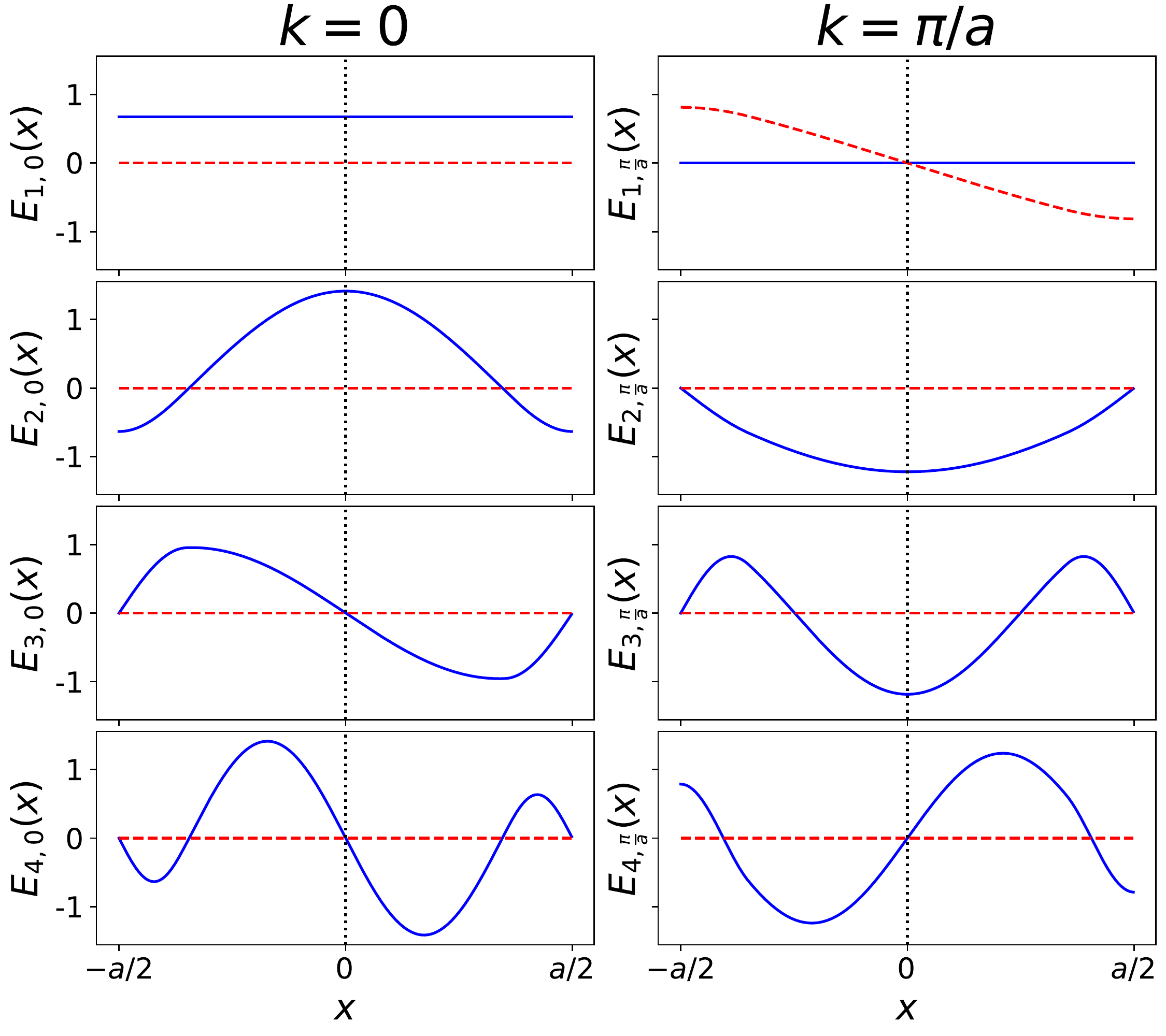}
    \caption{The real (solid blue) and imaginary (dashed red) parts of the electric field eigenmodes $E_{nk}(x)$ (in units of $a^{-1/2}$) for the lowest four bands of the photonic Kronig Penney model with $\epsilon=5$ and $b=0.3$. We see that the parity of the modes is consistent with the second column of Table~\ref{tab:kpwfprops}.}
    \label{fig:phkpmodes}
\end{figure}

\begin{figure}[ht]
    \centering
    \includegraphics[width=0.48\textwidth]{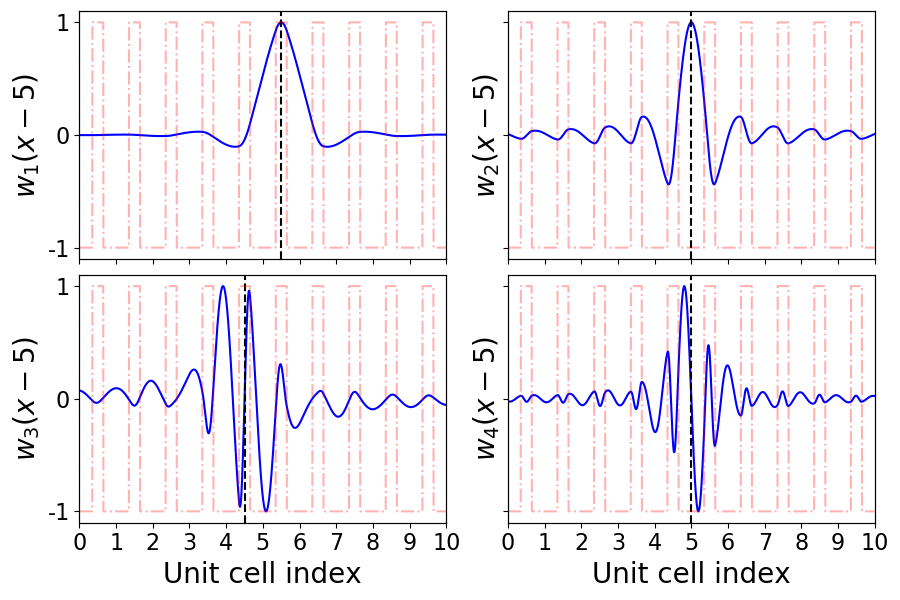}
    \caption{Wannier functions for the Kronig-Penney model for $\varepsilon = 5$ and $b = 0.3$ (normalized such that the maximum value is 1). 
    The black dashed lines shows the Wannier centers of the respective curves. 
    The red dashdot background curve is $\epsilon \pqty{x}$. 
    The parities match those given in Table~\ref{tab:kpwfprops}.}
    \label{fig:kronig_penney_wann_funcs}
\end{figure}

\begin{figure}[ht]
\includegraphics[width=0.48\textwidth]{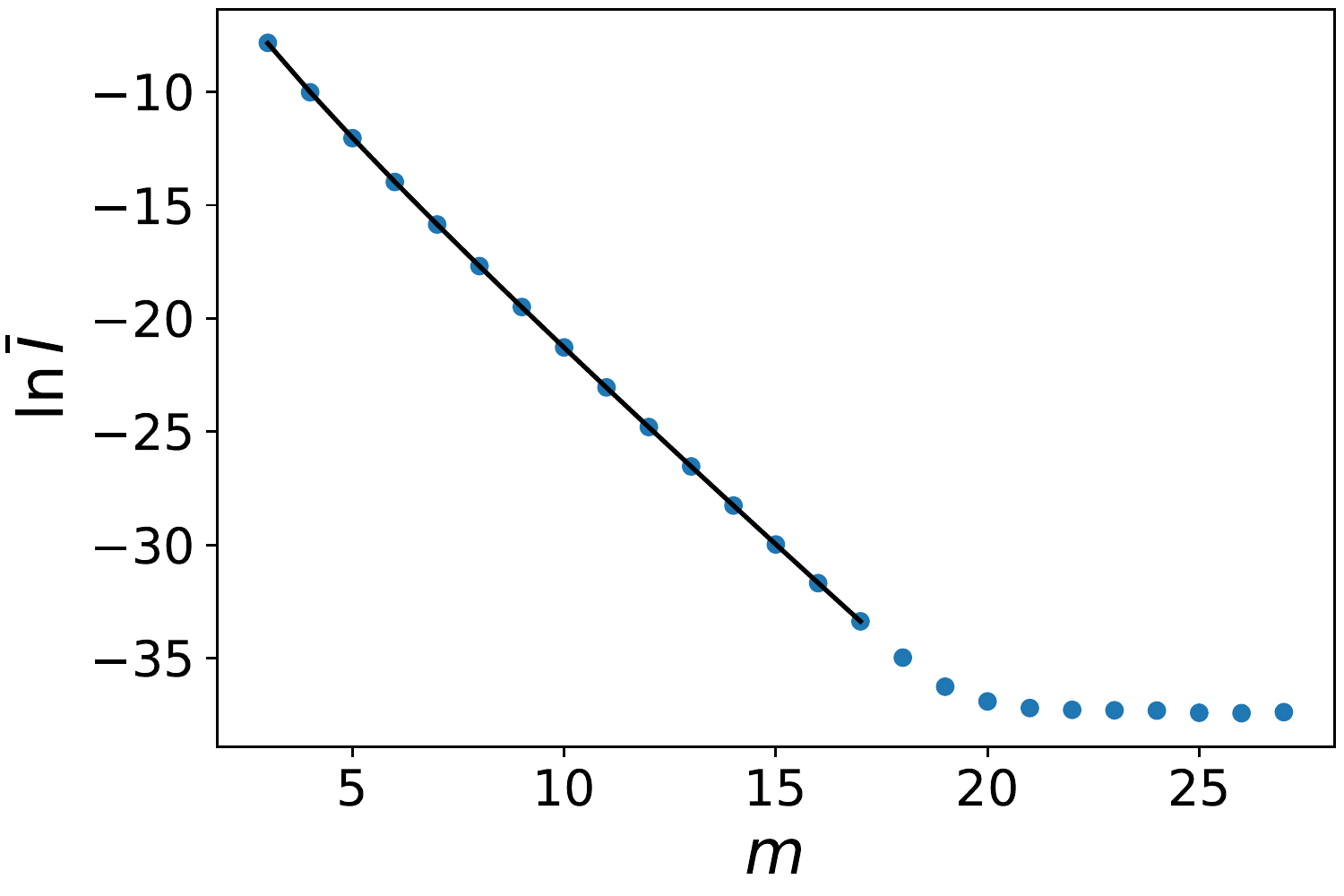}
\caption{Decay of the average (dimensionless) energy $\bar{I}$ per unit cell for the lowest-band Wannier function in the PhKP model plotted as a function of unit cell index $m$. 
The blue dots are the results computed from the Wannier function in the top panel of Fig.~\ref{fig:kronig_penney_wann_funcs}. 
The energy levels off at $m\approx 18$ due to the noise floor of the numerical computation. 
The black curve is the best-fit curve $\ln\bar{I} = C - 2h x -2\alpha\ln x$. 
We find $h=0.804\pm0.001$ and $\alpha=0.737\pm0.008$, in good agreement with analytic results.}\label{fig:wf_decay}
\end{figure}

Using MPB, we can numerically solve the PhKP model and construct the MLWFs. 
In Fig. \ref{fig:kronig_penney_band_diagram}, we show the spectrum of the PhKP model with $\varepsilon = 5$ and $b = 0.3$. 
Note the singularity of the lowest band at the $\Gamma$ point. 
We see from the figure that the first four bands are all isolated, and so we can use Eq.~\eqref{eq:isol_wann_funcs} to construct their Wannier functions as show in Fig.~\ref{fig:kronig_penney_wann_funcs}. 
The Wannier functions for the first four bands are centered at $R+1/2, R, R+1/2,$ and $R$, respectively, consistent with the Berry phases of the bands as computed in Table.~\ref{tab:kpwfprops}. 
Furthermore, the parity and center of the Wannier functions are determined by the inversion eigenvalues of the Bloch band used to construct them. 
In particular, if the parity of the field eigenmodes are equal at $\Gamma$ and $X$, then the Wannier functions have a Berry phase $0$ and are centered at $R$, whereas if the parities are opposite, then the Wannier functions have a Berry phase of $\pi$ and are centered at $R+1/2$. 
The parity of the MLWFs under inversion about their center can be read off from the inversion eigenvalue of the field eigenmode at $\Gamma$. 
To determine the inversion eigenvalue we show in Fig.~\ref{fig:phkpmodes} the field eigenmodes at $k=0$ and $k=\pi/a$; functions even about $x=0$ have $+$ parity, and functions odd about $x=0$ have $-$ parity.
We report these results in Table~\ref{tab:kpwfprops}, and see that they are visually consistent with the MLWFs in Fig.~\ref{fig:kronig_penney_wann_funcs}. 
We see also that our Wannier functions are exponentially localized. 
Following Refs.~\cite{Romano_2010,he2001exponential}, in Fig.~\ref{fig:wf_decay} we show the average (dimensionless) energy 
\begin{equation}\bar{I}(m)=\int_{m}^{m+1}\epsilon(r)|w_{nR}|^2 dr
\end{equation} 
in unit cell $m$ for the lowest band Wannier function, fit to an asymptotic form $\bar{I}(m)=m^{2\alpha}e^{-2hm}$. 
Our result of $h=0.804\pm0.001$ matches the exact analytic expressions derived from $\mathrm{arccosh}|f(\omega_*^2)| =0.804$ at the first minimum $\omega_*^2$ of $f$. 
Additionally, our fit of $\alpha=0.737\pm0.008$ compares favorably with the exact value of $\alpha=3/4$. 
A similar analysis of the decay can be carried out for the higher bands, although the small gap between the second and third bands means that high resolution and high numerical precision is necessary to obtain precise agreement with the analytic results.

\begin{table}
    \centering
    \begin{tabular}{c|c|c|c}
         Band index& Parity at $(\Gamma, X)$ & Berry phase & MLWF parity  \\
         \hline
         1 & $(+,-)$ & $\pi$& $ +$ \\
         2 & $(+,+)$ & $0$ & $+$ \\
         3 & $(-,+)$ & $\pi$ & $-$ \\
         4 & $(-,-)$ & $0$ & $-$
    \end{tabular}
    \caption{Symmetry and topological properties of the first four bands of the PhKP model with $\varepsilon=5, b=0.3$. 
    The first column gives the band index. 
    The second column gives the eigenvalue under inversion about the position space origin (parity) for the Bloch states at $\Gamma$ ($k=0$) and $X$ ($k=\pi/a$) in the Brillouin zone. 
    Third column gives the single-band Berry phase for each band. 
    The last column gives the parity of the Wannier functions constructed from the band, which coincides with the inversion eigenvalue of the Bloch states at $\Gamma$.}
    \label{tab:kpwfprops}
\end{table}

We would now like to apply these tools to analyze a topologically interesting 1D photonic crystal, for which exact analytical methods are more cumbersome. 
To do so, we will take inspiration from polyacetylene, and consider a twofold modulation of the PhKP model. 
This will allow us to construct a photonic analog of the Su-Schrieffer-Heeger (SSH) chain.

\section{Example: The Photonic SSH Chain}\label{sec:1dmodel}

\begin{figure}[htbp]
\subfloat[]{
        \includegraphics[width=0.22\textwidth]{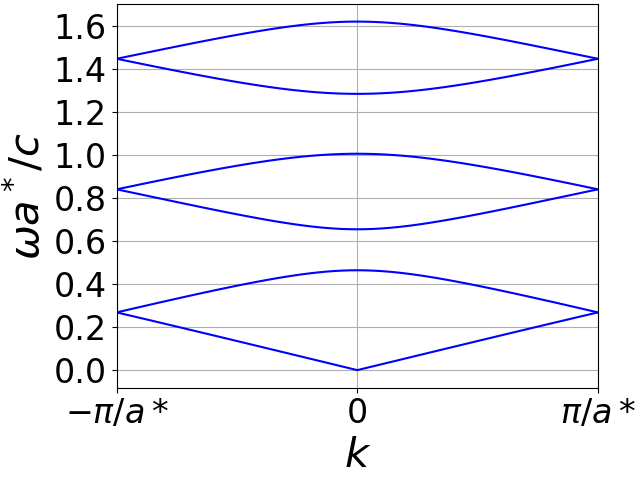}
}
\subfloat[]{
        \includegraphics[width=0.22\textwidth]{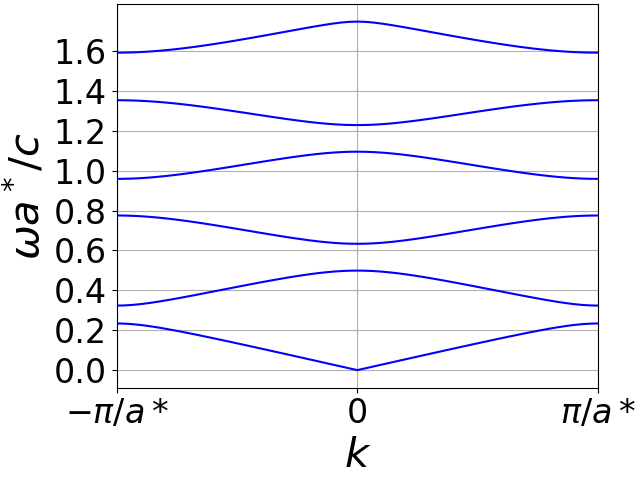}
}

\caption{Band structure for the photonic SSH model generated with MPB. 
(a) shows the first six bands at the critical point $\Delta \varepsilon = 0$. 
The emergent half-translation symmetry at this point leads to degeneracies between pairs of bands at the Brillouin zone boundary. 
(b) shows the first six bands with $|\Delta\varepsilon|=2$, where all bands are isolated.}
\label{fig:band_diagrams}
\end{figure}

Let us introduce a modulation of the dielectric constant in the PhKP model that doubles the length of the unit cell. 
Specifically, let us imagine modulating the dielectric constant of each slab, such that the slabs in even numbered unit cells have dielectric constant $\varepsilon_e=\varepsilon+\Delta\varepsilon$, and the slabs in odd-numbered unit cells have dielectric constant $\varepsilon_o=\varepsilon-\Delta\varepsilon$. 
This doubles the periodicity of the system in position space (i.e. $\epsilon(x+2a)=\epsilon(x)$), and hence folds the Brillouin zone into a reduced zone with $k\in[-\pi/(2a),\pi/(2a)]$. 
When $\Delta\varepsilon=0$, the band structure in the reduced zone can be obtained by backfolding the bands in Fig.~\ref{fig:kronig_penney_band_diagram} into the reduced Brillouin zone; this leads to band degeneracies at the $X$ point of the reduced zone, as shown in Fig.~\ref{fig:band_diagrams}(a). 
Just like in the electronic Peierls transition, increasing $|\Delta\varepsilon|$ opens a gap between the degenerate sets of bands. 
In Fig.~\ref{fig:band_diagrams}(b) we show the band structure when $\Delta\varepsilon=2$. 
In both figures, we use $a^*=2a$ to represent the lattice constant of the enlarged unit cell. 
This model realizes a photonic analogue of the Su-Schreiffer-Heeger model of polyacetylene. 
Much like that electronic model, we expect topologically distinct phases depending on the sign of $\Delta\varepsilon$, where the topology is protected by inversion symmetry. 
We will first use our photonic Wannier functions to study the topology of this model in the two phases. 
Then, we will show how we can use the photonic Wannier functions to recover an effective tight-binding description for two bands of the system, and show how this can help us predict the properties of defect states at the interface between two topologically distinct phases.

\begin{figure}[ht]
    \centering
    \includegraphics[width=0.48\textwidth]{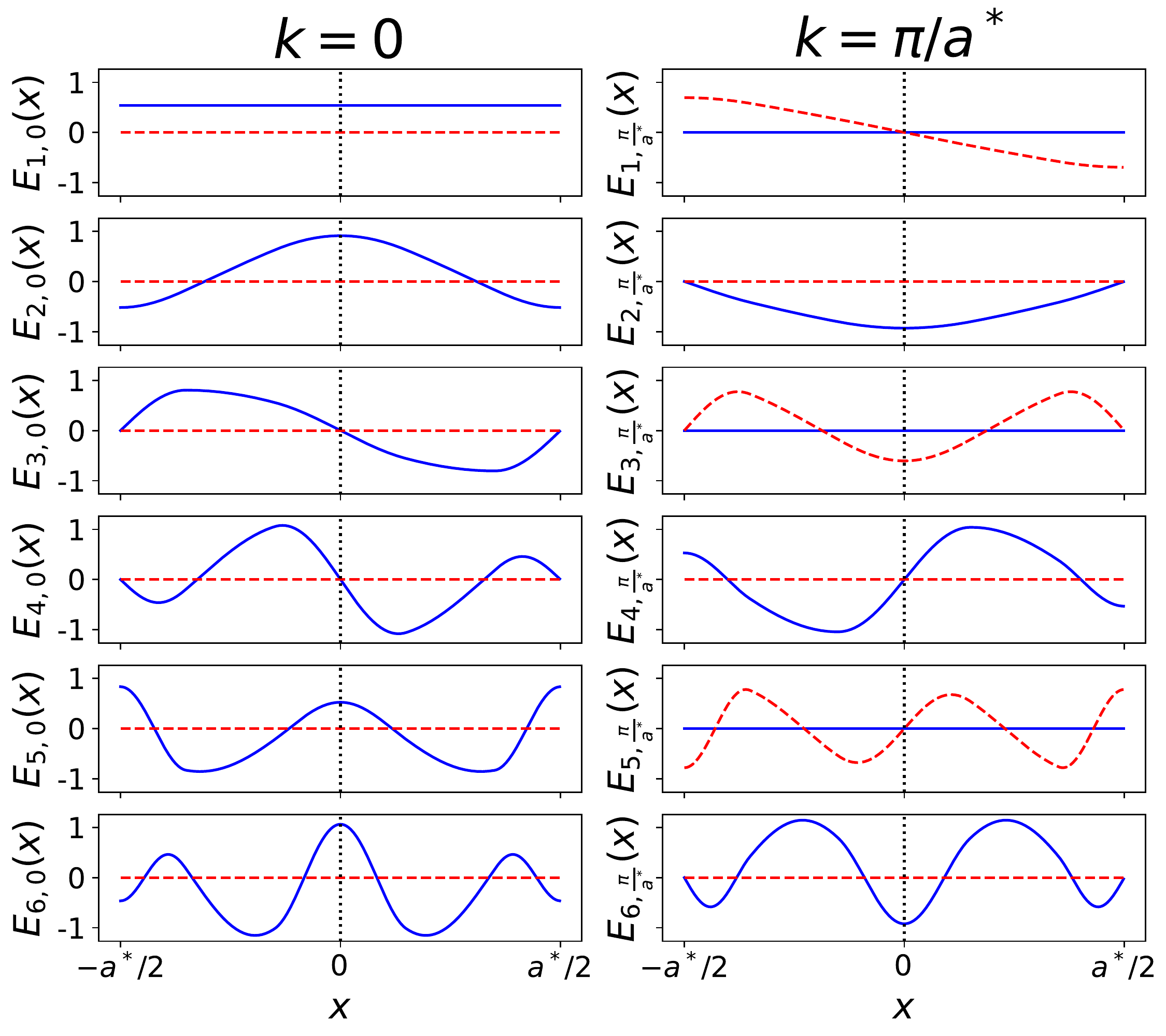}
    \caption{The real (solid blue) and imaginary (dashed red) parts of the electric field eigenmodes $E_{nk}(x)$ (in units of $a^{*-1/2}$) for the lowest four bands of the photonic SSH model with $\varepsilon=5,\Delta\varepsilon=2$ and $b=0.3$. We see that the parity of the modes is consistent with the second column of Table~\ref{tab:sshwfprops}}
    \label{fig:sshpmodes}
\end{figure}

\begin{figure}
    \centering
    \includegraphics[width=0.48\textwidth]{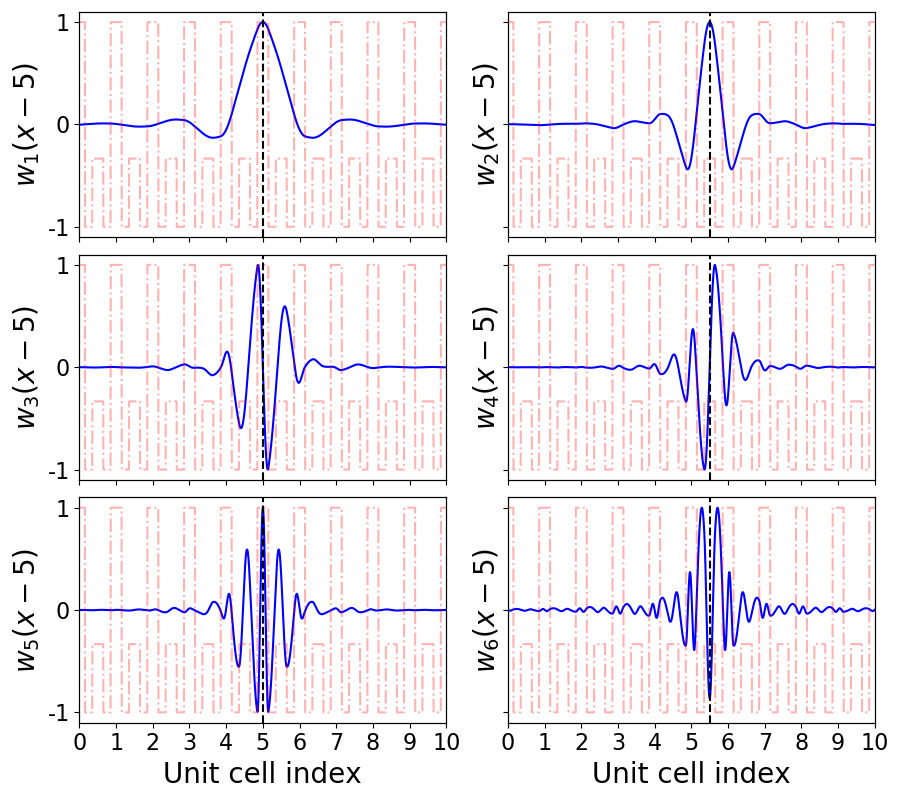}
    \caption{Maximally localized Wannier functions for the photonic SSH chain with $\varepsilon = 5$, $\Delta \varepsilon = 2$ and $b = 0.3$ (normalized to have the maximum value of 1). 
    The black dashed lines show the Wannier centers of the respective functions. 
    The red dashdot background curve is $\epsilon \pqty{x}$.}
    \label{fig:mlwfs}
\end{figure}

We begin by numerically solving the photonic SSH model using MPB, and computing the Berry phases for the bands. 
We focus first on the case where $\Delta\varepsilon > 0$. 
We show the electric field eigenmodes for the first six bands in Fig.~\ref{fig:sshpmodes}, from which the parity of the eigenstates at $k=0$ and $k=\pi/a^*$ (with $a^*=2a$ being the periodicity of the system in position space) can be seen.
The Berry phases and parities of the first six bands are shown in Table~\ref{tab:sshwfprops}. 
We see that the Berry phases alternate between $\pi$ and $0$, indicating that the MLWFs for each band are localized alternatively at $r=R+a^*/2$ and $r=R$. We can verify this by explicitly constructing the Wannier functions for the first four bands, as shown in Fig. \ref{fig:mlwfs}. 
As for the PhKP model in Sec.~\ref{sec:1dmodel}, we numerically construct the Wannier functions using the Wilson loop method in MPB. 
Finally, we can deduce the properties of states and Wannier functions for $\Delta\varepsilon<0$ by noting that changing the sign of $\Delta\varepsilon$ is equivalent to translating our model by half a lattice constant $a^*/2$, thus translating the Wannier functions in Fig.~\ref{fig:mlwfs} by $a^*/2$. 
Additionally, a half-lattice translation interchanges the two inversion centers in the position space unit cell. 
Hence the parity of Bloch states at $X$ is reversed relative to that shown in Table~\ref{tab:sshwfprops}.

We see then that the set of Wannier functions $\ket{nR}$ for the (generically) isolated bands of our photonic SSH model carry an elementary band representation of the line group $p\bar{1}1'$. 
We will follow the convention of \cite{cano2021band} to denote these band representations.
When $\Delta\varepsilon>0$ we see from Table \ref{tab:sshwfprops} that Wannier functions for bands $2$ and $6$ carry the band representation $(A\uparrow G)_{1a}$ corresponding to even-parity Wannier functions centered at the $1a$ Wykcoff position (origin of the unit cell). 
Band $4$ carries the $(B\uparrow G)_{1a}$ band represnetaiton, corresponding to an odd parity functions centered at the $1a$ position. 
Similarly, bands $1$ and $5$ carry the $(A\uparrow G)_{1b}$ band representation, corresponding to even-parity Wannier functions centered at the $1b$ Wyckoff position (the inversion center at $a^*/2$). 
Finally, band $3$ carries the $(B\uparrow G)_{1b}$ band representation, corresponding to odd-parity Wannier functions centered at the $1b$ position. 
Based on our analysis above, when $\Delta\varepsilon < 0$, each band still carries an elementary band representation, but centered at the opposite Wyckoff position ($1a\leftrightarrow 1b$).
 
To procede further, we would like to use our Wannier functions to obtain a quantitatively accurate effective model for the topological phase transition that occurs when $\Delta\varepsilon$ changes sign. 
To do so, we will focus on bands 2 and 3, which remain energetically isolated from the rest of the bands in the spectrum for small $|\Delta\varepsilon|$. 
We will show how we can derive such a model from our Wilson loop approach. 
Using both the tight-binding model and the full MPB calculation, we will study the properties of defect states that emerge at the boundary between distinct topological phases. 

\begin{table}[b]
    \centering
    \begin{tabular}{c|c|c|c}
        Band index& Parity at $(\Gamma, X)$ & Berry phase & MLWF parity  \\
        \hline
        1 & $(+,-)$ & $\pi$& $ +$ \\
        2 & $(+,+)$ & $0$ & $+$ \\
        3 & $(-,+)$ & $\pi$ & $-$ \\
        4 & $(-,-)$ & $0$ & $-$ \\
        5 & $(+,-)$ & $\pi$ & $+$ \\
        6 & $(+,+)$ & $0$ & $+$
    \end{tabular}
    \caption{Symmetry and topological properties of the first six bands of the photonic SSH model with $\varepsilon=5, b=0.3,$ and $\Delta\varepsilon=2$. 
    The first column gives the band index. 
    The second column gives the eigenvalue under inversion about the position space origin (parity) for the Bloch states at $\Gamma$ ($k=0$) and $X$ ($k=\pi/a$) in the Brillouin zone. 
    Third column gives the single-band Berry phase for each band. 
    The last column gives the parity of the Wannier functions constructed from the band, which coincides with the inversion eigenvalue of the Bloch states at $\Gamma$.}
    \label{tab:sshwfprops}
\end{table}

\section{Tight Binding Approximations for Photonic Systems}\label{sec:tb}

The exponentially-localized Wannier functions allow the construction of accurate tight-binding (TB) models for 1d photonic crystals just like in the case of electronic systems. 
We will first outline the general construction, after which we will apply it to our photonic SSH model. 
Assume we have a finite system with $N$ unit cells, with periodic boundary conditions. 
Let us suppose that for each band $m$ of interest we have constructed a set of MLWF's $\Bqty{\ket{m R}}$ where $R = 0, 1, \ldots, N - 1$.  Using these Wannier functions, we can write down an equivalent position space tight-binding model for the photonic system. 
Let $c_{mR}$ be the (bosonic) annihilation operator for a photon in the rescaled Wannier state $\ket{\widetilde{mR}}$. 
Then we can project the Hermitian Maxwell operator Eq.~\eqref{eq:maxwellops} into the space of Wannier functions to define the tight-binding operator
\begin{equation}\label{eq:tbdef}
    \mathcal{O}_E^\mathrm{tb} = \sum_{m,n,R_1,R_2}\bra{\widetilde{m R_2}}\mathcal{O}_E\ket{\widetilde{n R_1}}c^\dag_{mR_2}c_{nR_1}
\end{equation}
We can re-express this tight-binding model in momentum space by introducing the Bloch-like functions
\beq
    \ket{E^\prime_{m k}} = \sum_{R} e^{i k (R+r_m)} \ket{\widetilde{m R}},
\eeq
where $r_m=a\theta_m/(2\pi)$ is the center of the Wannier function $\ket{mR}$.  Defining the associated Fourier-transformed second-quantized operators
\begin{equation}\label{eq:optransform}
    c_{nk} = \sum_{R} e^{-ik(R+r_n)}c_{mR},
\end{equation}
we can write
\begin{equation}
    \mathcal{O}_E^\mathrm{tb} = \sum_k o^{mn}_E(k)c^\dag_{mk}c_{nk},
\end{equation}
where the matrix $o_E(k)$ is the photonic equivalent of a tight-binding Bloch ``Hamiltonian" for the bands of interest and is given by
\begin{align}
   o^{mn}_E(k)&=\mel{E^\prime_{m k}}{\mathcal{O}_E}{E^\prime_{n k}}\nonumber \\
   &= \frac{1}{N} \sum_{R_1, R_2} e^{i k \qty{R_1 - R_2 + r_n-r_m}} \mel{\widetilde{m R_2}}{\mathcal{O}_E}{\widetilde{n R_1}}.
\end{align}
The eigenvalues of $o_E(k)$ coincide with the eigenfrequencies $\omega^2(k)/c^2$ of the full Maxwell operator restricted to the bands of interest, and the eigenvectors of $o_E(k)$ are the expansion coefficients of the eigenstates of $\mathcal{O}_E$ in the basis of Wannier functions. 
Note that we have included the Wannier function centers explicitly in our convention for the Fourier transform. 
This ensures that our tight-binding operator satisfies the Brillouin zone boundary conditions commonly used in the study of electronic systems\cite{vanderbilt2018berry}. 
The eigenfunctions of the tight-binding operator $o_E(k)$ then allow us to reconstruct the cell-periodic functions $\widetilde{v}_{nk}$. 
For this choice of boundary conditions, the Berry phase computed from the tight-binding eigenfunctions can be related to (a contribution to\cite{alexandradinata2014wilsonloop}) the matrix elements of the position operator.

We thus see that we can use our Wilson-loop approach to numerically construct Wannier functions in order to derive quantitatively accurate tight-binding models for 1D photonic crystals. 
By truncating the set of matrix elements retained in Eq~\eqref{eq:tbdef} to retain only those with less than a fixed distance between $R_1$ and $R_2$, we can construct systematic approximations to the Hermitian Maxwell operator and its eigenfunctions for a small set of bands. 
There are several advantages of studying topological photonic crystals using the quantitative tight-binding approach that we will explore here. 
First, short-ranged tight-binding models allow us to make analytical statements about topological indices and invariants that may be tedious to compute from full photonic field simulations. 
Additionally, by studying the dependence of tight-binding matrix elements on model parameters, we will be able to more efficiently control the topological properties of photonic crystals. 
Lastly, we will see that the tight-binding approach will allow us to compute localization properties of topological defect modes using only the results of bulk photonic simulations.

As a proof of concept, we will construct and analyze a tight-binding model for our photonic SSH chain. 
We focus on the third and fourth bands, which remain well-separated from other bands over a wide range of $\Delta\varepsilon$. 
Each unit cell has a Wannier state A corresponding to band 3 centered at $x = R$ and a Wannier state B corresponding to band 4 centered at $x = R+a^*/2$. 
We use non-abelian Wilson-loop to construct Wannier functions for the two bands, which allows us to study models in which the gap between the two bands goes to zero. 
Truncating to hopping between adjacent unit cells only, the photonic tight-binding operator takes the general form
\begin{widetext}
\begin{multline}
    \mathcal{O}^{\mathrm{tb}}_E = \frac{1}{a^{*2}} \sum_{R = 0}^{N - 1} [\frac{1}{2} t_0^{A A} c^\dagger_{A R} c_{A R} + \frac{1}{2} t_0^{B B} c^\dagger_{B R} c_{B R} + t_0^{A B} c^\dagger_{A R} c_{B R} + t_1^{A A} c^\dagger_{A R - 1} c_{A R} \\
        + t_1^{B B} c^\dagger_{B R - 1} c_{B R} + t_1^{A B} c^\dagger_{A R - 1} c_{B R} + t_1^{B A} c^\dagger_{B R - 1} c_{A R} + \text{h.c.}]
\end{multline}
\end{widetext}
where $t_l^{\beta \alpha} \equiv a^{*2} \mel{\beta R - l}{\mathcal{O}_E}{\alpha R}$ is the ``hopping amplitude'' connecting the Wannier state $\ket{\alpha R}$ to the Wannier state $\ket{\beta R - l}$ and $c^\dagger_{\alpha R}$ ($c_{\alpha R}$) is the creation (annihilation) operator associated with the state $\ket{\alpha R}$. Here, we have pulled out an overall factor of $a^{*2}$ to make the hopping amplitudes dimensionless.
Using the Fourier transforms Eq.~\eqref{eq:optransform} of the creation and annihilation operators, we can write
\beq
    \mathcal{O}^{tb}_E = \sum_{k, \alpha, \beta} c^\dagger_{\alpha k} o^{\alpha\beta}_E \pqty{k} c_{\beta k}
\eeq
where the Bloch matrix $o_E(k)$ is given by
\begin{widetext}
\beq\label{eq:sshblochop}
    o_E \pqty{k} \equiv \frac{1}{a^{*2}}
        \mqty(t_0^{A A} + 2 t_1^{A A} \cos \pqty{k a^*} & t_0^{A B} e^{i k a^*/2} + t_1^{A B} e^{-i k a^*/2} + t_1^{B A} e^{i 3 k a^*/2} \\
        t_0^{A B} e^{-i k a^*/2} + t_1^{A B} e^{i k a^*/2} + t_1^{B A} e^{-i 3 k a^*/2} & t_0^{B B} + 2 t_1^{B B} \cos \pqty{k a^*})
\eeq
\end{widetext}

The properties of the photonic SSH model place several constraints on the hopping parameters in Eq.~\eqref{eq:sshblochop}. 
First, $t_0^{A B} = t_1^{A B}$ due to inversion symmetry. 
Second, $t_1^{A A}$ and $t_1^{B B}$ represent next-nearest neighbor hoppings and are small compared to $t_0^{A A}$ and $t_0^{B B}$ respectively, provided the gap separating bands 3 and 4 from bands 2 and 5 remains large. 
Provided this gap is large, $t_1^{B A}$ will also be small compared to $t_0^{A B}$. 
We can thus, as a first approximation, set $t^{AA}_1=t^{BB}_1=t^{BA}_1=0$. 
The Bloch matrix then takes the simplified form
\beq \label{eq:Bloch_operator}
    o_E \pqty{k} \equiv \frac{1}{a^{*2}} \pqty{\Delta t_0 \tau_z + 2 t_0^{A B} \cos \pqty{k a^* / 2} \tau_x + \overline{t_0} \tau_0},
\eeq
where $\Delta t_0 \equiv \frac{t_0^{A A} - t_0^{B B}}{2}$, $\overline{t_0} \equiv \frac{t_0^{A A} + t_0^{B B}}{2}$, $\tau_{x/z}$ are Pauli matrices acting in the space of Wannier functions, and $\tau_0$ is the $2\times 2$ identity matrix. 
The eigenfrequencies for our tight-binding model are then obtained by diagonalizing $o_E(k)$, yielding
\beq
    \pqty{\omega / c}^2 = \frac{1}{a^{*2}} \pqty{\overline{t_0} \pm \sqrt{\pqty{\Delta t_0}^2 + 4 \pqty{t_0^{A B}}^2 \cos^2 \pqty{k a^*/2}}}.
\eeq

\begin{figure}
    \subfloat[]{
        \includegraphics[width=0.22\textwidth]{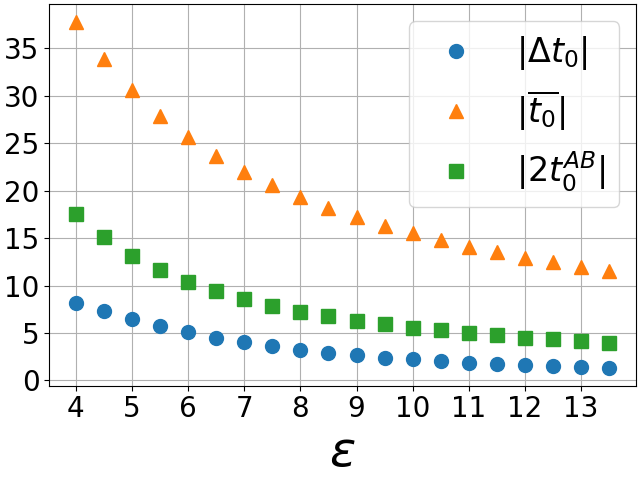}
}
\subfloat[]{
        \includegraphics[width=0.22\textwidth]{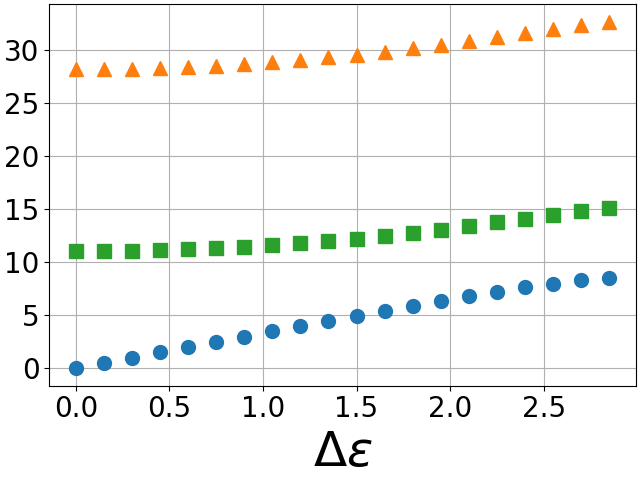}
}
    \caption{(a) Hopping terms vs $\varepsilon$ with $\Delta\varepsilon=2$ and $b=0.3$. (b) Hopping terms vs $\Delta \varepsilon$ with $\varepsilon=5$ and $b=0.3$. $\Delta t_0$ linearly goes to 0 with $\Delta\varepsilon$ and other parameters are flat (Both plots share the same legend)}
    \label{fig:hopping_terms}
\end{figure}

In Fig. \ref{fig:hopping_terms} we show the hopping parameters computed in MPB as a function of $\varepsilon$ and $\Delta\varepsilon$, with $b=0.3$. 
Since the spectrum and Wannier function centers do not depend on the sign of $t_0^{AB}$, we can choose a gauge to make $t_0^{AB}$ always positive. 
Thus, we plot only $|t_0^{AB}|$. 
Note that for $\varepsilon = 5$, there is an accidental degeneracy between bands 4 and 5 when $\Delta\varepsilon \approx 2.95$. 
We see that for small $\Delta\varepsilon,$ $\bar{t}_0$ and $t_0^{AB}$ are approximately constant, while $\Delta t_0$ depends approximately linearly on $\Delta\varepsilon$. 
Furthermore, $\Delta t_0\rightarrow 0$ as $\Delta\varepsilon\rightarrow 0$, indicating that the gap between bands 3 and 4 closes; this is consistent with the fact that when $\Delta\varepsilon=0$, the SSH model has an additional half-translation symmetry. 
We can see from Eq.~\ref{eq:Bloch_operator} that $\Delta t_0$ plays the role of the on-site mass term in the photonic SSH model; a change in sign of $\Delta t_0$ reflects the topological phase transition of the SSH model. 

As $\Delta t_0$ goes from positive to zero and then negative, alternate band gaps at the $X$ point become smaller, vanish (as shown in Fig.~\ref{fig:band_diagrams}) and then again become larger. 
This leads to band inversions between bands $2n-1$ and $2n$ ($n=0,1,\dots$). 
As we discussed in Sec.~\ref{sec:1dmodel}, the system with $\Delta t_0<0$ differs from the system with $\Delta t_0>0$ (with the same absolute value) by a shift of half a unit cell. 
Correspondingly, the Wannier centers for band 3 are located at $R + a^*/2$ for $\Delta t_0>0$ (as can be seen from Tab~\ref{tab:sshwfprops}), and shift to $R$ when $\Delta t_0 < 0$.

This can be quickly seen from the tight-binding Hamiltonian, by noting that when $\Delta t_0 \neq 0 $, we can adiabatically deform the model to the flat-band limit $t_0^{AB}=0$ without closing a gap. 
In this limit the Bloch eigenfunctions coincide with the Fourier transformed Wannier functions. 
Since the adiabatic deformation preserves inversion symmetry, it cannot change the center of localization of the Wannier functions; we can thus read off that when $\Delta t_0 >0 (<0)$ the Wannier centers (and hence the tight-binding Berry phase) for band 3 are given by $r_3 = R+ a^*/2$ $(R)$.

\begin{figure}
    \centering
    \includegraphics[width=0.48\textwidth]{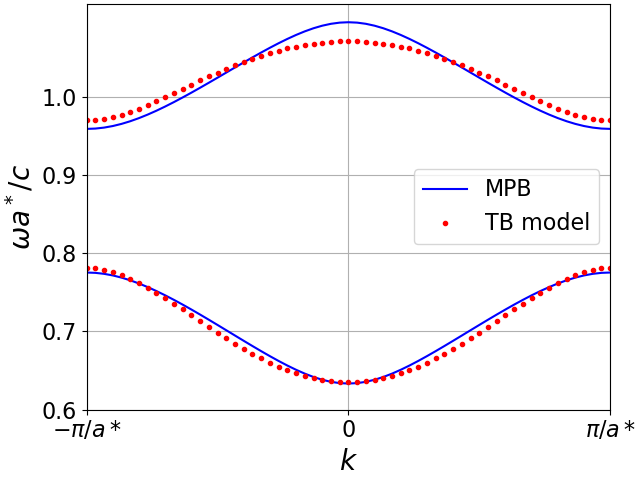}
    \caption{Comparison of the nearest-neighbor tight-binding spectrum with the photonic SSH band structure from MPB, for $\varepsilon=5,\Delta\varepsilon=2$, and $b=0.3$. 
    Frequencies for the third and fourth bands obtained from MPB are shown in solid blue; eigenvalues of the $2\times2$ nearest-neighbor tight-binding model are shown in dotted red.}
    \label{fig:TB_vs_MPB}
\end{figure}

In what follows, we will focus on the particular values $\varepsilon=5$, $|\Delta\varepsilon|=2$. 
In Fig.~\ref{fig:TB_vs_MPB} we show the spectrum of our tight-binding Bloch matrix for bands 3 and 4 compared with the exact spectrum computed from MPB. 
We see that our nearest-neighbor tight-binding model provides a quantitatively accurate spectrum, correct to within $\approx 10\%$. 
We can systematically improve the accuracy of the model by including more distant hoppings, which we show in Appendix~\ref{app:more_hoppings}. 
We will now analyze the boundary that arises between the topologically distinct phases with $\Delta\varepsilon=\pm 2$. 
We will see that the nearest neighbor model is sufficient to develop a quantitative understanding of modes trapped at the boundary.

\section{Defect bound states and the Jackiw-Rebbi model}

\begin{figure*}
    \centering
    \includegraphics[width=\textwidth]{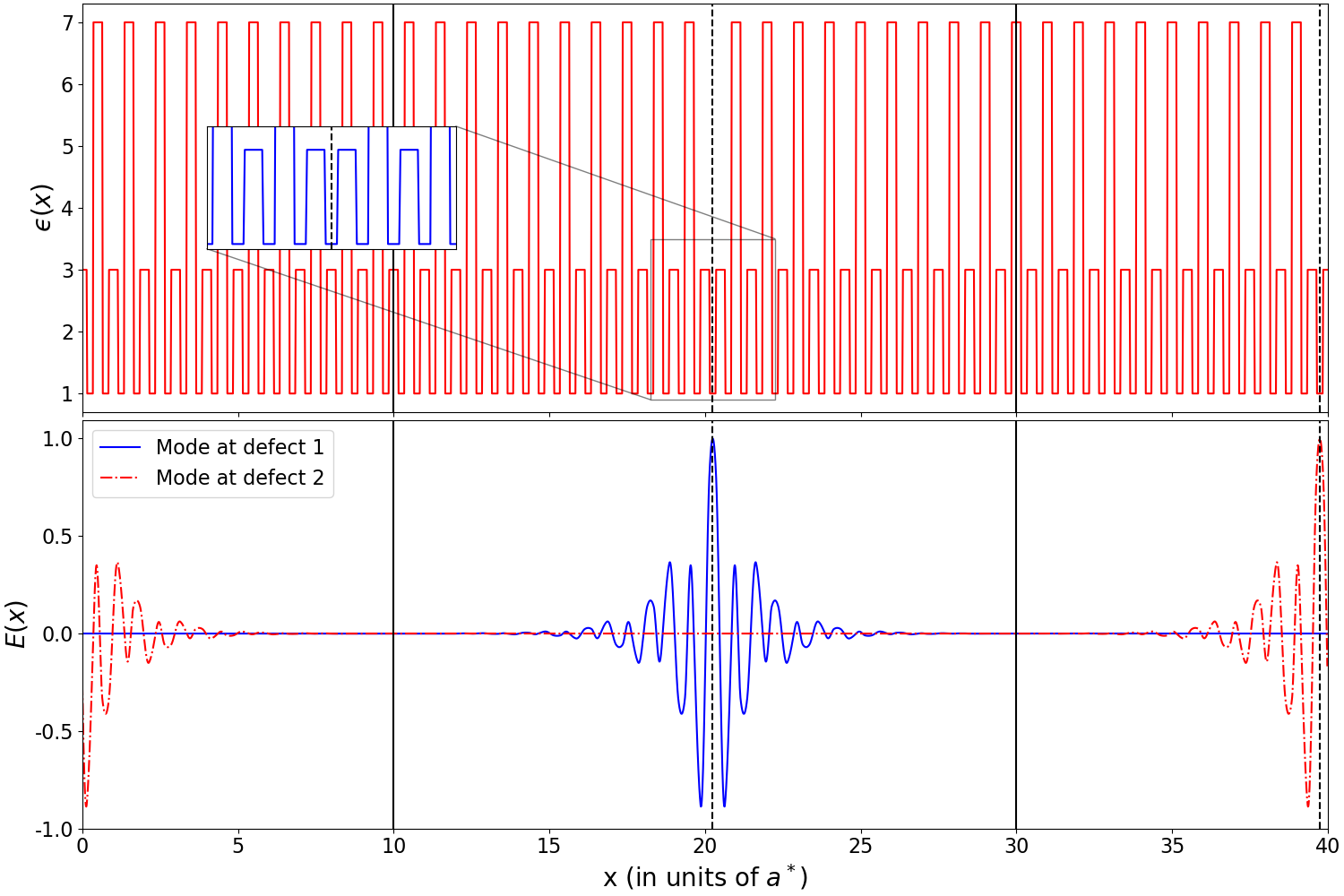}
    \caption{Domain wall states (normalized to have the maximum value of 1) in the photonic SSH model. 
    Top shows the dielectric function for a periodic photonic SSH chain with two domain walls at $x=0$ and $x=20$. 
    The inset shows a closeup of the domain wall, represented here as the absence of a slab of dielectric constant $\varepsilon+|\Delta\varepsilon|$. 
    The domain walls are thus the boundary between two systems with $\varepsilon=5, b=0.3$, and $\Delta\varepsilon=\pm2$. 
    The dashed vertical lines show the positions of the defects, while the solid vertical lines show two inversion centers of the finite size system. 
    Note, crucially, that inversion symmetry about these lines interchanges the defects. 
    Bottom panel shows the electric field eigenmodes for two states (blue solid and red dashdot) with eigenfrequency in the bandgap between bands 3 and 4. 
    The eigenmodes have been normalized such that their maximum value is $1$.  
    We see that there is a single state exponentially localized near each defect.}
    \label{fig:interface}
\end{figure*}

Let us consider a ``phase-slip'' defect in our photonic SSH model, where  $\Delta\varepsilon$ abruptly changes sign. 
Physically, this corresponds to an interface between two identical photonic SSH systems with a relative half unit cell shift between them. 
In any system with periodic boundary conditions, these phase slip defects must come in pairs. 
In Fig. \ref{fig:interface} we show the dielectric function for a pair of defects; We see that physically the defects manifest as a ``missing'' $\varepsilon+|\Delta\varepsilon|$ dielectric slab. 
In this system, as we move from left to right $\Delta\varepsilon$ changes the sign at the interface. 
From our tight-binding mapping Eq.~\eqref{eq:sshblochop}, this corresponds to a change in the sign of the mass $\Delta t_0$ at the defect. 
Thus, we expect to find well-localized mid-gap modes at each defect, in analogy with solitons in the electronic SSH model. 
Note that similar models were considered in Refs.~\cite{henriques2020topological,jiang2021topological,tan2014photonic,slobozhanyuk2015subwavelength}, although those works analyzed defect states with different boundary conditions, using a different set of tools.
 
\begin{figure*}
    \centering
    \includegraphics[width=\textwidth]{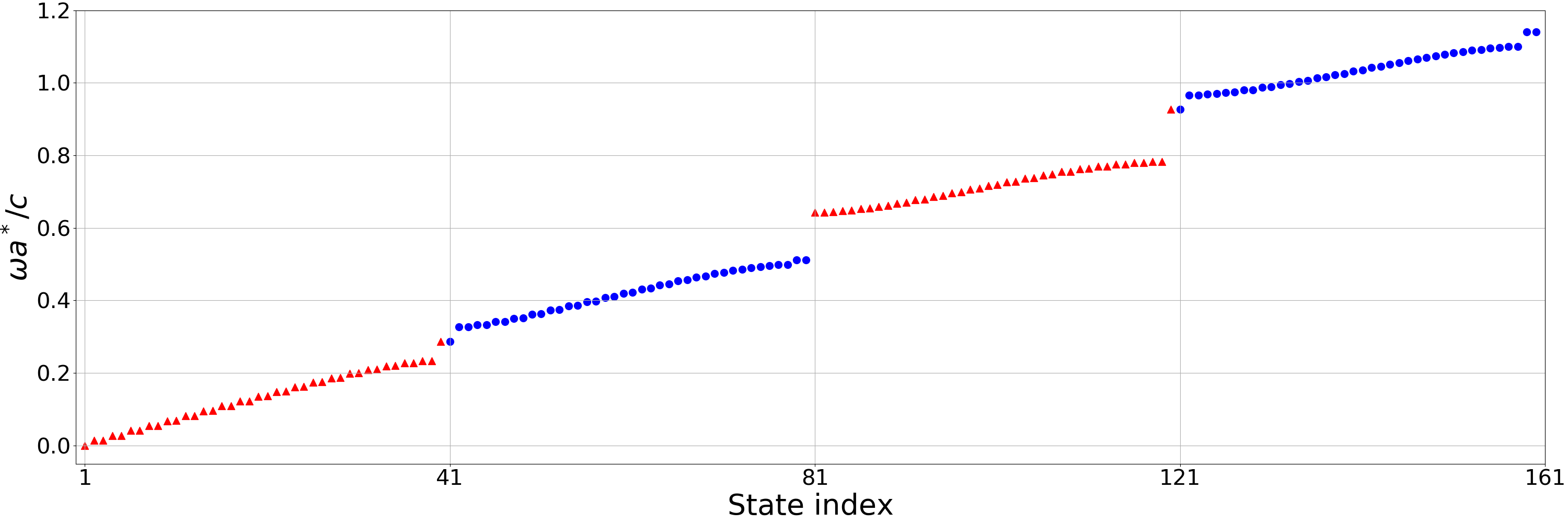}
    \caption{Spectrum for the finite size photonic SSH model with 40 unit cells, and two inversion-symmetric domain walls. 
    The parameters of the model are $\varepsilon=5, b=0.3,$ and $|\Delta\varepsilon|=2$, where the sign of $\Delta\varepsilon$ flips at each domain wall. 
    Groups of 40 states are colored alternatively with red triangles and blue circles, corresponding to states originating from each isolated bulk band. 
    We see that the gap between bands 1 and 2, and the gap between bands 3 and 4 feature a pairs of midgap states. 
    These states are filling anomalous, in the sense that for frequencies $\mu$ in the gap between bands 3 and 4, there are $N= \pm 1\mod 40$ states with frequency less than $\mu$.}
    \label{fig:filling_anomaly}
\end{figure*}

By solving Maxwell's equations numerically in MPB for a large supercell, we can verify the existence of these defect modes. 
We consider a system with 20 unit cells with $\Delta\varepsilon >0$, 20 unit cells with $\Delta\varepsilon <0$, two phase slip defects, and periodic boundary conditions. 
The number of unit cells is chosen to ensure that the finite system has inversion symmetry, and that inversion acts to interchange the two defects. 
In Fig.~\ref{fig:filling_anomaly} we show the eigenfrequencies for the lowest four bands of states in this finite system. 
We denote by red triangles the groups of 40 states originating from the first and third bulk bands, while we denote by blue circles the groups of 40 states originating from the second and fourth bulk bands. 
We find a pair of midgap states between bands one and two, and a second pair of midgap states between bands three and four. 
As the frequencies of these states correspond to regions of the bulk gap, we deduce that they must be states bound to the defects. 
We verify this directly by plotting in Fig.~\ref{fig:interface} the eigenmodes for the two states in the gap between bands three and four.

We now argue that the defect bound states in our photonic SSH model are of topological origin. 
We will show this in two ways, first using our numerical results, and then analytically from our tight-binding model. 
Recall from Fig.~\ref{fig:filling_anomaly} that for the pairs of defect modes, by counting alone one mode must have originated from the lower frequency band and one mode must have originated from the upper frequency band. 
Since the two modes are related by inversion symmetry, however, this cannot be - instead each defect mode consists of a superposition of states originating in the lower and upper frequency band. 
This represents a photonic realization of the ``filling anomaly'' from condensed matter physics\cite{niemi1986fermion,benalcazar2019quantization,fang2021filling,wieder2020strong,he2020quadrupole,proctor2020robustness}. 
In a system with $N$ unit cells and no defects, if we ``fill'' $m$ bands, then we would have $m\times N$ photons in the system. 
Here, however, we see that if we fill $N$ bands we will have either $m\times N -1$ or $m\times N +1$ photons. 
This $\mathcal{O}(1)$ correction is the filling anomaly, and is directly related to the change of bulk Berry phase between the two phases on either side of the defect.

Just as in the condensed matter context, the filling anomaly and defect states can be studied analytically using a Jackiw-Rebbi approach starting from our tight-binding model\cite{jackiw1976solitons}. 
Let us focus on bands three and four, so that we can make use of our nearest-neighbor tight-binding model from Eq.~\eqref{eq:Bloch_operator}. 
Each defect can be represented by a region where $\Delta\varepsilon$ changes sign, so we focus on small $|\Delta\varepsilon|$. 
In this limit, the gap between bands three and four near $k=\pi/a^*$ is small. 
We can linearize the tight-binding operator Eq. \eqref{eq:Bloch_operator} about $k=\pi/a^*$ to find
\beq
    o_E \pqty{k} = \frac{1}{a^{*2}} \Delta t_0 \tau_z - \frac{1}{a^*} t_0^{A B} k \tau_x
\eeq
where we have ignored the term proportional to $\tau_0$ as it merely provides a constant shift to the spectrum. 
We want to consider a situation where $\Delta\varepsilon\rightarrow\Delta\varepsilon(x)$ becomes a position dependent function. 
From our fit for the tight-binding parameters in Fig.~\ref{fig:hopping_terms}, we see that $\Delta t_0\propto\Delta\varepsilon$ and $t_0^{AB}\propto (\Delta\varepsilon)^0$ for small $\Delta\varepsilon$. 
We thus allow $\Delta t_0\rightarrow\Delta t_0(x)$ to depend on position to model the defect. 
Since this breaks translation symmetry, we must go to position space and replace $k$ by the operator $-i\pdv{x}$. 
We thus look for solutions $\psi \pqty{x}$ to the envelope function approximation (EFA) equation
\beq
    o_E^{\text{(EFA)}} \psi \pqty{x} = 0
\eeq
where $o_E^{\text{(EFA)}} \equiv \frac{1}{a^{*2}} \Delta t_0(x) \tau_z + i \frac{1}{a^*} t_0^{AB} \tau_x \pdv{x}$ is the EFA operator. 
We take $\Delta t_0(x\rightarrow-\infty)=-|\Delta t_0|$, $\Delta t_0(x\rightarrow\infty)>+|\Delta t_0|$, and $\Delta t_0(0)=0$ to model our defect at $x=0$. 
The EFA equation has a zero-mode solution in this case, given by
\beq 
    \psi \pqty{x} = \psi \pqty{0} \exp \pqty{- \int_0^x \dd x^\prime  \frac{|\Delta t_0(x)|}{a^* t_0^{A B}}} \mqty(1 & i)^{\mathrm{T}}.\label{eq:efasol}
\eeq

\begin{figure}[htbp]
    \centering
    \includegraphics[width=0.48\textwidth]{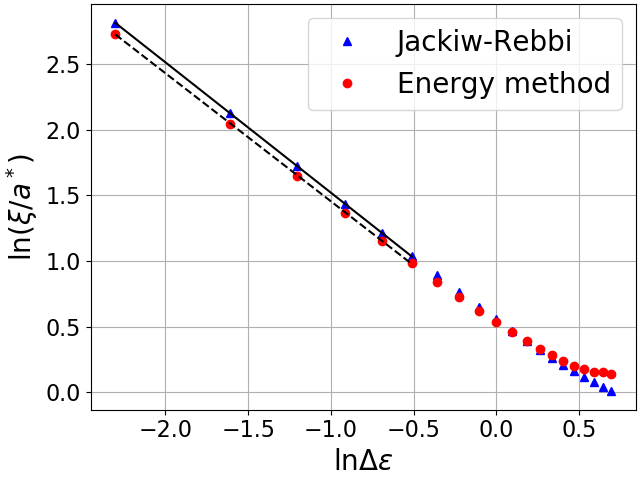}
    \caption{Comparison of the decay lengths obtained directly from MPB simulations using the energy method (as discussed in Sec.~\ref{sec:construction}) and the Jackiw-Rebbi model. 
    The solid and the dashed black lines are linear fits for the Jackiw-Rebbiw method and the MPB simulations with the slopes being $\nu_{\mathrm{JR}}= 0.975 \pm 0.004$ and $\nu_\mathrm{MPB}= 0.992 \pm 0.002$ respectively which agree to within 2\%.}
    \label{fig:model_comparison}
\end{figure}

In the simple situation which we model in MPB, $\Delta t_0(x)$ is approximately a step function; in this case, our solution Eq.~\eqref{eq:efasol} becomes
\begin{equation}
    \psi \pqty{x} \rightarrow \psi \pqty{0} \exp \pqty{-\frac{|x|}{\xi}} \mqty(1 & i)^{\mathrm{T}},\label{eq:efasolstep}
\end{equation}
where the decay length is 
\beq\label{eq:xiexpr}
\xi = a^* t_0^{A B} / |\Delta t_0|.
\eeq 
This thus gives our mode bound to the defect We see from the vector structure of $\psi\pqty{x}$ that it is an equal superposition of states from bands 3 and 4, in line with our arguments based on the filling anomaly. 
The nearest-neighbor tight-binding model has an emergent chiral symmetry which pins the frequency of the bound mode to be exactly at the middle of the gap (taken to be $0$ here). 
This is not a symmetry of the real system, and will be broken by next-nearest neighbor terms in the tight-binding model. 
However, for an inversion-symmetry related pair of defects, the corresponding pair of defect modes must always have the same frequency. 
This explains the filling anomaly - chiral symmetry breaking pushes the frequency of both defect modes in the same direction.

Since $|\Delta t_0|$ and $t_0^{AB}$ are \emph{bulk} tight-binding parameters, The Jackiw-Rebbi estimate Eq.~\eqref{eq:xiexpr} allows us to estimate the localization length of defect bound states using only infinite-system calculations. 
In Fig. \ref{fig:model_comparison}, we compare the Jackiw-Rebbi estimate of the defect mode decay length to the exact decay length obtained from a much more computationally intensive MPB supercell calculation, as a function of $|\Delta\varepsilon|$.
The decay length is extracted from the MPB eigenfunction by computing the average energy per unit cell, as in Sec.~\ref{sec:PhKP}.  
We see that the decay length extracted from the Jackiw-Rebbi analysis agrees to within $10\%$ with the decay length extracted from exact diagonalization in MPB.  
The small discrepancy between the Jackiw-Rebbi analysis and the exact decay length come from several sources.

First, we know from Fig.~\ref{fig:TB_vs_MPB} and our discussion in Sec.~\ref{sec:tb} that our nearest neighbor tight-binding model matches the MPB spectrum to within about $10\%$. 
These deviations are due to next-nearest-neighbor and longer hoppings that we have neglected to obtain our simple model. 
This can also be seen in Fig.~\ref{fig:filling_anomaly}, where the presence of non-negligible next-nearest-neighbor hopping matrix elements shifts the defect modes away from the center of the bulk band gap (i.~e., these hopping matrix elements break chiral symmetry). 
For large $|\Delta\varepsilon|$, we see that the Jackiw-Rebbi analysis breaks down, as expected for a perturbative treatment. 
For small $|\Delta\varepsilon|$, we note that the longer-range hopping terms remain non-negligible compared to the nearest-neighbor hopping as $|\Delta\varepsilon|\rightarrow 0$, explaining the discrepancies between the decay lengths in this regime.

 We can compare the scaling of the decay length with $\Delta\varepsilon$ by extracting the best-fit exponent $\xi\propto (\Delta\varepsilon)^{-\nu}$. 
 From the Jackiw-Rebbi analysis we find the decay length to be $\nu_{\mathrm{JR}}= 0.975 \pm 0.004$, while from MPB we find $\nu_\mathrm{MPB}= 0.992 \pm 0.002$. 
 We see that the Jackiw-Rebbi decay exponent agrees with the exact MPB exponent to within approximately $2\%$. 
 Thus, we see that even the coarse nearest-neighbor tight-binding approximation can quantitatively capture universal features of topological defect modes.

\section{Conclusions}
In this work, we reviewed how Wilson loop techniques can be applied to construct exponentially localized Wannier functions for one-dimensional photonic crystals. 
We then showed how the Wannier functions allow us to derive quantitatively accurate tight-binding models for any set of bands of a 1D photonic crystal separated from others by a frequency gap, in analogy to the tight-binding method in electronic systems. 
As an example of this method, we examined a photonic analog of the SSH chain, where the unit cell consists of dimerized regions of high and low dielectric constant. 
We showed that our Wannier- and tight-binding based methods can efficiently extract topological information about the model. 
First, we saw that the Berry phase of a band coincided with the center of the Wannier functions in the unit cell. 
Second, we showed that our tight-binding method could be used to analyze topological defect states at the boundary between the two dimerizations of the 1D chain. 
By deriving an effective Jackiw-Rebbi model from the bulk tight-binding Hamiltonians, we could compute the universal scaling of the localization length of the defect mode. 
Finally, We showed how our Berry phase analysis allowed us to classify defect modes of topological and nontopological origin due to the presence of a filling anomaly.

While we are not the first to apply Wannier function based methods to photonic systems, we have shown that these methods allow us to map Maxwell's equations for photonic crystals to quantitatively accurate tight-binding models, from which topological information can be extracted. 
In particular, we can predict the presence of topological defect states and the associate filling anomaly from knowledge only of the bulk Wannier functions. 
Furthermore, we have shown that the universal contribution to the localization length of topological defect states can be computed only using knowledge of the bulk tight-binding parameters, via the mapping to a Jackiw-Rebbi problem. 

Going forward our work paves the way to study topologically nontrivial photonic crystals in two and three dimensions using Wannier function based methods. 
First, the algorithm we outlined here for finding 1D exponentially localized Wannier functions extends trivially to higher dimensions, allowing us to compute \emph{hybrid} Wannier functions. 
By considering a 2d interpolation between the two dimerized phases of our photonic SSH chain from the hybrid Wannier perspective, we arrive at a photonic generalization of the Thouless charge pumps\cite{thouless1983quantization,ozawa2019topological}. 
Beyond well-studied examples such as this, extending the ideas presented here to derive variationally maximally localized Wannier functions for 2D photonic crystals would allow for the extension of our tight-binding and Jackiw-Rebbi methods to design and study photonic higher-order topology. 
This would allow us to make contact with the continuum Jackiw-Rossi methods of Ref.~\cite{kovsata2021second},
providing quantitative support to the qualitative tight-binding models used to design and understand these systems. 

Finally, on the theoretical side, the introduction of Wannier function methods to the study of topological photonic crystals gives the final ingredient to a complete theory of photonic topological quantum chemistry. 
Using the methods outlined in this work, we are able to determine not just the symmetry indicators and Wannier centers for photonic bands, but also the full \emph{band representation} spanned by the set of Wannier functions in position space. 
We have seen how the inversion symmetry eigenvalues of the Bloch states in momentum space determine the centers of the Wannier functions, as well as their parity under inversion symmetry in position space. 
By extending these approaches to higher dimensions, the properties of localized defect, edge, and corner states can be explored from a position space, band representation perspective.

\begin{acknowledgments}
This material is based upon work supported by the Air Force Office of Scientific Research under award number FA9550-21-1-0131.
\end{acknowledgments}
\appendix

\section{Some properties of the energy function}
\label{app:energy_function}

In this appendix, we will show that the extrema of the energy function $f(\omega^2)$ for the photonic Kronig-Penney model occur at points $\omega_n^2$ satisfying three key properties:
\begin{align}
|f(\omega_n^2)|^2 &\ge 1, \label{eq:fproperty1}\\
f(0)&=1 \label{eq:fproperty2}\\
\left.\frac{df}{d\omega^2}\right|_{\omega=0} &<0.\label{eq:fproperty3}
\end{align}
The first condition ensures that the bands are analytic functions of $k$ except for isolated band touching points at the Brillouin zone edges where $|\cos ka|=1$. 
These touching points are unstable with respect to perturbations, and do not present an obstacle to finding exponentially localized Wannier functions provided we use the composite band formalism. 
The second condition ensures that the lowest band has a zero frequency state at $\omega=0$, as required by Maxwell's equations. 
Finally, the third condition ensures that this lowest band is analytic at $\omega^2=0$; this ensures that we can find an exponentially localized Wannier function for this band.

To prove Eqs.~(\ref{eq:fproperty1}-\ref{eq:fproperty3}), let us first rewrite $f(\omega^2)$ as
\begin{align}
f \pqty{\omega^2} &= \cos A\omega\cos B\omega -\cosh \gamma\sin A\omega\sin B\omega \\
&=\cos(A+B)\omega\cosh^2\frac{\gamma}{2} - \cos(A-B)\omega\sinh^2\frac{\gamma}{2}
\end{align}
where we have introduced the shorthand
\begin{align}
A&=1-b >0,\\
B&=b\sqrt{\varepsilon},\\
\gamma &= \frac{1}{4}\log\varepsilon.
\end{align}
We immediately verify that since $\cosh^2\gamma/2 - \sinh^2\gamma/2 =1$, that $f(0)=1$. 
This proves Eq.~\eqref{eq:fproperty2}. 
Next, note that
\begin{equation}
\frac{\partial f}{\partial\omega^2} = \frac{1}{2\omega}\left(
\sin(A-B)\omega\sinh^2\frac{\gamma}{2}-\sin(A+B)\omega\cosh^2\frac{\gamma}{2}
\right),\label{eq:fderiv}
\end{equation}
where the factor of $(2\omega)^{-1}$ comes from $d\omega/d\omega^2$. 
Taking the limit as $\omega\rightarrow 0$, we find
\begin{align}
\left.\frac{df}{d\omega^2}\right|_{\omega=0} &= \frac{1}{2}\left[
(A-B)\cosh^2\frac{\gamma}{2} -(A+B)\sinh^2\frac{\gamma}{2}
\right] \\
&=-(A+B\cosh\gamma) < 0.
\end{align}
This proves Eq.~\eqref{eq:fproperty3}. 
This means that when we take the inverse of $f$ to find the dispersion $\omega^2(k)$, the dispersion relation will be analytic as $\omega,k\rightarrow 0$. 
This ensures that we can find exponentially localized Wannier functions even for the lowest band.

Finally, we will prove Eq.~\eqref{eq:fproperty1}. 
To do so, let us note that if $f(\omega^2)<1=\cosh^2\gamma/2-\sinh^2\gamma/2$ then we find after using a half-angle identity that
\begin{equation} 
{\sin^2\left(\frac{A-B}{2}\omega\right)}\tanh^2\frac{\gamma}{2} < {\sin^2\left(\frac{A+B}{2}\omega\right)}.\label{eq:ineq1}
\end{equation}
Similarly, we find that if $f(\omega^2)>-1$ then
\begin{equation}
{\cos^2\left(\frac{A-B}{2}\omega\right)}\tanh^2\frac{\gamma}{2} < {\cos^2\left(\frac{A+B}{2}\omega\right)}.\label{eq:ineq2}
\end{equation}
Let us then assume there exists $\omega_n$ such that $df/d\omega^2|_{\omega=\omega_n}=0$ while $|f(\omega_n^2)|<1$. 
Setting Eq.~\eqref{eq:fderiv} equal to zero gives
\begin{equation}
\sin(A-B)\omega_n\sinh^2\frac{\gamma}{2} = \sin(A+B)\omega_n\cosh^2\frac{\gamma}{2},
\end{equation}
which we can rearrange as
\begin{align}
\sin&\left(\frac{A-B}{2}\omega_n\right)\cos\left(\frac{A-B}{2}\omega_n\right)\tanh^2\frac{\gamma}{2}\nonumber \\
&=\sin\left(\frac{A+B}{2}\omega_n\right)\cos\left(\frac{A+B}{2}\omega_n\right).\label{eq:zerocondition}
\end{align}
Additionally, our assumption that $|f(\omega_n^2)|<1$ implies that the inequalities \eqref{eq:ineq1} and \eqref{eq:ineq2} are both simultaneously satisfied. 
Squaring Eq.~(\ref{eq:zerocondition}) and substiuting in the two inequalities yields
\begin{align}
\sin^2&\left(\frac{A+B}{2}\omega_n\right)\cos^2\left(\frac{A+B}{2}\omega_n \right)\nonumber \\
&= \sin^2\left(\frac{A-B}{2}\omega_n\right)\cos^2\left(\frac{A-B}{2}\omega_n\right)\tanh^4\frac{\gamma}{2} \nonumber \\
&< \sin^2\left(\frac{A+B}{2}\omega_n\right)\cos^2\left(\frac{A+B}{2}\omega_n\right),
\end{align}
which is a contradiction. 
Thus, we conclude that $|f(\omega_n^2)|\ge 1$. 
Whenever $|f(\omega_n^2)|> 1$, the branch points of $f^{-1}$ will occur in the complex $k$ plane, and bands will be analytic. 
The points where $|f(\omega_n^2)|=1$, on the other hand, correspond to points at which bands touch at one of the two inversion-symmetric momenta $k=0$ or $k=\pi$. 
These are accidental band touching points, which can arise in our model if $a,b$ and $\sqrt{\varepsilon}$ are all commensurate. 
While these degeneracies lead to a failure of the single-band formulas for computing Wannier functions, we can still use the composite band formulas to obtain exponentially localized Wannier functions for the degenerate bands\cite{cornean2016construction}.

Finally, let us examine the properties of $f(\omega^2)$ for $\omega^2 < 0$. 
In particular, we will show that $df/d\omega^2$ is a monotonically decreasing function for $\omega^2 < 0$. 
The consequence of this is that all extrema of $f(\omega^2)$--and hence all branch points of $f^{-1}$--occur for positive (physical) values of $\omega^2$. 
Since the distance of the branch points of $f^{-1}$ from the physical strip in momentum space determines the decay length of the Wannier functions for a band, the monotonicity of $df/d\omega^2$ at negative argument ensures that the decay length of the lowest band Wannier function is not determined by properties of $f$ in the unphysical negative region.

To prove this assertion, let us return to Eq.~\eqref{eq:fderiv} for the derivative of $f$. 
Let us take $\omega^2=-x^2$ for $x>0$. 
Then $\omega=\pm ix$, where the sign is determined by our choice of branch for the square root. 
Crucially, we have that
\begin{align}
    \frac{\sin q\sqrt{-x^2}}{\sqrt{-x^2}} &= \frac{\pm\sin iqx}{\pm i x} \\
    &=\frac{\sinh{qx}}{x}
\end{align}
regardless of our branch choice for the square root. 
Substiuting this into Eq.~\eqref{eq:fderiv} we have
\begin{align}
\left.\frac{d f}{d\omega^2}\right|&_{\omega^2=-x^2}\nonumber \\
&= \frac{1}{2x}\left(
\sinh(A-B)x\sinh^2\frac{\gamma}{2}-\sin(A+B)x\cosh^2\frac{\gamma}{2}\right)
\end{align}
Now since $A,B$ and $\gamma$ are larger than zero, the term in brackets above is always less than zero. 
Thus we have
\begin{equation}
    \left.\frac{d f}{d\omega^2}\right|_{\omega^2=-x^2} < 0
\end{equation}
as required.

\section{Kronig-Penney model and comparison of the energy functions}

The Kronig-Penney model in solid-state physics is a toy-model to demonstrate the existence of bands and band gaps in 1D crystals. 
The Hamiltonian for this model is given as\cite{griffiths2016introduction} 
\beq
    H = \frac{p^2}{2 m} + V \pqty{x}        
\eeq
where $V \pqty{x} = V \pqty{x + a}$ and within the first unit cell spanning $x \in [0, a)$, we have
\beq
    V \pqty{x} =
            \begin{cases}
                V_0   &  0 \leq x \leq b \\
                0     &  b \leq x < a
            \end{cases}
\eeq    

Bloch wave solutions for this model with crystal momentum $k$ and energy $E$ exist if $\cos \pqty{k b} = f \pqty{E}$ where the energy function $f \pqty{E} \equiv \cos \pqty{\pqty{a - b} \alpha} \cos \pqty{b \beta} - \frac{\alpha^2 + \beta^2}{2 \alpha \beta} \sin \pqty{\pqty{a - b} \alpha} \sin \pqty{b \beta}$ with $\alpha \equiv \sqrt{2 m E} / \hbar$ and $\beta \equiv \sqrt{2 m \pqty{E + V_0}} / \hbar$. 
The energy function is analytic in $E$ and satisfies the following properties \cite{kohn1959analytic}
\beq
\bal
    \lim_{E \rightarrow - \infty} f \pqty{E} &= \infty \\
    \lim_{E \rightarrow \infty} \pqty{f \pqty{E} - \cos \pqty{\sqrt{E} b}} &= 0 \\
    \dv{f \pqty{E}}{E} = 0 \text{ at } E = E_n
\eal
\eeq
where $E_n$ alternate in sign and $\abs{f \pqty{E_n}} \geq 1$.

The energy function for the photonic Kronig-Penney model $f \pqty{\omega^2}$ shares almost all of these features except the $\omega^2 \rightarrow \infty$ limit. 
When we substitute $\alpha = \sqrt{\omega^2}$ and $\beta = \sqrt{\varepsilon \omega^2}$ in $f \pqty{\omega^2}$, we get
\beq
\bal
f \pqty{\omega^2} = &\cos \pqty{\pqty{{a-b}} \sqrt{\omega^2}} \cos \pqty{b \sqrt{\varepsilon \omega^2}} \\ 
&- \frac{1 + \varepsilon}{2 \sqrt{\varepsilon}} \sin \pqty{\pqty{a-b} \sqrt{\omega^2}} \sin \pqty{b \sqrt{\varepsilon \omega^2}}.
\eal
\eeq
The coefficient of the second term is independent of $\omega^2$ which means that the even in the high frequency limit, there are band gaps in the photonic Kronig-Penney model. 
This doesn't limit us from constructing MLFWs for the finite frequency bands.

\section{Construction of Wannier Functions in MPB}
\label{app:numerical_MLWF}

In MPB, we first create a 1D mesh in real space by specifying the mesh resolution $\Delta x = a / N_1$ such that $x_n = n \Delta x$ ($n = 0, 1, \cdots, N_1-1$). 
Then we also create a 1D mesh spanning the Brillouin zone with a mesh resolution $\Delta k = 2 \pi / (a N_2)$ such that $k_m = -\pi/a + m \Delta k$ ($m = 0, 1, \cdots, N_2-1$). 
After solving for the cell-periodic functions, we first set $v_{00} \pqty{x} = 1 / \sqrt{\mathcal{N}}$ where $\mathcal{N} \equiv \sum^{N_1 - 1}_{n = 0} \epsilon \pqty{x_n} \Delta x$. 
This is to satisfy the normalization convention followed by MPB where the fields are normalized in the unit cell. 
MPB can make the fields as real as possible and we choose to do so. 
Finally, we manually set the phases of the cell-periodic fields such that the Bloch waves satisfy the condition $E_{k=2 \pi/a } \pqty{x} = E_{k=0} \pqty{x}$.  

We then use the composite bands formalism to find the Wannier functions for the bands under consideration. 
First we find the overlap matrix defined as $S_{ij} \pqty{k_1, k_2} = \braket{\widetilde{v}_{i k_1}}{\widetilde{v}_{j k_2}}$. 
To ensure the unitarity of the overlap matrices, we use the Singular Value Decomposition (SVD) to obtain the unitary part of the overlap matrices denoted by $\mathcal{S} \pqty{k_1, k_2}$. 
Then we find the Wilson line operator which is defined as $W_{k_f \leftarrow k_i} \equiv \mathcal{S} \pqty{k_f, k_f-\Delta k} \mathcal{S} \pqty{k_f-\Delta k, k_f-2\Delta k} \cdots \mathcal{S} \pqty{k_i + \Delta k, k_i}$. 
If $k_f = k_i + 2 \pi / a$, then the Wilson line becomes the Wilson loop. 
Instead of finding $Q \pqty{k}$ by diagonalizing $W_{k + 2 \pi / a \leftarrow k}$ for every $k$, we use the identity $Q(k_1+k_2) = W_{k_2\leftarrow k_1}Q(k_1)$. 
We perform a Schur decomposition on $W_{\pi / a \leftarrow -\pi/a}$ to obtain the $Q \pqty{-\pi/a}$, ensuring unitarity. 
Then, $W_{\pi / a \leftarrow -\pi/a}$ is diagonalized to obtain $\theta_n$. 
Now having all the components, we implement the discrete version of Eq.~\eqref{eq:comp_wann_funcs} to obtain the Wannier functions as
\begin{align}
    \ket{n R} \equiv \sum_{j = m}^{m + l} \sum_{l = 0}^{N_2 - 1} \frac{\Delta k}{2 \pi} & \exp \qty[-i k_l \qty(R + \frac{\theta_n}{2 \pi})] \nonumber \\ 
    & \times Q_{j n} \pqty{k_l} \ket{{E}_{j k_l}}
\end{align}

\section{Including more hopping terms in the tight-binding model}
\label{app:more_hoppings}

\begin{figure}
    \centering
    \includegraphics[width=0.48\textwidth]{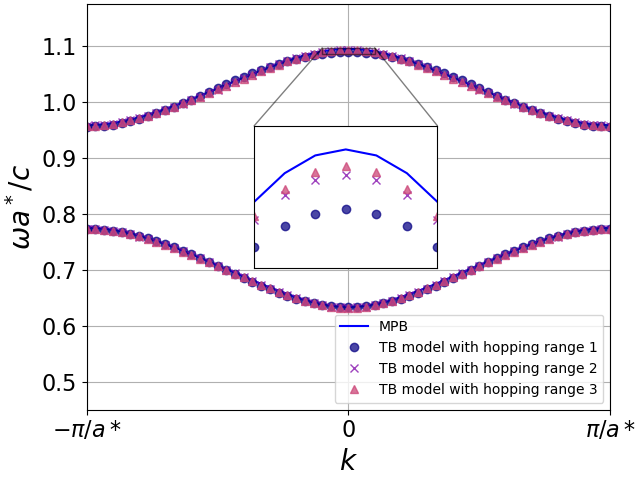}
    \caption{Convergence of the tight-binding spectrum to the MPB band structure for the photonic SSH model. 
    Solid blue shows the third and fourth bands for the photonic SSH model with $\varepsilon=5,\Delta\varepsilon=2$, and $b=0.3$. 
    Circles, crosses, and triangles show the spectrum of the tight-binding model retaining hopping of length one, two, and three lattice constants, respectively. 
    We see that as the hopping range increases, the tight-binding spectrum more accurately captures the behavior of the MPB spectrum.}
    \label{fig:more_hopping_terms}
\end{figure}

Fig. \ref{fig:more_hopping_terms} shows the how the tight-binding approximation to the photonic SSH spectrum converges to the actual spectrum as we keep including hopping terms between more distant unit cells. 
We make sure that while including additional terms we preserve inversion symmetry in the models. Note that our nearest-neighbor model from Sec.~\ref{sec:tb} corresponds to a range of $1/2$, since it only includes hoppings between neighboring Wannier functions.

\bibliography{references}

\begin{thebibliography}{91}%
\makeatletter
\providecommand \@ifxundefined [1]{%
 \@ifx{#1\undefined}
}%
\providecommand \@ifnum [1]{%
 \ifnum #1\expandafter \@firstoftwo
 \else \expandafter \@secondoftwo
 \fi
}%
\providecommand \@ifx [1]{%
 \ifx #1\expandafter \@firstoftwo
 \else \expandafter \@secondoftwo
 \fi
}%
\providecommand \natexlab [1]{#1}%
\providecommand \enquote  [1]{``#1''}%
\providecommand \bibnamefont  [1]{#1}%
\providecommand \bibfnamefont [1]{#1}%
\providecommand \citenamefont [1]{#1}%
\providecommand \href@noop [0]{\@secondoftwo}%
\providecommand \href [0]{\begingroup \@sanitize@url \@href}%
\providecommand \@href[1]{\@@startlink{#1}\@@href}%
\providecommand \@@href[1]{\endgroup#1\@@endlink}%
\providecommand \@sanitize@url [0]{\catcode `\\12\catcode `\$12\catcode
  `\&12\catcode `\#12\catcode `\^12\catcode `\_12\catcode `\%12\relax}%
\providecommand \@@startlink[1]{}%
\providecommand \@@endlink[0]{}%
\providecommand \url  [0]{\begingroup\@sanitize@url \@url }%
\providecommand \@url [1]{\endgroup\@href {#1}{\urlprefix }}%
\providecommand \urlprefix  [0]{URL }%
\providecommand \Eprint [0]{\href }%
\providecommand \doibase [0]{https://doi.org/}%
\providecommand \selectlanguage [0]{\@gobble}%
\providecommand \bibinfo  [0]{\@secondoftwo}%
\providecommand \bibfield  [0]{\@secondoftwo}%
\providecommand \translation [1]{[#1]}%
\providecommand \BibitemOpen [0]{}%
\providecommand \bibitemStop [0]{}%
\providecommand \bibitemNoStop [0]{.\EOS\space}%
\providecommand \EOS [0]{\spacefactor3000\relax}%
\providecommand \BibitemShut  [1]{\csname bibitem#1\endcsname}%
\let\auto@bib@innerbib\@empty
\bibitem [{\citenamefont {Haldane}\ and\ \citenamefont
  {Raghu}(2008)}]{haldane2008possible}%
  \BibitemOpen
  \bibfield  {author} {\bibinfo {author} {\bibfnamefont {F.}~\bibnamefont
  {Haldane}}\ and\ \bibinfo {author} {\bibfnamefont {S.}~\bibnamefont
  {Raghu}},\ }\bibfield  {title} {\bibinfo {title} {Possible realization of
  directional optical waveguides in photonic crystals with broken time-reversal
  symmetry},\ }\href@noop {} {\bibfield  {journal} {\bibinfo  {journal}
  {Physical review letters}\ }\textbf {\bibinfo {volume} {100}},\ \bibinfo
  {pages} {013904} (\bibinfo {year} {2008})}\BibitemShut {NoStop}%
\bibitem [{\citenamefont {Raghu}\ and\ \citenamefont
  {Haldane}(2008)}]{raghu2008analogs}%
  \BibitemOpen
  \bibfield  {author} {\bibinfo {author} {\bibfnamefont {S.}~\bibnamefont
  {Raghu}}\ and\ \bibinfo {author} {\bibfnamefont {F.~D.~M.}\ \bibnamefont
  {Haldane}},\ }\bibfield  {title} {\bibinfo {title} {Analogs of
  quantum-hall-effect edge states in photonic crystals},\ }\href@noop {}
  {\bibfield  {journal} {\bibinfo  {journal} {Physical Review A}\ }\textbf
  {\bibinfo {volume} {78}},\ \bibinfo {pages} {033834} (\bibinfo {year}
  {2008})}\BibitemShut {NoStop}%
\bibitem [{\citenamefont {Joannopoulos}\ \emph {et~al.}(2008)\citenamefont
  {Joannopoulos}, \citenamefont {Johnson}, \citenamefont {Winn},\ and\
  \citenamefont {Meade}}]{Joannopoulos08Book}%
  \BibitemOpen
  \bibfield  {author} {\bibinfo {author} {\bibfnamefont {J.~D.}\ \bibnamefont
  {Joannopoulos}}, \bibinfo {author} {\bibfnamefont {S.~G.}\ \bibnamefont
  {Johnson}}, \bibinfo {author} {\bibfnamefont {J.~N.}\ \bibnamefont {Winn}},\
  and\ \bibinfo {author} {\bibfnamefont {R.~D.}\ \bibnamefont {Meade}},\
  }\href@noop {} {\emph {\bibinfo {title} {Photonic Crystals: Molding the Flow
  of Light (Second Edition)}}},\ \bibinfo {edition} {2nd}\ ed.\ (\bibinfo
  {publisher} {Princeton University Press},\ \bibinfo {year}
  {2008})\BibitemShut {NoStop}%
\bibitem [{\citenamefont {Wang}\ \emph {et~al.}(2009)\citenamefont {Wang},
  \citenamefont {Chong}, \citenamefont {Joannopoulos},\ and\ \citenamefont
  {Solja{\v{c}}i{\'c}}}]{wang2009observation}%
  \BibitemOpen
  \bibfield  {author} {\bibinfo {author} {\bibfnamefont {Z.}~\bibnamefont
  {Wang}}, \bibinfo {author} {\bibfnamefont {Y.}~\bibnamefont {Chong}},
  \bibinfo {author} {\bibfnamefont {J.~D.}\ \bibnamefont {Joannopoulos}},\ and\
  \bibinfo {author} {\bibfnamefont {M.}~\bibnamefont {Solja{\v{c}}i{\'c}}},\
  }\bibfield  {title} {\bibinfo {title} {Observation of unidirectional
  backscattering-immune topological electromagnetic states},\ }\href@noop {}
  {\bibfield  {journal} {\bibinfo  {journal} {Nature}\ }\textbf {\bibinfo
  {volume} {461}},\ \bibinfo {pages} {772} (\bibinfo {year}
  {2009})}\BibitemShut {NoStop}%
\bibitem [{\citenamefont {Wang}\ \emph {et~al.}(2008)\citenamefont {Wang},
  \citenamefont {Chong}, \citenamefont {Joannopoulos},\ and\ \citenamefont
  {Solja{\v{c}}i{\'c}}}]{wang2008reflection}%
  \BibitemOpen
  \bibfield  {author} {\bibinfo {author} {\bibfnamefont {Z.}~\bibnamefont
  {Wang}}, \bibinfo {author} {\bibfnamefont {Y.}~\bibnamefont {Chong}},
  \bibinfo {author} {\bibfnamefont {J.~D.}\ \bibnamefont {Joannopoulos}},\ and\
  \bibinfo {author} {\bibfnamefont {M.}~\bibnamefont {Solja{\v{c}}i{\'c}}},\
  }\bibfield  {title} {\bibinfo {title} {Reflection-free one-way edge modes in
  a gyromagnetic photonic crystal},\ }\href@noop {} {\bibfield  {journal}
  {\bibinfo  {journal} {Physical review letters}\ }\textbf {\bibinfo {volume}
  {100}},\ \bibinfo {pages} {013905} (\bibinfo {year} {2008})}\BibitemShut
  {NoStop}%
\bibitem [{\citenamefont {Lu}\ \emph {et~al.}(2013)\citenamefont {Lu},
  \citenamefont {Fu}, \citenamefont {Joannopoulos},\ and\ \citenamefont
  {Solja{\v{c}}i{\'c}}}]{lu2013weyl}%
  \BibitemOpen
  \bibfield  {author} {\bibinfo {author} {\bibfnamefont {L.}~\bibnamefont
  {Lu}}, \bibinfo {author} {\bibfnamefont {L.}~\bibnamefont {Fu}}, \bibinfo
  {author} {\bibfnamefont {J.~D.}\ \bibnamefont {Joannopoulos}},\ and\ \bibinfo
  {author} {\bibfnamefont {M.}~\bibnamefont {Solja{\v{c}}i{\'c}}},\ }\bibfield
  {title} {\bibinfo {title} {Weyl points and line nodes in gyroid photonic
  crystals},\ }\href@noop {} {\bibfield  {journal} {\bibinfo  {journal} {Nature
  photonics}\ }\textbf {\bibinfo {volume} {7}},\ \bibinfo {pages} {294}
  (\bibinfo {year} {2013})}\BibitemShut {NoStop}%
\bibitem [{\citenamefont {Slobozhanyuk}\ \emph {et~al.}(2017)\citenamefont
  {Slobozhanyuk}, \citenamefont {Mousavi}, \citenamefont {Ni}, \citenamefont
  {Smirnova}, \citenamefont {Kivshar},\ and\ \citenamefont
  {Khanikaev}}]{slobozhanyuk2017three}%
  \BibitemOpen
  \bibfield  {author} {\bibinfo {author} {\bibfnamefont {A.}~\bibnamefont
  {Slobozhanyuk}}, \bibinfo {author} {\bibfnamefont {S.~H.}\ \bibnamefont
  {Mousavi}}, \bibinfo {author} {\bibfnamefont {X.}~\bibnamefont {Ni}},
  \bibinfo {author} {\bibfnamefont {D.}~\bibnamefont {Smirnova}}, \bibinfo
  {author} {\bibfnamefont {Y.~S.}\ \bibnamefont {Kivshar}},\ and\ \bibinfo
  {author} {\bibfnamefont {A.~B.}\ \bibnamefont {Khanikaev}},\ }\bibfield
  {title} {\bibinfo {title} {Three-dimensional all-dielectric photonic
  topological insulator},\ }\href@noop {} {\bibfield  {journal} {\bibinfo
  {journal} {Nature Photonics}\ }\textbf {\bibinfo {volume} {11}},\ \bibinfo
  {pages} {130} (\bibinfo {year} {2017})}\BibitemShut {NoStop}%
\bibitem [{\citenamefont {Khanikaev}\ and\ \citenamefont
  {Shvets}(2017)}]{khanikaev2017two}%
  \BibitemOpen
  \bibfield  {author} {\bibinfo {author} {\bibfnamefont {A.~B.}\ \bibnamefont
  {Khanikaev}}\ and\ \bibinfo {author} {\bibfnamefont {G.}~\bibnamefont
  {Shvets}},\ }\bibfield  {title} {\bibinfo {title} {Two-dimensional
  topological photonics},\ }\href@noop {} {\bibfield  {journal} {\bibinfo
  {journal} {Nature photonics}\ }\textbf {\bibinfo {volume} {11}},\ \bibinfo
  {pages} {763} (\bibinfo {year} {2017})}\BibitemShut {NoStop}%
\bibitem [{\citenamefont {Sun}\ \emph {et~al.}(2017)\citenamefont {Sun},
  \citenamefont {He}, \citenamefont {Liu}, \citenamefont {Lu}, \citenamefont
  {Zhu},\ and\ \citenamefont {Chen}}]{sun2017photonics}%
  \BibitemOpen
  \bibfield  {author} {\bibinfo {author} {\bibfnamefont {X.}~\bibnamefont
  {Sun}}, \bibinfo {author} {\bibfnamefont {C.}~\bibnamefont {He}}, \bibinfo
  {author} {\bibfnamefont {X.}~\bibnamefont {Liu}}, \bibinfo {author}
  {\bibfnamefont {M.}~\bibnamefont {Lu}}, \bibinfo {author} {\bibfnamefont
  {S.}~\bibnamefont {Zhu}},\ and\ \bibinfo {author} {\bibfnamefont
  {Y.}~\bibnamefont {Chen}},\ }\bibfield  {title} {\bibinfo {title} {Photonics
  meets topology},\ }\href@noop {} {\bibfield  {journal} {\bibinfo  {journal}
  {Prog. Quantum Electron}\ }\textbf {\bibinfo {volume} {55}},\ \bibinfo
  {pages} {52} (\bibinfo {year} {2017})}\BibitemShut {NoStop}%
\bibitem [{\citenamefont {Rider}\ \emph {et~al.}(2019)\citenamefont {Rider},
  \citenamefont {Palmer}, \citenamefont {Pocock}, \citenamefont {Xiao},
  \citenamefont {Arroyo~Huidobro},\ and\ \citenamefont
  {Giannini}}]{rider2019perspective}%
  \BibitemOpen
  \bibfield  {author} {\bibinfo {author} {\bibfnamefont {M.~S.}\ \bibnamefont
  {Rider}}, \bibinfo {author} {\bibfnamefont {S.~J.}\ \bibnamefont {Palmer}},
  \bibinfo {author} {\bibfnamefont {S.~R.}\ \bibnamefont {Pocock}}, \bibinfo
  {author} {\bibfnamefont {X.}~\bibnamefont {Xiao}}, \bibinfo {author}
  {\bibfnamefont {P.}~\bibnamefont {Arroyo~Huidobro}},\ and\ \bibinfo {author}
  {\bibfnamefont {V.}~\bibnamefont {Giannini}},\ }\bibfield  {title} {\bibinfo
  {title} {A perspective on topological nanophotonics: current status and
  future challenges},\ }\href@noop {} {\bibfield  {journal} {\bibinfo
  {journal} {Journal of Applied Physics}\ }\textbf {\bibinfo {volume} {125}},\
  \bibinfo {pages} {120901} (\bibinfo {year} {2019})}\BibitemShut {NoStop}%
\bibitem [{\citenamefont {Ozawa}\ \emph {et~al.}(2019)\citenamefont {Ozawa},
  \citenamefont {Price}, \citenamefont {Amo}, \citenamefont {Goldman},
  \citenamefont {Hafezi}, \citenamefont {Lu}, \citenamefont {Rechtsman},
  \citenamefont {Schuster}, \citenamefont {Simon}, \citenamefont {Zilberberg}
  \emph {et~al.}}]{ozawa2019topological}%
  \BibitemOpen
  \bibfield  {author} {\bibinfo {author} {\bibfnamefont {T.}~\bibnamefont
  {Ozawa}}, \bibinfo {author} {\bibfnamefont {H.~M.}\ \bibnamefont {Price}},
  \bibinfo {author} {\bibfnamefont {A.}~\bibnamefont {Amo}}, \bibinfo {author}
  {\bibfnamefont {N.}~\bibnamefont {Goldman}}, \bibinfo {author} {\bibfnamefont
  {M.}~\bibnamefont {Hafezi}}, \bibinfo {author} {\bibfnamefont
  {L.}~\bibnamefont {Lu}}, \bibinfo {author} {\bibfnamefont {M.~C.}\
  \bibnamefont {Rechtsman}}, \bibinfo {author} {\bibfnamefont {D.}~\bibnamefont
  {Schuster}}, \bibinfo {author} {\bibfnamefont {J.}~\bibnamefont {Simon}},
  \bibinfo {author} {\bibfnamefont {O.}~\bibnamefont {Zilberberg}}, \emph
  {et~al.},\ }\bibfield  {title} {\bibinfo {title} {Topological photonics},\
  }\href@noop {} {\bibfield  {journal} {\bibinfo  {journal} {Reviews of Modern
  Physics}\ }\textbf {\bibinfo {volume} {91}},\ \bibinfo {pages} {015006}
  (\bibinfo {year} {2019})}\BibitemShut {NoStop}%
\bibitem [{\citenamefont {Yang}\ \emph {et~al.}(2019)\citenamefont {Yang},
  \citenamefont {Gao}, \citenamefont {Xue}, \citenamefont {Zhang},
  \citenamefont {He}, \citenamefont {Yang}, \citenamefont {Singh},
  \citenamefont {Chong}, \citenamefont {Zhang},\ and\ \citenamefont
  {Chen}}]{yang2019realization}%
  \BibitemOpen
  \bibfield  {author} {\bibinfo {author} {\bibfnamefont {Y.}~\bibnamefont
  {Yang}}, \bibinfo {author} {\bibfnamefont {Z.}~\bibnamefont {Gao}}, \bibinfo
  {author} {\bibfnamefont {H.}~\bibnamefont {Xue}}, \bibinfo {author}
  {\bibfnamefont {L.}~\bibnamefont {Zhang}}, \bibinfo {author} {\bibfnamefont
  {M.}~\bibnamefont {He}}, \bibinfo {author} {\bibfnamefont {Z.}~\bibnamefont
  {Yang}}, \bibinfo {author} {\bibfnamefont {R.}~\bibnamefont {Singh}},
  \bibinfo {author} {\bibfnamefont {Y.}~\bibnamefont {Chong}}, \bibinfo
  {author} {\bibfnamefont {B.}~\bibnamefont {Zhang}},\ and\ \bibinfo {author}
  {\bibfnamefont {H.}~\bibnamefont {Chen}},\ }\bibfield  {title} {\bibinfo
  {title} {Realization of a three-dimensional photonic topological insulator},\
  }\href@noop {} {\bibfield  {journal} {\bibinfo  {journal} {Nature}\ }\textbf
  {\bibinfo {volume} {565}},\ \bibinfo {pages} {622} (\bibinfo {year}
  {2019})}\BibitemShut {NoStop}%
\bibitem [{\citenamefont {Kim}\ \emph {et~al.}(2020)\citenamefont {Kim},
  \citenamefont {Jacob},\ and\ \citenamefont {Rho}}]{kim2020recent}%
  \BibitemOpen
  \bibfield  {author} {\bibinfo {author} {\bibfnamefont {M.}~\bibnamefont
  {Kim}}, \bibinfo {author} {\bibfnamefont {Z.}~\bibnamefont {Jacob}},\ and\
  \bibinfo {author} {\bibfnamefont {J.}~\bibnamefont {Rho}},\ }\bibfield
  {title} {\bibinfo {title} {Recent advances in {2D}, {3D} and higher-order
  topological photonics},\ }\href {https://doi.org/10.1038/s41377-020-0331-y}
  {\bibfield  {journal} {\bibinfo  {journal} {Light: Science \& Applications}\
  }\textbf {\bibinfo {volume} {9}},\ \bibinfo {pages} {130} (\bibinfo {year}
  {2020})}\BibitemShut {NoStop}%
\bibitem [{Note1()}]{Note1}%
  \BibitemOpen
  \bibinfo {note} {Throughout this work, we will refer to gapped photonic
  systems as insulators in analogy with electronic systems, even though there
  is no analogous filling of states}\BibitemShut {NoStop}%
\bibitem [{\citenamefont {Liu}\ \emph {et~al.}(2021)\citenamefont {Liu},
  \citenamefont {Gao}, \citenamefont {Zhou}, \citenamefont {Wang},
  \citenamefont {Hu}, \citenamefont {Wang}, \citenamefont {Liu}, \citenamefont
  {Lin}, \citenamefont {Yang}, \citenamefont {Yang} \emph
  {et~al.}}]{liu2021observation}%
  \BibitemOpen
  \bibfield  {author} {\bibinfo {author} {\bibfnamefont {G.-G.}\ \bibnamefont
  {Liu}}, \bibinfo {author} {\bibfnamefont {Z.}~\bibnamefont {Gao}}, \bibinfo
  {author} {\bibfnamefont {P.}~\bibnamefont {Zhou}}, \bibinfo {author}
  {\bibfnamefont {Q.}~\bibnamefont {Wang}}, \bibinfo {author} {\bibfnamefont
  {Y.-H.}\ \bibnamefont {Hu}}, \bibinfo {author} {\bibfnamefont
  {M.}~\bibnamefont {Wang}}, \bibinfo {author} {\bibfnamefont {C.}~\bibnamefont
  {Liu}}, \bibinfo {author} {\bibfnamefont {X.}~\bibnamefont {Lin}}, \bibinfo
  {author} {\bibfnamefont {S.~A.}\ \bibnamefont {Yang}}, \bibinfo {author}
  {\bibfnamefont {Y.}~\bibnamefont {Yang}}, \emph {et~al.},\ }\bibfield
  {title} {\bibinfo {title} {Observation of weyl point pair annihilation in a
  gyromagnetic photonic crystal},\ }\href@noop {} {\bibfield  {journal}
  {\bibinfo  {journal} {arXiv preprint arXiv:2106.02461}\ } (\bibinfo {year}
  {2021})}\BibitemShut {NoStop}%
\bibitem [{\citenamefont {Wieder}\ \emph
  {et~al.}(2020{\natexlab{a}})\citenamefont {Wieder}, \citenamefont {Lin},\
  and\ \citenamefont {Bradlyn}}]{wieder2020axionic}%
  \BibitemOpen
  \bibfield  {author} {\bibinfo {author} {\bibfnamefont {B.~J.}\ \bibnamefont
  {Wieder}}, \bibinfo {author} {\bibfnamefont {K.-S.}\ \bibnamefont {Lin}},\
  and\ \bibinfo {author} {\bibfnamefont {B.}~\bibnamefont {Bradlyn}},\
  }\bibfield  {title} {\bibinfo {title} {Axionic band topology in
  inversion-symmetric weyl-charge-density waves},\ }\href@noop {} {\bibfield
  {journal} {\bibinfo  {journal} {Physical Review Research}\ }\textbf {\bibinfo
  {volume} {2}},\ \bibinfo {pages} {042010} (\bibinfo {year}
  {2020}{\natexlab{a}})}\BibitemShut {NoStop}%
\bibitem [{\citenamefont {Devescovi}\ \emph {et~al.}(2021)\citenamefont
  {Devescovi}, \citenamefont {{Garc{\'i}a-D{\'i}ez}}, \citenamefont {Robredo},
  \citenamefont {{Blanco de Paz}}, \citenamefont {{Lasa-Alonso}}, \citenamefont
  {Bradlyn}, \citenamefont {Ma{\~n}es}, \citenamefont {G.~Vergniory},\ and\
  \citenamefont {{Garc{\'i}a-Etxarri}}}]{devescovi2021cubic}%
  \BibitemOpen
  \bibfield  {author} {\bibinfo {author} {\bibfnamefont {C.}~\bibnamefont
  {Devescovi}}, \bibinfo {author} {\bibfnamefont {M.}~\bibnamefont
  {{Garc{\'i}a-D{\'i}ez}}}, \bibinfo {author} {\bibfnamefont {I.}~\bibnamefont
  {Robredo}}, \bibinfo {author} {\bibfnamefont {M.}~\bibnamefont {{Blanco de
  Paz}}}, \bibinfo {author} {\bibfnamefont {J.}~\bibnamefont {{Lasa-Alonso}}},
  \bibinfo {author} {\bibfnamefont {B.}~\bibnamefont {Bradlyn}}, \bibinfo
  {author} {\bibfnamefont {J.~L.}\ \bibnamefont {Ma{\~n}es}}, \bibinfo {author}
  {\bibfnamefont {M.}~\bibnamefont {G.~Vergniory}},\ and\ \bibinfo {author}
  {\bibfnamefont {A.}~\bibnamefont {{Garc{\'i}a-Etxarri}}},\ }\bibfield
  {title} {\bibinfo {title} {Cubic {{3D Chern}} photonic insulators with
  orientable large {{Chern}} vectors},\ }\href
  {https://doi.org/10.1038/s41467-021-27168-w} {\bibfield  {journal} {\bibinfo
  {journal} {Nature Communications}\ }\textbf {\bibinfo {volume} {12}},\
  \bibinfo {pages} {7330} (\bibinfo {year} {2021})}\BibitemShut {NoStop}%
\bibitem [{\citenamefont {He}\ \emph {et~al.}(2020)\citenamefont {He},
  \citenamefont {Addison}, \citenamefont {Mele},\ and\ \citenamefont
  {Zhen}}]{he2020quadrupole}%
  \BibitemOpen
  \bibfield  {author} {\bibinfo {author} {\bibfnamefont {L.}~\bibnamefont
  {He}}, \bibinfo {author} {\bibfnamefont {Z.}~\bibnamefont {Addison}},
  \bibinfo {author} {\bibfnamefont {E.~J.}\ \bibnamefont {Mele}},\ and\
  \bibinfo {author} {\bibfnamefont {B.}~\bibnamefont {Zhen}},\ }\bibfield
  {title} {\bibinfo {title} {Quadrupole topological photonic crystals},\
  }\href@noop {} {\bibfield  {journal} {\bibinfo  {journal} {Nature
  communications}\ }\textbf {\bibinfo {volume} {11}},\ \bibinfo {pages} {1}
  (\bibinfo {year} {2020})}\BibitemShut {NoStop}%
\bibitem [{\citenamefont {Kim}\ \emph {et~al.}(2021)\citenamefont {Kim},
  \citenamefont {Lee}, \citenamefont {Nguyen}, \citenamefont {Lee},
  \citenamefont {Byun},\ and\ \citenamefont {Rho}}]{kim2021total}%
  \BibitemOpen
  \bibfield  {author} {\bibinfo {author} {\bibfnamefont {M.}~\bibnamefont
  {Kim}}, \bibinfo {author} {\bibfnamefont {D.}~\bibnamefont {Lee}}, \bibinfo
  {author} {\bibfnamefont {T.~H.-Y.}\ \bibnamefont {Nguyen}}, \bibinfo {author}
  {\bibfnamefont {H.-J.}\ \bibnamefont {Lee}}, \bibinfo {author} {\bibfnamefont
  {G.}~\bibnamefont {Byun}},\ and\ \bibinfo {author} {\bibfnamefont
  {J.}~\bibnamefont {Rho}},\ }\bibfield  {title} {\bibinfo {title} {Total
  reflection-induced efficiency enhancement of the spin hall effect of light},\
  }\href {https://doi.org/10.1021/acsphotonics.1c00727} {\bibfield  {journal}
  {\bibinfo  {journal} {ACS Photonics}\ }\textbf {\bibinfo {volume} {8}},\
  \bibinfo {pages} {2705} (\bibinfo {year} {2021})}\BibitemShut {NoStop}%
\bibitem [{\citenamefont {Jin}\ \emph {et~al.}(2021)\citenamefont {Jin},
  \citenamefont {He}, \citenamefont {Lu}, \citenamefont {Mele},\ and\
  \citenamefont {Zhen}}]{jin2021floquet}%
  \BibitemOpen
  \bibfield  {author} {\bibinfo {author} {\bibfnamefont {J.}~\bibnamefont
  {Jin}}, \bibinfo {author} {\bibfnamefont {L.}~\bibnamefont {He}}, \bibinfo
  {author} {\bibfnamefont {J.}~\bibnamefont {Lu}}, \bibinfo {author}
  {\bibfnamefont {E.~J.}\ \bibnamefont {Mele}},\ and\ \bibinfo {author}
  {\bibfnamefont {B.}~\bibnamefont {Zhen}},\ }\href@noop {} {\bibinfo {title}
  {Floquet quadrupole photonic crystals protected by space-time symmetry}}
  (\bibinfo {year} {2021}),\ \Eprint {https://arxiv.org/abs/2103.01198}
  {arXiv:2103.01198 [physics.optics]} \BibitemShut {NoStop}%
\bibitem [{\citenamefont {Proctor}\ \emph {et~al.}(2020)\citenamefont
  {Proctor}, \citenamefont {Huidobro}, \citenamefont {Bradlyn}, \citenamefont
  {de~Paz}, \citenamefont {Vergniory}, \citenamefont {Bercioux},\ and\
  \citenamefont {Garc{\'\i}a-Etxarri}}]{proctor2020robustness}%
  \BibitemOpen
  \bibfield  {author} {\bibinfo {author} {\bibfnamefont {M.}~\bibnamefont
  {Proctor}}, \bibinfo {author} {\bibfnamefont {P.~A.}\ \bibnamefont
  {Huidobro}}, \bibinfo {author} {\bibfnamefont {B.}~\bibnamefont {Bradlyn}},
  \bibinfo {author} {\bibfnamefont {M.~B.}\ \bibnamefont {de~Paz}}, \bibinfo
  {author} {\bibfnamefont {M.~G.}\ \bibnamefont {Vergniory}}, \bibinfo {author}
  {\bibfnamefont {D.}~\bibnamefont {Bercioux}},\ and\ \bibinfo {author}
  {\bibfnamefont {A.}~\bibnamefont {Garc{\'\i}a-Etxarri}},\ }\bibfield  {title}
  {\bibinfo {title} {Robustness of topological corner modes in photonic
  crystals},\ }\href@noop {} {\bibfield  {journal} {\bibinfo  {journal}
  {Physical Review Research}\ }\textbf {\bibinfo {volume} {2}},\ \bibinfo
  {pages} {042038} (\bibinfo {year} {2020})}\BibitemShut {NoStop}%
\bibitem [{\citenamefont {Zhang}\ \emph {et~al.}(2021)\citenamefont {Zhang},
  \citenamefont {Ren}, \citenamefont {Li},\ and\ \citenamefont
  {Ye}}]{zhang2021topological}%
  \BibitemOpen
  \bibfield  {author} {\bibinfo {author} {\bibfnamefont {Y.}~\bibnamefont
  {Zhang}}, \bibinfo {author} {\bibfnamefont {B.}~\bibnamefont {Ren}}, \bibinfo
  {author} {\bibfnamefont {Y.}~\bibnamefont {Li}},\ and\ \bibinfo {author}
  {\bibfnamefont {F.}~\bibnamefont {Ye}},\ }\bibfield  {title} {\bibinfo
  {title} {Topological states in the super-ssh model},\ }\href@noop {}
  {\bibfield  {journal} {\bibinfo  {journal} {Optics Express}\ }\textbf
  {\bibinfo {volume} {29}},\ \bibinfo {pages} {42827} (\bibinfo {year}
  {2021})}\BibitemShut {NoStop}%
\bibitem [{\citenamefont {Liu}\ and\ \citenamefont
  {Wakabayashi}(2017)}]{liu2017novel}%
  \BibitemOpen
  \bibfield  {author} {\bibinfo {author} {\bibfnamefont {F.}~\bibnamefont
  {Liu}}\ and\ \bibinfo {author} {\bibfnamefont {K.}~\bibnamefont
  {Wakabayashi}},\ }\bibfield  {title} {\bibinfo {title} {Novel topological
  phase with a zero berry curvature},\ }\href@noop {} {\bibfield  {journal}
  {\bibinfo  {journal} {Physical review letters}\ }\textbf {\bibinfo {volume}
  {118}},\ \bibinfo {pages} {076803} (\bibinfo {year} {2017})}\BibitemShut
  {NoStop}%
\bibitem [{\citenamefont {Xie}\ \emph {et~al.}(2018)\citenamefont {Xie},
  \citenamefont {Wang}, \citenamefont {Wang}, \citenamefont {Zhu},
  \citenamefont {Jiang}, \citenamefont {Lu},\ and\ \citenamefont
  {Chen}}]{xie2018second}%
  \BibitemOpen
  \bibfield  {author} {\bibinfo {author} {\bibfnamefont {B.-Y.}\ \bibnamefont
  {Xie}}, \bibinfo {author} {\bibfnamefont {H.-F.}\ \bibnamefont {Wang}},
  \bibinfo {author} {\bibfnamefont {H.-X.}\ \bibnamefont {Wang}}, \bibinfo
  {author} {\bibfnamefont {X.-Y.}\ \bibnamefont {Zhu}}, \bibinfo {author}
  {\bibfnamefont {J.-H.}\ \bibnamefont {Jiang}}, \bibinfo {author}
  {\bibfnamefont {M.-H.}\ \bibnamefont {Lu}},\ and\ \bibinfo {author}
  {\bibfnamefont {Y.-F.}\ \bibnamefont {Chen}},\ }\bibfield  {title} {\bibinfo
  {title} {Second-order photonic topological insulator with corner states},\
  }\href@noop {} {\bibfield  {journal} {\bibinfo  {journal} {Physical Review
  B}\ }\textbf {\bibinfo {volume} {98}},\ \bibinfo {pages} {205147} (\bibinfo
  {year} {2018})}\BibitemShut {NoStop}%
\bibitem [{\citenamefont {Ota}\ \emph {et~al.}(2019)\citenamefont {Ota},
  \citenamefont {Liu}, \citenamefont {Katsumi}, \citenamefont {Watanabe},
  \citenamefont {Wakabayashi}, \citenamefont {Arakawa},\ and\ \citenamefont
  {Iwamoto}}]{ota2019photonic}%
  \BibitemOpen
  \bibfield  {author} {\bibinfo {author} {\bibfnamefont {Y.}~\bibnamefont
  {Ota}}, \bibinfo {author} {\bibfnamefont {F.}~\bibnamefont {Liu}}, \bibinfo
  {author} {\bibfnamefont {R.}~\bibnamefont {Katsumi}}, \bibinfo {author}
  {\bibfnamefont {K.}~\bibnamefont {Watanabe}}, \bibinfo {author}
  {\bibfnamefont {K.}~\bibnamefont {Wakabayashi}}, \bibinfo {author}
  {\bibfnamefont {Y.}~\bibnamefont {Arakawa}},\ and\ \bibinfo {author}
  {\bibfnamefont {S.}~\bibnamefont {Iwamoto}},\ }\bibfield  {title} {\bibinfo
  {title} {Photonic crystal nanocavity based on a topological corner state},\
  }\href@noop {} {\bibfield  {journal} {\bibinfo  {journal} {Optica}\ }\textbf
  {\bibinfo {volume} {6}},\ \bibinfo {pages} {786} (\bibinfo {year}
  {2019})}\BibitemShut {NoStop}%
\bibitem [{\citenamefont {Chen}\ \emph {et~al.}(2019)\citenamefont {Chen},
  \citenamefont {Deng}, \citenamefont {Shi}, \citenamefont {Zhao},
  \citenamefont {Chen},\ and\ \citenamefont {Dong}}]{chen2019direct}%
  \BibitemOpen
  \bibfield  {author} {\bibinfo {author} {\bibfnamefont {X.-D.}\ \bibnamefont
  {Chen}}, \bibinfo {author} {\bibfnamefont {W.-M.}\ \bibnamefont {Deng}},
  \bibinfo {author} {\bibfnamefont {F.-L.}\ \bibnamefont {Shi}}, \bibinfo
  {author} {\bibfnamefont {F.-L.}\ \bibnamefont {Zhao}}, \bibinfo {author}
  {\bibfnamefont {M.}~\bibnamefont {Chen}},\ and\ \bibinfo {author}
  {\bibfnamefont {J.-W.}\ \bibnamefont {Dong}},\ }\bibfield  {title} {\bibinfo
  {title} {Direct observation of corner states in second-order topological
  photonic crystal slabs},\ }\href@noop {} {\bibfield  {journal} {\bibinfo
  {journal} {Physical Review Letters}\ }\textbf {\bibinfo {volume} {122}},\
  \bibinfo {pages} {233902} (\bibinfo {year} {2019})}\BibitemShut {NoStop}%
\bibitem [{\citenamefont {Xie}\ \emph {et~al.}(2019)\citenamefont {Xie},
  \citenamefont {Su}, \citenamefont {Wang}, \citenamefont {Su}, \citenamefont
  {Shen}, \citenamefont {Zhan}, \citenamefont {Lu}, \citenamefont {Wang},\ and\
  \citenamefont {Chen}}]{xie2019visualization}%
  \BibitemOpen
  \bibfield  {author} {\bibinfo {author} {\bibfnamefont {B.-Y.}\ \bibnamefont
  {Xie}}, \bibinfo {author} {\bibfnamefont {G.-X.}\ \bibnamefont {Su}},
  \bibinfo {author} {\bibfnamefont {H.-F.}\ \bibnamefont {Wang}}, \bibinfo
  {author} {\bibfnamefont {H.}~\bibnamefont {Su}}, \bibinfo {author}
  {\bibfnamefont {X.-P.}\ \bibnamefont {Shen}}, \bibinfo {author}
  {\bibfnamefont {P.}~\bibnamefont {Zhan}}, \bibinfo {author} {\bibfnamefont
  {M.-H.}\ \bibnamefont {Lu}}, \bibinfo {author} {\bibfnamefont {Z.-L.}\
  \bibnamefont {Wang}},\ and\ \bibinfo {author} {\bibfnamefont {Y.-F.}\
  \bibnamefont {Chen}},\ }\bibfield  {title} {\bibinfo {title} {Visualization
  of higher-order topological insulating phases in two-dimensional dielectric
  photonic crystals},\ }\href@noop {} {\bibfield  {journal} {\bibinfo
  {journal} {Physical review letters}\ }\textbf {\bibinfo {volume} {122}},\
  \bibinfo {pages} {233903} (\bibinfo {year} {2019})}\BibitemShut {NoStop}%
\bibitem [{\citenamefont {Noh}\ \emph {et~al.}(2018)\citenamefont {Noh},
  \citenamefont {Benalcazar}, \citenamefont {Huang}, \citenamefont {Collins},
  \citenamefont {Chen}, \citenamefont {Hughes},\ and\ \citenamefont
  {Rechtsman}}]{noh2018topological}%
  \BibitemOpen
  \bibfield  {author} {\bibinfo {author} {\bibfnamefont {J.}~\bibnamefont
  {Noh}}, \bibinfo {author} {\bibfnamefont {W.~A.}\ \bibnamefont {Benalcazar}},
  \bibinfo {author} {\bibfnamefont {S.}~\bibnamefont {Huang}}, \bibinfo
  {author} {\bibfnamefont {M.~J.}\ \bibnamefont {Collins}}, \bibinfo {author}
  {\bibfnamefont {K.~P.}\ \bibnamefont {Chen}}, \bibinfo {author}
  {\bibfnamefont {T.~L.}\ \bibnamefont {Hughes}},\ and\ \bibinfo {author}
  {\bibfnamefont {M.~C.}\ \bibnamefont {Rechtsman}},\ }\bibfield  {title}
  {\bibinfo {title} {Topological protection of photonic mid-gap defect modes},\
  }\href@noop {} {\bibfield  {journal} {\bibinfo  {journal} {Nature Photonics}\
  }\textbf {\bibinfo {volume} {12}},\ \bibinfo {pages} {408} (\bibinfo {year}
  {2018})}\BibitemShut {NoStop}%
\bibitem [{\citenamefont {Mittal}\ \emph {et~al.}(2019)\citenamefont {Mittal},
  \citenamefont {Orre}, \citenamefont {Zhu}, \citenamefont {Gorlach},
  \citenamefont {Poddubny},\ and\ \citenamefont
  {Hafezi}}]{mittal2019photonica}%
  \BibitemOpen
  \bibfield  {author} {\bibinfo {author} {\bibfnamefont {S.}~\bibnamefont
  {Mittal}}, \bibinfo {author} {\bibfnamefont {V.~V.}\ \bibnamefont {Orre}},
  \bibinfo {author} {\bibfnamefont {G.}~\bibnamefont {Zhu}}, \bibinfo {author}
  {\bibfnamefont {M.~A.}\ \bibnamefont {Gorlach}}, \bibinfo {author}
  {\bibfnamefont {A.}~\bibnamefont {Poddubny}},\ and\ \bibinfo {author}
  {\bibfnamefont {M.}~\bibnamefont {Hafezi}},\ }\bibfield  {title} {\bibinfo
  {title} {Photonic quadrupole topological phases},\ }\href@noop {} {\bibfield
  {journal} {\bibinfo  {journal} {Nature Photonics}\ }\textbf {\bibinfo
  {volume} {13}},\ \bibinfo {pages} {692} (\bibinfo {year} {2019})}\BibitemShut
  {NoStop}%
\bibitem [{\citenamefont {El~Hassan}\ \emph {et~al.}(2019)\citenamefont
  {El~Hassan}, \citenamefont {Kunst}, \citenamefont {Moritz}, \citenamefont
  {Andler}, \citenamefont {Bergholtz},\ and\ \citenamefont
  {Bourennane}}]{el2019corner}%
  \BibitemOpen
  \bibfield  {author} {\bibinfo {author} {\bibfnamefont {A.}~\bibnamefont
  {El~Hassan}}, \bibinfo {author} {\bibfnamefont {F.~K.}\ \bibnamefont
  {Kunst}}, \bibinfo {author} {\bibfnamefont {A.}~\bibnamefont {Moritz}},
  \bibinfo {author} {\bibfnamefont {G.}~\bibnamefont {Andler}}, \bibinfo
  {author} {\bibfnamefont {E.~J.}\ \bibnamefont {Bergholtz}},\ and\ \bibinfo
  {author} {\bibfnamefont {M.}~\bibnamefont {Bourennane}},\ }\bibfield  {title}
  {\bibinfo {title} {Corner states of light in photonic waveguides},\
  }\href@noop {} {\bibfield  {journal} {\bibinfo  {journal} {Nature Photonics}\
  }\textbf {\bibinfo {volume} {13}},\ \bibinfo {pages} {697} (\bibinfo {year}
  {2019})}\BibitemShut {NoStop}%
\bibitem [{\citenamefont {Li}\ \emph {et~al.}(2020)\citenamefont {Li},
  \citenamefont {Zhirihin}, \citenamefont {Gorlach}, \citenamefont {Ni},
  \citenamefont {Filonov}, \citenamefont {Slobozhanyuk}, \citenamefont
  {Al{\`u}},\ and\ \citenamefont {Khanikaev}}]{li2020higher}%
  \BibitemOpen
  \bibfield  {author} {\bibinfo {author} {\bibfnamefont {M.}~\bibnamefont
  {Li}}, \bibinfo {author} {\bibfnamefont {D.}~\bibnamefont {Zhirihin}},
  \bibinfo {author} {\bibfnamefont {M.}~\bibnamefont {Gorlach}}, \bibinfo
  {author} {\bibfnamefont {X.}~\bibnamefont {Ni}}, \bibinfo {author}
  {\bibfnamefont {D.}~\bibnamefont {Filonov}}, \bibinfo {author} {\bibfnamefont
  {A.}~\bibnamefont {Slobozhanyuk}}, \bibinfo {author} {\bibfnamefont
  {A.}~\bibnamefont {Al{\`u}}},\ and\ \bibinfo {author} {\bibfnamefont {A.~B.}\
  \bibnamefont {Khanikaev}},\ }\bibfield  {title} {\bibinfo {title}
  {Higher-order topological states in photonic kagome crystals with long-range
  interactions},\ }\href@noop {} {\bibfield  {journal} {\bibinfo  {journal}
  {Nature Photonics}\ }\textbf {\bibinfo {volume} {14}},\ \bibinfo {pages} {89}
  (\bibinfo {year} {2020})}\BibitemShut {NoStop}%
\bibitem [{\citenamefont {Cerjan}\ \emph {et~al.}(2020)\citenamefont {Cerjan},
  \citenamefont {J{\"u}rgensen}, \citenamefont {Benalcazar}, \citenamefont
  {Mukherjee},\ and\ \citenamefont {Rechtsman}}]{cerjan2020observation}%
  \BibitemOpen
  \bibfield  {author} {\bibinfo {author} {\bibfnamefont {A.}~\bibnamefont
  {Cerjan}}, \bibinfo {author} {\bibfnamefont {M.}~\bibnamefont
  {J{\"u}rgensen}}, \bibinfo {author} {\bibfnamefont {W.~A.}\ \bibnamefont
  {Benalcazar}}, \bibinfo {author} {\bibfnamefont {S.}~\bibnamefont
  {Mukherjee}},\ and\ \bibinfo {author} {\bibfnamefont {M.~C.}\ \bibnamefont
  {Rechtsman}},\ }\bibfield  {title} {\bibinfo {title} {Observation of a
  higher-order topological bound state in the continuum},\ }\href@noop {}
  {\bibfield  {journal} {\bibinfo  {journal} {Physical review letters}\
  }\textbf {\bibinfo {volume} {125}},\ \bibinfo {pages} {213901} (\bibinfo
  {year} {2020})}\BibitemShut {NoStop}%
\bibitem [{\citenamefont {Zhou}\ \emph {et~al.}(2020)\citenamefont {Zhou},
  \citenamefont {Lin}, \citenamefont {Lu}, \citenamefont {Lai}, \citenamefont
  {Hou},\ and\ \citenamefont {Jiang}}]{zhou2020photonic}%
  \BibitemOpen
  \bibfield  {author} {\bibinfo {author} {\bibfnamefont {X.}~\bibnamefont
  {Zhou}}, \bibinfo {author} {\bibfnamefont {Z.-K.}\ \bibnamefont {Lin}},
  \bibinfo {author} {\bibfnamefont {W.}~\bibnamefont {Lu}}, \bibinfo {author}
  {\bibfnamefont {Y.}~\bibnamefont {Lai}}, \bibinfo {author} {\bibfnamefont
  {B.}~\bibnamefont {Hou}},\ and\ \bibinfo {author} {\bibfnamefont {J.-H.}\
  \bibnamefont {Jiang}},\ }\bibfield  {title} {\bibinfo {title} {Photonic
  crystals: {{Twisted}} quadrupole topological photonic crystals (laser
  photonics rev. 14 (8)/2020)},\ }\href@noop {} {\bibfield  {journal} {\bibinfo
   {journal} {Laser \& Photonics Reviews}\ }\textbf {\bibinfo {volume} {14}},\
  \bibinfo {pages} {2070046} (\bibinfo {year} {2020})}\BibitemShut {NoStop}%
\bibitem [{\citenamefont {Wang}\ \emph {et~al.}(2021)\citenamefont {Wang},
  \citenamefont {Liang}, \citenamefont {Jiang}, \citenamefont {Hu},
  \citenamefont {Lu},\ and\ \citenamefont {Jiang}}]{wang2021higher}%
  \BibitemOpen
  \bibfield  {author} {\bibinfo {author} {\bibfnamefont {H.-X.}\ \bibnamefont
  {Wang}}, \bibinfo {author} {\bibfnamefont {L.}~\bibnamefont {Liang}},
  \bibinfo {author} {\bibfnamefont {B.}~\bibnamefont {Jiang}}, \bibinfo
  {author} {\bibfnamefont {J.}~\bibnamefont {Hu}}, \bibinfo {author}
  {\bibfnamefont {X.}~\bibnamefont {Lu}},\ and\ \bibinfo {author}
  {\bibfnamefont {J.-H.}\ \bibnamefont {Jiang}},\ }\bibfield  {title} {\bibinfo
  {title} {Higher-order topological phases in tunable {{C}} 3 symmetric
  photonic crystals},\ }\href@noop {} {\bibfield  {journal} {\bibinfo
  {journal} {Photonics Research}\ }\textbf {\bibinfo {volume} {9}},\ \bibinfo
  {pages} {1854} (\bibinfo {year} {2021})}\BibitemShut {NoStop}%
\bibitem [{\citenamefont {Lin}\ and\ \citenamefont
  {Jiang}(2022)}]{lin2022dirac}%
  \BibitemOpen
  \bibfield  {author} {\bibinfo {author} {\bibfnamefont {Z.-K.}\ \bibnamefont
  {Lin}}\ and\ \bibinfo {author} {\bibfnamefont {J.-H.}\ \bibnamefont
  {Jiang}},\ }\bibfield  {title} {\bibinfo {title} {Dirac cones and
  higher-order topology in quasi-continuous media},\ }\href@noop {} {\bibfield
  {journal} {\bibinfo  {journal} {Europhysics Letters}\ }\textbf {\bibinfo
  {volume} {137}},\ \bibinfo {pages} {15001} (\bibinfo {year}
  {2022})}\BibitemShut {NoStop}%
\bibitem [{\citenamefont {Xie}\ \emph {et~al.}(2021)\citenamefont {Xie},
  \citenamefont {Wang}, \citenamefont {Zhang}, \citenamefont {Zhan},
  \citenamefont {Jiang}, \citenamefont {Lu},\ and\ \citenamefont
  {Chen}}]{xie2021higher}%
  \BibitemOpen
  \bibfield  {author} {\bibinfo {author} {\bibfnamefont {B.}~\bibnamefont
  {Xie}}, \bibinfo {author} {\bibfnamefont {H.-X.}\ \bibnamefont {Wang}},
  \bibinfo {author} {\bibfnamefont {X.}~\bibnamefont {Zhang}}, \bibinfo
  {author} {\bibfnamefont {P.}~\bibnamefont {Zhan}}, \bibinfo {author}
  {\bibfnamefont {J.-H.}\ \bibnamefont {Jiang}}, \bibinfo {author}
  {\bibfnamefont {M.}~\bibnamefont {Lu}},\ and\ \bibinfo {author}
  {\bibfnamefont {Y.}~\bibnamefont {Chen}},\ }\bibfield  {title} {\bibinfo
  {title} {Higher-order band topology},\ }\href@noop {} {\bibfield  {journal}
  {\bibinfo  {journal} {Nature Reviews Physics}\ }\textbf {\bibinfo {volume}
  {3}},\ \bibinfo {pages} {520} (\bibinfo {year} {2021})}\BibitemShut {NoStop}%
\bibitem [{\citenamefont {Blanco~de Paz}\ \emph {et~al.}(2020)\citenamefont
  {Blanco~de Paz}, \citenamefont {Devescovi}, \citenamefont {Giedke},
  \citenamefont {Saenz}, \citenamefont {Vergniory}, \citenamefont {Bradlyn},
  \citenamefont {Bercioux},\ and\ \citenamefont
  {Garc{\'\i}a-Etxarri}}]{blanco2020tutorial}%
  \BibitemOpen
  \bibfield  {author} {\bibinfo {author} {\bibfnamefont {M.}~\bibnamefont
  {Blanco~de Paz}}, \bibinfo {author} {\bibfnamefont {C.}~\bibnamefont
  {Devescovi}}, \bibinfo {author} {\bibfnamefont {G.}~\bibnamefont {Giedke}},
  \bibinfo {author} {\bibfnamefont {J.~J.}\ \bibnamefont {Saenz}}, \bibinfo
  {author} {\bibfnamefont {M.~G.}\ \bibnamefont {Vergniory}}, \bibinfo {author}
  {\bibfnamefont {B.}~\bibnamefont {Bradlyn}}, \bibinfo {author} {\bibfnamefont
  {D.}~\bibnamefont {Bercioux}},\ and\ \bibinfo {author} {\bibfnamefont
  {A.}~\bibnamefont {Garc{\'\i}a-Etxarri}},\ }\bibfield  {title} {\bibinfo
  {title} {Tutorial: computing topological invariants in 2d photonic
  crystals},\ }\href@noop {} {\bibfield  {journal} {\bibinfo  {journal}
  {Advanced Quantum Technologies}\ }\textbf {\bibinfo {volume} {3}},\ \bibinfo
  {pages} {1900117} (\bibinfo {year} {2020})}\BibitemShut {NoStop}%
\bibitem [{\citenamefont {Lu}\ \emph {et~al.}(2016)\citenamefont {Lu},
  \citenamefont {Fang}, \citenamefont {Fu}, \citenamefont {Johnson},
  \citenamefont {Joannopoulos},\ and\ \citenamefont {Solja{\v
  c}i{\'c}}}]{lu2016symmetryprotected}%
  \BibitemOpen
  \bibfield  {author} {\bibinfo {author} {\bibfnamefont {L.}~\bibnamefont
  {Lu}}, \bibinfo {author} {\bibfnamefont {C.}~\bibnamefont {Fang}}, \bibinfo
  {author} {\bibfnamefont {L.}~\bibnamefont {Fu}}, \bibinfo {author}
  {\bibfnamefont {S.~G.}\ \bibnamefont {Johnson}}, \bibinfo {author}
  {\bibfnamefont {J.~D.}\ \bibnamefont {Joannopoulos}},\ and\ \bibinfo {author}
  {\bibfnamefont {M.}~\bibnamefont {Solja{\v c}i{\'c}}},\ }\bibfield  {title}
  {\bibinfo {title} {Symmetry-protected topological photonic crystal in three
  dimensions},\ }\href@noop {} {\bibfield  {journal} {\bibinfo  {journal}
  {Nature Physics}\ }\textbf {\bibinfo {volume} {12}},\ \bibinfo {pages} {337}
  (\bibinfo {year} {2016})}\BibitemShut {NoStop}%
\bibitem [{\citenamefont {Wang}\ \emph {et~al.}(2019)\citenamefont {Wang},
  \citenamefont {Guo},\ and\ \citenamefont {Jiang}}]{wang2019banda}%
  \BibitemOpen
  \bibfield  {author} {\bibinfo {author} {\bibfnamefont {H.-X.}\ \bibnamefont
  {Wang}}, \bibinfo {author} {\bibfnamefont {G.-Y.}\ \bibnamefont {Guo}},\ and\
  \bibinfo {author} {\bibfnamefont {J.-H.}\ \bibnamefont {Jiang}},\ }\bibfield
  {title} {\bibinfo {title} {Band topology in classical waves: {{Wilson-loop}}
  approach to topological numbers and fragile topology},\ }\href@noop {}
  {\bibfield  {journal} {\bibinfo  {journal} {New Journal of Physics}\ }\textbf
  {\bibinfo {volume} {21}},\ \bibinfo {pages} {093029} (\bibinfo {year}
  {2019})}\BibitemShut {NoStop}%
\bibitem [{\citenamefont {Xiao}\ \emph {et~al.}(2014)\citenamefont {Xiao},
  \citenamefont {Zhang},\ and\ \citenamefont {Chan}}]{xiao2014surface}%
  \BibitemOpen
  \bibfield  {author} {\bibinfo {author} {\bibfnamefont {M.}~\bibnamefont
  {Xiao}}, \bibinfo {author} {\bibfnamefont {Z.}~\bibnamefont {Zhang}},\ and\
  \bibinfo {author} {\bibfnamefont {C.~T.}\ \bibnamefont {Chan}},\ }\bibfield
  {title} {\bibinfo {title} {Surface impedance and bulk band geometric phases
  in one-dimensional systems},\ }\href@noop {} {\bibfield  {journal} {\bibinfo
  {journal} {Physical Review X}\ }\textbf {\bibinfo {volume} {4}},\ \bibinfo
  {pages} {021017} (\bibinfo {year} {2014})}\BibitemShut {NoStop}%
\bibitem [{\citenamefont {Po}\ \emph {et~al.}(2017)\citenamefont {Po},
  \citenamefont {Vishwanath},\ and\ \citenamefont {Watanabe}}]{po2017symmetry}%
  \BibitemOpen
  \bibfield  {author} {\bibinfo {author} {\bibfnamefont {H.~C.}\ \bibnamefont
  {Po}}, \bibinfo {author} {\bibfnamefont {A.}~\bibnamefont {Vishwanath}},\
  and\ \bibinfo {author} {\bibfnamefont {H.}~\bibnamefont {Watanabe}},\
  }\bibfield  {title} {\bibinfo {title} {Symmetry-based indicators of band
  topology in the 230 space groups},\ }\href@noop {} {\bibfield  {journal}
  {\bibinfo  {journal} {Nature communications}\ }\textbf {\bibinfo {volume}
  {8}},\ \bibinfo {pages} {1} (\bibinfo {year} {2017})}\BibitemShut {NoStop}%
\bibitem [{\citenamefont {de~Paz}\ \emph {et~al.}(2019)\citenamefont {de~Paz},
  \citenamefont {Vergniory}, \citenamefont {Bercioux}, \citenamefont
  {Garc{\'\i}a-Etxarri},\ and\ \citenamefont {Bradlyn}}]{de2019engineering}%
  \BibitemOpen
  \bibfield  {author} {\bibinfo {author} {\bibfnamefont {M.~B.}\ \bibnamefont
  {de~Paz}}, \bibinfo {author} {\bibfnamefont {M.~G.}\ \bibnamefont
  {Vergniory}}, \bibinfo {author} {\bibfnamefont {D.}~\bibnamefont {Bercioux}},
  \bibinfo {author} {\bibfnamefont {A.}~\bibnamefont {Garc{\'\i}a-Etxarri}},\
  and\ \bibinfo {author} {\bibfnamefont {B.}~\bibnamefont {Bradlyn}},\
  }\bibfield  {title} {\bibinfo {title} {Engineering fragile topology in
  photonic crystals: Topological quantum chemistry of light},\ }\href@noop {}
  {\bibfield  {journal} {\bibinfo  {journal} {Physical Review Research}\
  }\textbf {\bibinfo {volume} {1}},\ \bibinfo {pages} {032005} (\bibinfo {year}
  {2019})}\BibitemShut {NoStop}%
\bibitem [{\citenamefont {Alexandradinata}\ \emph {et~al.}(2020)\citenamefont
  {Alexandradinata}, \citenamefont {H{\"o}ller}, \citenamefont {Wang},
  \citenamefont {Cheng},\ and\ \citenamefont
  {Lu}}]{alexandradinata2020crystallographic}%
  \BibitemOpen
  \bibfield  {author} {\bibinfo {author} {\bibfnamefont {A.}~\bibnamefont
  {Alexandradinata}}, \bibinfo {author} {\bibfnamefont {J.}~\bibnamefont
  {H{\"o}ller}}, \bibinfo {author} {\bibfnamefont {C.}~\bibnamefont {Wang}},
  \bibinfo {author} {\bibfnamefont {H.}~\bibnamefont {Cheng}},\ and\ \bibinfo
  {author} {\bibfnamefont {L.}~\bibnamefont {Lu}},\ }\bibfield  {title}
  {\bibinfo {title} {Crystallographic splitting theorem for band
  representations and fragile topological photonic crystals},\ }\href@noop {}
  {\bibfield  {journal} {\bibinfo  {journal} {Physical Review B}\ }\textbf
  {\bibinfo {volume} {102}},\ \bibinfo {pages} {115117} (\bibinfo {year}
  {2020})}\BibitemShut {NoStop}%
\bibitem [{\citenamefont {Christensen}\ \emph {et~al.}(2021)\citenamefont
  {Christensen}, \citenamefont {Po}, \citenamefont {Joannopoulos},\ and\
  \citenamefont {Solja{\v{c}}i{\'c}}}]{christensen2021location}%
  \BibitemOpen
  \bibfield  {author} {\bibinfo {author} {\bibfnamefont {T.}~\bibnamefont
  {Christensen}}, \bibinfo {author} {\bibfnamefont {H.~C.}\ \bibnamefont {Po}},
  \bibinfo {author} {\bibfnamefont {J.~D.}\ \bibnamefont {Joannopoulos}},\ and\
  \bibinfo {author} {\bibfnamefont {M.}~\bibnamefont {Solja{\v{c}}i{\'c}}},\
  }\bibfield  {title} {\bibinfo {title} {Location and topology of the
  fundamental gap in photonic crystals},\ }\href@noop {} {\bibfield  {journal}
  {\bibinfo  {journal} {arXiv preprint arXiv:2106.10267}\ } (\bibinfo {year}
  {2021})}\BibitemShut {NoStop}%
\bibitem [{\citenamefont {Soluyanov}\ and\ \citenamefont
  {Vanderbilt}(2011{\natexlab{a}})}]{soluyanov2011wannier}%
  \BibitemOpen
  \bibfield  {author} {\bibinfo {author} {\bibfnamefont {A.~A.}\ \bibnamefont
  {Soluyanov}}\ and\ \bibinfo {author} {\bibfnamefont {D.}~\bibnamefont
  {Vanderbilt}},\ }\bibfield  {title} {\bibinfo {title} {Wannier representation
  of z 2 topological insulators},\ }\href@noop {} {\bibfield  {journal}
  {\bibinfo  {journal} {Physical Review B}\ }\textbf {\bibinfo {volume} {83}},\
  \bibinfo {pages} {035108} (\bibinfo {year} {2011}{\natexlab{a}})}\BibitemShut
  {NoStop}%
\bibitem [{\citenamefont {Soluyanov}\ and\ \citenamefont
  {Vanderbilt}(2011{\natexlab{b}})}]{z2packvanderbilt}%
  \BibitemOpen
  \bibfield  {author} {\bibinfo {author} {\bibfnamefont {A.~A.}\ \bibnamefont
  {Soluyanov}}\ and\ \bibinfo {author} {\bibfnamefont {D.}~\bibnamefont
  {Vanderbilt}},\ }\bibfield  {title} {\bibinfo {title} {Computing topological
  invariants without inversion symmetry},\ }\href
  {https://doi.org/10.1103/PhysRevB.83.235401} {\bibfield  {journal} {\bibinfo
  {journal} {Phys. Rev. B}\ }\textbf {\bibinfo {volume} {83}},\ \bibinfo
  {pages} {235401} (\bibinfo {year} {2011}{\natexlab{b}})}\BibitemShut
  {NoStop}%
\bibitem [{\citenamefont {Vanderbilt}(2018)}]{vanderbilt2018berry}%
  \BibitemOpen
  \bibfield  {author} {\bibinfo {author} {\bibfnamefont {D.}~\bibnamefont
  {Vanderbilt}},\ }\href@noop {} {\emph {\bibinfo {title} {Berry Phases in
  Electronic Structure Theory: Electric Polarization, Orbital Magnetization and
  Topological Insulators}}}\ (\bibinfo  {publisher} {Cambridge University
  Press},\ \bibinfo {year} {2018})\BibitemShut {NoStop}%
\bibitem [{\citenamefont {Bradlyn}\ \emph {et~al.}(2017)\citenamefont
  {Bradlyn}, \citenamefont {Elcoro}, \citenamefont {Cano}, \citenamefont
  {Vergniory}, \citenamefont {Wang}, \citenamefont {Felser}, \citenamefont
  {Aroyo},\ and\ \citenamefont {Bernevig}}]{bradlyn2017topological}%
  \BibitemOpen
  \bibfield  {author} {\bibinfo {author} {\bibfnamefont {B.}~\bibnamefont
  {Bradlyn}}, \bibinfo {author} {\bibfnamefont {L.}~\bibnamefont {Elcoro}},
  \bibinfo {author} {\bibfnamefont {J.}~\bibnamefont {Cano}}, \bibinfo {author}
  {\bibfnamefont {M.}~\bibnamefont {Vergniory}}, \bibinfo {author}
  {\bibfnamefont {Z.}~\bibnamefont {Wang}}, \bibinfo {author} {\bibfnamefont
  {C.}~\bibnamefont {Felser}}, \bibinfo {author} {\bibfnamefont {M.~I.}\
  \bibnamefont {Aroyo}},\ and\ \bibinfo {author} {\bibfnamefont {B.~A.}\
  \bibnamefont {Bernevig}},\ }\bibfield  {title} {\bibinfo {title} {Topological
  quantum chemistry},\ }\href@noop {} {\bibfield  {journal} {\bibinfo
  {journal} {Nature}\ }\textbf {\bibinfo {volume} {547}},\ \bibinfo {pages}
  {298} (\bibinfo {year} {2017})}\BibitemShut {NoStop}%
\bibitem [{\citenamefont {Wieder}\ \emph
  {et~al.}(2020{\natexlab{b}})\citenamefont {Wieder}, \citenamefont {Wang},
  \citenamefont {Cano}, \citenamefont {Dai}, \citenamefont {Schoop},
  \citenamefont {Bradlyn},\ and\ \citenamefont {Bernevig}}]{wieder2020strong}%
  \BibitemOpen
  \bibfield  {author} {\bibinfo {author} {\bibfnamefont {B.~J.}\ \bibnamefont
  {Wieder}}, \bibinfo {author} {\bibfnamefont {Z.}~\bibnamefont {Wang}},
  \bibinfo {author} {\bibfnamefont {J.}~\bibnamefont {Cano}}, \bibinfo {author}
  {\bibfnamefont {X.}~\bibnamefont {Dai}}, \bibinfo {author} {\bibfnamefont
  {L.~M.}\ \bibnamefont {Schoop}}, \bibinfo {author} {\bibfnamefont
  {B.}~\bibnamefont {Bradlyn}},\ and\ \bibinfo {author} {\bibfnamefont {B.~A.}\
  \bibnamefont {Bernevig}},\ }\bibfield  {title} {\bibinfo {title} {Strong and
  fragile topological {Dirac} semimetals with higher-order {Fermi} arcs},\
  }\href {https://doi.org/10.1038/s41467-020-14443-5} {\bibfield  {journal}
  {\bibinfo  {journal} {Nature Communications}\ }\textbf {\bibinfo {volume}
  {11}},\ \bibinfo {pages} {627} (\bibinfo {year}
  {2020}{\natexlab{b}})}\BibitemShut {NoStop}%
\bibitem [{\citenamefont {Fang}\ and\ \citenamefont
  {Cano}(2021)}]{fang2021filling}%
  \BibitemOpen
  \bibfield  {author} {\bibinfo {author} {\bibfnamefont {Y.}~\bibnamefont
  {Fang}}\ and\ \bibinfo {author} {\bibfnamefont {J.}~\bibnamefont {Cano}},\
  }\bibfield  {title} {\bibinfo {title} {Filling anomaly for general two- and
  three-dimensional c4 symmetric lattices},\ }\href
  {https://doi.org/10.1103/PhysRevB.103.165109} {\bibfield  {journal} {\bibinfo
   {journal} {Physical Review B}\ }\textbf {\bibinfo {volume} {103}},\ \bibinfo
  {pages} {165109} (\bibinfo {year} {2021})}\BibitemShut {NoStop}%
\bibitem [{\citenamefont {Benalcazar}\ \emph {et~al.}(2019)\citenamefont
  {Benalcazar}, \citenamefont {Li},\ and\ \citenamefont
  {Hughes}}]{benalcazar2019quantization}%
  \BibitemOpen
  \bibfield  {author} {\bibinfo {author} {\bibfnamefont {W.~A.}\ \bibnamefont
  {Benalcazar}}, \bibinfo {author} {\bibfnamefont {T.}~\bibnamefont {Li}},\
  and\ \bibinfo {author} {\bibfnamefont {T.~L.}\ \bibnamefont {Hughes}},\
  }\bibfield  {title} {\bibinfo {title} {Quantization of fractional corner
  charge in {{C}} n-symmetric higher-order topological crystalline
  insulators},\ }\href@noop {} {\bibfield  {journal} {\bibinfo  {journal}
  {Physical Review B}\ }\textbf {\bibinfo {volume} {99}},\ \bibinfo {pages}
  {245151} (\bibinfo {year} {2019})}\BibitemShut {NoStop}%
\bibitem [{\citenamefont {Longhi}(2019)}]{longhi2019probing}%
  \BibitemOpen
  \bibfield  {author} {\bibinfo {author} {\bibfnamefont {S.}~\bibnamefont
  {Longhi}},\ }\bibfield  {title} {\bibinfo {title} {Probing topological phases
  in waveguide superlattices},\ }\href@noop {} {\bibfield  {journal} {\bibinfo
  {journal} {Optics letters}\ }\textbf {\bibinfo {volume} {44}},\ \bibinfo
  {pages} {2530} (\bibinfo {year} {2019})}\BibitemShut {NoStop}%
\bibitem [{\citenamefont {Romano}\ \emph {et~al.}(2010)\citenamefont {Romano},
  \citenamefont {Nacbar},\ and\ \citenamefont {Bruno-Alfonso}}]{Romano_2010}%
  \BibitemOpen
  \bibfield  {author} {\bibinfo {author} {\bibfnamefont {M.~C.}\ \bibnamefont
  {Romano}}, \bibinfo {author} {\bibfnamefont {D.~R.}\ \bibnamefont {Nacbar}},\
  and\ \bibinfo {author} {\bibfnamefont {A.}~\bibnamefont {Bruno-Alfonso}},\
  }\bibfield  {title} {\bibinfo {title} {Wannier functions of a one-dimensional
  photonic crystal with inversion symmetry},\ }\href
  {https://doi.org/10.1088/0953-4075/43/21/215403} {\bibfield  {journal}
  {\bibinfo  {journal} {Journal of Physics B: Atomic, Molecular and Optical
  Physics}\ }\textbf {\bibinfo {volume} {43}},\ \bibinfo {pages} {215403}
  (\bibinfo {year} {2010})}\BibitemShut {NoStop}%
\bibitem [{\citenamefont {Leung}(1993)}]{leung1993defect}%
  \BibitemOpen
  \bibfield  {author} {\bibinfo {author} {\bibfnamefont {K.~M.}\ \bibnamefont
  {Leung}},\ }\bibfield  {title} {\bibinfo {title} {Defect modes in photonic
  band structures: A {{Green}}'s function approach using vector {{Wannier}}
  functions},\ }\href@noop {} {\bibfield  {journal} {\bibinfo  {journal} {JOSA
  B}\ }\textbf {\bibinfo {volume} {10}},\ \bibinfo {pages} {303} (\bibinfo
  {year} {1993})}\BibitemShut {NoStop}%
\bibitem [{\citenamefont {Albert}\ \emph {et~al.}(2000)\citenamefont {Albert},
  \citenamefont {Jouanin}, \citenamefont {Cassagne},\ and\ \citenamefont
  {Bertho}}]{albert2000generalized}%
  \BibitemOpen
  \bibfield  {author} {\bibinfo {author} {\bibfnamefont {J.}~\bibnamefont
  {Albert}}, \bibinfo {author} {\bibfnamefont {C.}~\bibnamefont {Jouanin}},
  \bibinfo {author} {\bibfnamefont {D.}~\bibnamefont {Cassagne}},\ and\
  \bibinfo {author} {\bibfnamefont {D.}~\bibnamefont {Bertho}},\ }\bibfield
  {title} {\bibinfo {title} {Generalized {{Wannier}} function method for
  photonic crystals},\ }\href@noop {} {\bibfield  {journal} {\bibinfo
  {journal} {Physical Review B}\ }\textbf {\bibinfo {volume} {61}},\ \bibinfo
  {pages} {4381} (\bibinfo {year} {2000})}\BibitemShut {NoStop}%
\bibitem [{\citenamefont {Albert}\ \emph {et~al.}(2002)\citenamefont {Albert},
  \citenamefont {Jouanin}, \citenamefont {Cassagne},\ and\ \citenamefont
  {Monge}}]{albert2002photonic}%
  \BibitemOpen
  \bibfield  {author} {\bibinfo {author} {\bibfnamefont {J.}~\bibnamefont
  {Albert}}, \bibinfo {author} {\bibfnamefont {C.}~\bibnamefont {Jouanin}},
  \bibinfo {author} {\bibfnamefont {D.}~\bibnamefont {Cassagne}},\ and\
  \bibinfo {author} {\bibfnamefont {D.}~\bibnamefont {Monge}},\ }\bibfield
  {title} {\bibinfo {title} {Photonic crystal modelling using a tight-binding
  {{Wannier}} function method},\ }\href@noop {} {\bibfield  {journal} {\bibinfo
   {journal} {Optical and quantum electronics}\ }\textbf {\bibinfo {volume}
  {34}},\ \bibinfo {pages} {251} (\bibinfo {year} {2002})}\BibitemShut
  {NoStop}%
\bibitem [{\citenamefont {Busch}\ \emph {et~al.}(2003)\citenamefont {Busch},
  \citenamefont {Mingaleev}, \citenamefont {{Garcia-Martin}}, \citenamefont
  {Schillinger},\ and\ \citenamefont {Hermann}}]{busch2003wannier}%
  \BibitemOpen
  \bibfield  {author} {\bibinfo {author} {\bibfnamefont {K.}~\bibnamefont
  {Busch}}, \bibinfo {author} {\bibfnamefont {S.~F.}\ \bibnamefont
  {Mingaleev}}, \bibinfo {author} {\bibfnamefont {A.}~\bibnamefont
  {{Garcia-Martin}}}, \bibinfo {author} {\bibfnamefont {M.}~\bibnamefont
  {Schillinger}},\ and\ \bibinfo {author} {\bibfnamefont {D.}~\bibnamefont
  {Hermann}},\ }\bibfield  {title} {\bibinfo {title} {The {{Wannier}} function
  approach to photonic crystal circuits},\ }\href@noop {} {\bibfield  {journal}
  {\bibinfo  {journal} {Journal of Physics: Condensed Matter}\ }\textbf
  {\bibinfo {volume} {15}},\ \bibinfo {pages} {R1233} (\bibinfo {year}
  {2003})}\BibitemShut {NoStop}%
\bibitem [{\citenamefont {Busch}\ \emph {et~al.}(2011)\citenamefont {Busch},
  \citenamefont {Blum}, \citenamefont {Graham}, \citenamefont {Hermann},
  \citenamefont {K{\"o}hl}, \citenamefont {Mack},\ and\ \citenamefont
  {Wolff}}]{busch2011photonic}%
  \BibitemOpen
  \bibfield  {author} {\bibinfo {author} {\bibfnamefont {K.}~\bibnamefont
  {Busch}}, \bibinfo {author} {\bibfnamefont {C.}~\bibnamefont {Blum}},
  \bibinfo {author} {\bibfnamefont {A.~M.}\ \bibnamefont {Graham}}, \bibinfo
  {author} {\bibfnamefont {D.}~\bibnamefont {Hermann}}, \bibinfo {author}
  {\bibfnamefont {M.}~\bibnamefont {K{\"o}hl}}, \bibinfo {author}
  {\bibfnamefont {P.}~\bibnamefont {Mack}},\ and\ \bibinfo {author}
  {\bibfnamefont {C.}~\bibnamefont {Wolff}},\ }\bibfield  {title} {\bibinfo
  {title} {The photonic {{Wannier}} function approach to photonic crystal
  simulations: Status and perspectives},\ }\href@noop {} {\bibfield  {journal}
  {\bibinfo  {journal} {Journal of Modern Optics}\ }\textbf {\bibinfo {volume}
  {58}},\ \bibinfo {pages} {365} (\bibinfo {year} {2011})}\BibitemShut
  {NoStop}%
\bibitem [{\citenamefont {Romano}\ \emph {et~al.}(2018)\citenamefont {Romano},
  \citenamefont {{Vellasco-Gomes}},\ and\ \citenamefont
  {{Bruno-Alfonso}}}]{romano2018wannier}%
  \BibitemOpen
  \bibfield  {author} {\bibinfo {author} {\bibfnamefont {M.~C.}\ \bibnamefont
  {Romano}}, \bibinfo {author} {\bibfnamefont {A.}~\bibnamefont
  {{Vellasco-Gomes}}},\ and\ \bibinfo {author} {\bibfnamefont {A.}~\bibnamefont
  {{Bruno-Alfonso}}},\ }\bibfield  {title} {\bibinfo {title} {Wannier functions
  and the calculation of localized modes in one-dimensional photonic
  crystals},\ }\href@noop {} {\bibfield  {journal} {\bibinfo  {journal} {JOSA
  B}\ }\textbf {\bibinfo {volume} {35}},\ \bibinfo {pages} {826} (\bibinfo
  {year} {2018})}\BibitemShut {NoStop}%
\bibitem [{\citenamefont {Tanaue}\ and\ \citenamefont
  {Bruno-Alfonso}(2020)}]{tanaue2020wannier}%
  \BibitemOpen
  \bibfield  {author} {\bibinfo {author} {\bibfnamefont {H.~B.}\ \bibnamefont
  {Tanaue}}\ and\ \bibinfo {author} {\bibfnamefont {A.}~\bibnamefont
  {Bruno-Alfonso}},\ }\bibfield  {title} {\bibinfo {title} {Wannier-function
  expansion of localized modes in 1d photonic crystals without inversion
  symmetry},\ }\href@noop {} {\bibfield  {journal} {\bibinfo  {journal} {JOSA
  B}\ }\textbf {\bibinfo {volume} {37}},\ \bibinfo {pages} {3698} (\bibinfo
  {year} {2020})}\BibitemShut {NoStop}%
\bibitem [{\citenamefont {Garcia-Martin}\ \emph {et~al.}(2003)\citenamefont
  {Garcia-Martin}, \citenamefont {Hermann}, \citenamefont {Hagmann},
  \citenamefont {Busch},\ and\ \citenamefont {W{\"o}lfle}}]{garcia2003defect}%
  \BibitemOpen
  \bibfield  {author} {\bibinfo {author} {\bibfnamefont {A.}~\bibnamefont
  {Garcia-Martin}}, \bibinfo {author} {\bibfnamefont {D.}~\bibnamefont
  {Hermann}}, \bibinfo {author} {\bibfnamefont {F.}~\bibnamefont {Hagmann}},
  \bibinfo {author} {\bibfnamefont {K.}~\bibnamefont {Busch}},\ and\ \bibinfo
  {author} {\bibfnamefont {P.}~\bibnamefont {W{\"o}lfle}},\ }\bibfield  {title}
  {\bibinfo {title} {Defect computations in photonic crystals: a solid state
  theoretical approach},\ }\href@noop {} {\bibfield  {journal} {\bibinfo
  {journal} {Nanotechnology}\ }\textbf {\bibinfo {volume} {14}},\ \bibinfo
  {pages} {177} (\bibinfo {year} {2003})}\BibitemShut {NoStop}%
\bibitem [{\citenamefont {Bruno-Alfonso}\ and\ \citenamefont
  {Guo-Qiang}(2003)}]{bruno2003bloch}%
  \BibitemOpen
  \bibfield  {author} {\bibinfo {author} {\bibfnamefont {A.}~\bibnamefont
  {Bruno-Alfonso}}\ and\ \bibinfo {author} {\bibfnamefont {H.}~\bibnamefont
  {Guo-Qiang}},\ }\bibfield  {title} {\bibinfo {title} {Bloch--kohn and
  wannier--kohn functions inone dimension},\ }\href@noop {} {\bibfield
  {journal} {\bibinfo  {journal} {Journal of Physics: Condensed Matter}\
  }\textbf {\bibinfo {volume} {15}},\ \bibinfo {pages} {6701} (\bibinfo {year}
  {2003})}\BibitemShut {NoStop}%
\bibitem [{\citenamefont {Hermann}\ \emph {et~al.}(2008)\citenamefont
  {Hermann}, \citenamefont {Schillinger}, \citenamefont {Mingaleev},\ and\
  \citenamefont {Busch}}]{hermann2008wannier}%
  \BibitemOpen
  \bibfield  {author} {\bibinfo {author} {\bibfnamefont {D.}~\bibnamefont
  {Hermann}}, \bibinfo {author} {\bibfnamefont {M.}~\bibnamefont
  {Schillinger}}, \bibinfo {author} {\bibfnamefont {S.~F.}\ \bibnamefont
  {Mingaleev}},\ and\ \bibinfo {author} {\bibfnamefont {K.}~\bibnamefont
  {Busch}},\ }\bibfield  {title} {\bibinfo {title} {Wannier-function based
  scattering-matrix formalism for photonic crystal circuitry},\ }\href@noop {}
  {\bibfield  {journal} {\bibinfo  {journal} {JOSA B}\ }\textbf {\bibinfo
  {volume} {25}},\ \bibinfo {pages} {202} (\bibinfo {year} {2008})}\BibitemShut
  {NoStop}%
\bibitem [{\citenamefont {Johnson}\ and\ \citenamefont
  {Joannopoulos}(2001)}]{johnson2001block}%
  \BibitemOpen
  \bibfield  {author} {\bibinfo {author} {\bibfnamefont {S.~G.}\ \bibnamefont
  {Johnson}}\ and\ \bibinfo {author} {\bibfnamefont {J.~D.}\ \bibnamefont
  {Joannopoulos}},\ }\bibfield  {title} {\bibinfo {title} {Block-iterative
  frequency-domain methods for maxwell’s equations in a planewave basis},\
  }\href@noop {} {\bibfield  {journal} {\bibinfo  {journal} {Optics express}\
  }\textbf {\bibinfo {volume} {8}},\ \bibinfo {pages} {173} (\bibinfo {year}
  {2001})}\BibitemShut {NoStop}%
\bibitem [{\citenamefont {Su}\ \emph {et~al.}(1979)\citenamefont {Su},
  \citenamefont {Schrieffer},\ and\ \citenamefont {Heeger}}]{su1979solitons}%
  \BibitemOpen
  \bibfield  {author} {\bibinfo {author} {\bibfnamefont {W.}~\bibnamefont
  {Su}}, \bibinfo {author} {\bibfnamefont {J.}~\bibnamefont {Schrieffer}},\
  and\ \bibinfo {author} {\bibfnamefont {A.~J.}\ \bibnamefont {Heeger}},\
  }\bibfield  {title} {\bibinfo {title} {Solitons in polyacetylene},\
  }\href@noop {} {\bibfield  {journal} {\bibinfo  {journal} {Physical review
  letters}\ }\textbf {\bibinfo {volume} {42}},\ \bibinfo {pages} {1698}
  (\bibinfo {year} {1979})}\BibitemShut {NoStop}%
\bibitem [{\citenamefont {Sipe}(2000)}]{sipe2000vector}%
  \BibitemOpen
  \bibfield  {author} {\bibinfo {author} {\bibfnamefont {J.}~\bibnamefont
  {Sipe}},\ }\bibfield  {title} {\bibinfo {title} {Vector k.p approach for
  photonic band structures},\ }\href@noop {} {\bibfield  {journal} {\bibinfo
  {journal} {Physical Review E}\ }\textbf {\bibinfo {volume} {62}},\ \bibinfo
  {pages} {5672} (\bibinfo {year} {2000})}\BibitemShut {NoStop}%
\bibitem [{\citenamefont {Watanabe}\ and\ \citenamefont
  {Lu}(2018)}]{watanabe2018space}%
  \BibitemOpen
  \bibfield  {author} {\bibinfo {author} {\bibfnamefont {H.}~\bibnamefont
  {Watanabe}}\ and\ \bibinfo {author} {\bibfnamefont {L.}~\bibnamefont {Lu}},\
  }\bibfield  {title} {\bibinfo {title} {Space group theory of photonic
  bands},\ }\href@noop {} {\bibfield  {journal} {\bibinfo  {journal} {Physical
  review letters}\ }\textbf {\bibinfo {volume} {121}},\ \bibinfo {pages}
  {263903} (\bibinfo {year} {2018})}\BibitemShut {NoStop}%
\bibitem [{\citenamefont {Zak}(1989)}]{zak1989berry}%
  \BibitemOpen
  \bibfield  {author} {\bibinfo {author} {\bibfnamefont {J.}~\bibnamefont
  {Zak}},\ }\bibfield  {title} {\bibinfo {title} {Berry's phase for energy
  bands in solids},\ }\href {https://doi.org/10.1103/PhysRevLett.62.2747}
  {\bibfield  {journal} {\bibinfo  {journal} {Phys. Rev. Lett.}\ }\textbf
  {\bibinfo {volume} {62}},\ \bibinfo {pages} {2747} (\bibinfo {year}
  {1989})}\BibitemShut {NoStop}%
\bibitem [{\citenamefont {Marzari}\ and\ \citenamefont
  {Vanderbilt}(1997)}]{marzari1997maximally}%
  \BibitemOpen
  \bibfield  {author} {\bibinfo {author} {\bibfnamefont {N.}~\bibnamefont
  {Marzari}}\ and\ \bibinfo {author} {\bibfnamefont {D.}~\bibnamefont
  {Vanderbilt}},\ }\bibfield  {title} {\bibinfo {title} {Maximally localized
  generalized {{Wannier}} functions for composite energy bands},\ }\href@noop
  {} {\bibfield  {journal} {\bibinfo  {journal} {Physical review B}\ }\textbf
  {\bibinfo {volume} {56}},\ \bibinfo {pages} {12847} (\bibinfo {year}
  {1997})}\BibitemShut {NoStop}%
\bibitem [{\citenamefont {Kivelson}(1982)}]{kivelson1982wannier}%
  \BibitemOpen
  \bibfield  {author} {\bibinfo {author} {\bibfnamefont {S.}~\bibnamefont
  {Kivelson}},\ }\bibfield  {title} {\bibinfo {title} {Wannier functions in
  one-dimensional disordered systems: Application to fractionally charged
  solitons},\ }\href {https://doi.org/10.1103/PhysRevB.26.4269} {\bibfield
  {journal} {\bibinfo  {journal} {Phys. Rev. B}\ }\textbf {\bibinfo {volume}
  {26}},\ \bibinfo {pages} {4269} (\bibinfo {year} {1982})}\BibitemShut
  {NoStop}%
\bibitem [{\citenamefont {Alexandradinata}\ \emph {et~al.}(2014)\citenamefont
  {Alexandradinata}, \citenamefont {Dai},\ and\ \citenamefont
  {Bernevig}}]{alexandradinata2014wilsonloop}%
  \BibitemOpen
  \bibfield  {author} {\bibinfo {author} {\bibfnamefont {A.}~\bibnamefont
  {Alexandradinata}}, \bibinfo {author} {\bibfnamefont {X.}~\bibnamefont
  {Dai}},\ and\ \bibinfo {author} {\bibfnamefont {B.~A.}\ \bibnamefont
  {Bernevig}},\ }\bibfield  {title} {\bibinfo {title} {Wilson-loop
  characterization of inversion-symmetric topological insulators},\ }\href@noop
  {} {\bibfield  {journal} {\bibinfo  {journal} {Phys. Rev. B}\ }\textbf
  {\bibinfo {volume} {89}},\ \bibinfo {pages} {155114} (\bibinfo {year}
  {2014})}\BibitemShut {NoStop}%
\bibitem [{\citenamefont {Yu}\ \emph {et~al.}(2011)\citenamefont {Yu},
  \citenamefont {Qi}, \citenamefont {Bernevig}, \citenamefont {Fang},\ and\
  \citenamefont {Dai}}]{yu2011equivalent}%
  \BibitemOpen
  \bibfield  {author} {\bibinfo {author} {\bibfnamefont {R.}~\bibnamefont
  {Yu}}, \bibinfo {author} {\bibfnamefont {X.~L.}\ \bibnamefont {Qi}}, \bibinfo
  {author} {\bibfnamefont {A.}~\bibnamefont {Bernevig}}, \bibinfo {author}
  {\bibfnamefont {Z.}~\bibnamefont {Fang}},\ and\ \bibinfo {author}
  {\bibfnamefont {X.}~\bibnamefont {Dai}},\ }\bibfield  {title} {\bibinfo
  {title} {Equivalent expression of \$\textbackslash{{mathbbZ}}\_2\$
  topological invariant for band insulators using the non-{{Abelian Berry}}
  connection},\ }\href {https://doi.org/10.1103/PhysRevB.84.075119} {\bibfield
  {journal} {\bibinfo  {journal} {Phys. Rev. B}\ }\textbf {\bibinfo {volume}
  {84}},\ \bibinfo {pages} {075119} (\bibinfo {year} {2011})}\BibitemShut
  {NoStop}%
\bibitem [{\citenamefont {Gresch}\ \emph {et~al.}(2017)\citenamefont {Gresch},
  \citenamefont {Aut\`es}, \citenamefont {Yazyev}, \citenamefont {Troyer},
  \citenamefont {Vanderbilt}, \citenamefont {Bernevig},\ and\ \citenamefont
  {Soluyanov}}]{z2packbernevig}%
  \BibitemOpen
  \bibfield  {author} {\bibinfo {author} {\bibfnamefont {D.}~\bibnamefont
  {Gresch}}, \bibinfo {author} {\bibfnamefont {G.}~\bibnamefont {Aut\`es}},
  \bibinfo {author} {\bibfnamefont {O.~V.}\ \bibnamefont {Yazyev}}, \bibinfo
  {author} {\bibfnamefont {M.}~\bibnamefont {Troyer}}, \bibinfo {author}
  {\bibfnamefont {D.}~\bibnamefont {Vanderbilt}}, \bibinfo {author}
  {\bibfnamefont {B.~A.}\ \bibnamefont {Bernevig}},\ and\ \bibinfo {author}
  {\bibfnamefont {A.~A.}\ \bibnamefont {Soluyanov}},\ }\bibfield  {title}
  {\bibinfo {title} {Z2pack: Numerical implementation of hybrid wannier centers
  for identifying topological materials},\ }\href
  {https://doi.org/10.1103/PhysRevB.95.075146} {\bibfield  {journal} {\bibinfo
  {journal} {Phys. Rev. B}\ }\textbf {\bibinfo {volume} {95}},\ \bibinfo
  {pages} {075146} (\bibinfo {year} {2017})}\BibitemShut {NoStop}%
\bibitem [{\citenamefont {Marzari}\ \emph {et~al.}(2012)\citenamefont
  {Marzari}, \citenamefont {Mostofi}, \citenamefont {Yates}, \citenamefont
  {Souza},\ and\ \citenamefont {Vanderbilt}}]{marzari2012maximally}%
  \BibitemOpen
  \bibfield  {author} {\bibinfo {author} {\bibfnamefont {N.}~\bibnamefont
  {Marzari}}, \bibinfo {author} {\bibfnamefont {A.~A.}\ \bibnamefont
  {Mostofi}}, \bibinfo {author} {\bibfnamefont {J.~R.}\ \bibnamefont {Yates}},
  \bibinfo {author} {\bibfnamefont {I.}~\bibnamefont {Souza}},\ and\ \bibinfo
  {author} {\bibfnamefont {D.}~\bibnamefont {Vanderbilt}},\ }\bibfield  {title}
  {\bibinfo {title} {Maximally localized wannier functions: Theory and
  applications},\ }\href@noop {} {\bibfield  {journal} {\bibinfo  {journal}
  {Reviews of Modern Physics}\ }\textbf {\bibinfo {volume} {84}},\ \bibinfo
  {pages} {1419} (\bibinfo {year} {2012})}\BibitemShut {NoStop}%
\bibitem [{\citenamefont {{King-Smith}}\ and\ \citenamefont
  {Vanderbilt}(1993)}]{kingsmith1993theory}%
  \BibitemOpen
  \bibfield  {author} {\bibinfo {author} {\bibfnamefont {R.~D.}\ \bibnamefont
  {{King-Smith}}}\ and\ \bibinfo {author} {\bibfnamefont {D.}~\bibnamefont
  {Vanderbilt}},\ }\bibfield  {title} {\bibinfo {title} {Theory of polarization
  of crystalline solids},\ }\href@noop {} {\bibfield  {journal} {\bibinfo
  {journal} {Phys. Rev. B}\ }\textbf {\bibinfo {volume} {47}},\ \bibinfo
  {pages} {1651(R)} (\bibinfo {year} {1993})}\BibitemShut {NoStop}%
\bibitem [{\citenamefont {Wieder}\ \emph {et~al.}(2018)\citenamefont {Wieder},
  \citenamefont {Bradlyn}, \citenamefont {Wang}, \citenamefont {Cano},
  \citenamefont {Kim}, \citenamefont {Kim}, \citenamefont {Rappe},
  \citenamefont {Kane},\ and\ \citenamefont {Bernevig}}]{wieder2018wallpaper}%
  \BibitemOpen
  \bibfield  {author} {\bibinfo {author} {\bibfnamefont {B.~J.}\ \bibnamefont
  {Wieder}}, \bibinfo {author} {\bibfnamefont {B.}~\bibnamefont {Bradlyn}},
  \bibinfo {author} {\bibfnamefont {Z.}~\bibnamefont {Wang}}, \bibinfo {author}
  {\bibfnamefont {J.}~\bibnamefont {Cano}}, \bibinfo {author} {\bibfnamefont
  {Y.}~\bibnamefont {Kim}}, \bibinfo {author} {\bibfnamefont {H.-S.~D.}\
  \bibnamefont {Kim}}, \bibinfo {author} {\bibfnamefont {A.~M.}\ \bibnamefont
  {Rappe}}, \bibinfo {author} {\bibfnamefont {C.}~\bibnamefont {Kane}},\ and\
  \bibinfo {author} {\bibfnamefont {B.~A.}\ \bibnamefont {Bernevig}},\
  }\bibfield  {title} {\bibinfo {title} {Wallpaper fermions and the
  nonsymmorphic {{Dirac}} insulator},\ }\href@noop {} {\bibfield  {journal}
  {\bibinfo  {journal} {Science (New York, N.Y.)}\ }\textbf {\bibinfo {volume}
  {361}},\ \bibinfo {pages} {246} (\bibinfo {year} {2018})}\BibitemShut
  {NoStop}%
\bibitem [{\citenamefont {Bradlyn}\ and\ \citenamefont
  {Iraola}(2022)}]{bradlyn2021lecture}%
  \BibitemOpen
  \bibfield  {author} {\bibinfo {author} {\bibfnamefont {B.}~\bibnamefont
  {Bradlyn}}\ and\ \bibinfo {author} {\bibfnamefont {M.}~\bibnamefont
  {Iraola}},\ }\bibfield  {title} {\bibinfo {title} {{Lecture Notes on Berry
  Phases and Topology}},\ }\href
  {https://doi.org/10.21468/SciPostPhysLectNotes.51} {\bibfield  {journal}
  {\bibinfo  {journal} {SciPost Phys. Lect. Notes}\ ,\ \bibinfo {pages} {51}}
  (\bibinfo {year} {2022})}\BibitemShut {NoStop}%
\bibitem [{\citenamefont {De~Nittis}\ and\ \citenamefont
  {Lein}(2020)}]{de2020equivalence}%
  \BibitemOpen
  \bibfield  {author} {\bibinfo {author} {\bibfnamefont {G.}~\bibnamefont
  {De~Nittis}}\ and\ \bibinfo {author} {\bibfnamefont {M.}~\bibnamefont
  {Lein}},\ }\bibfield  {title} {\bibinfo {title} {Equivalence of electric,
  magnetic, and electromagnetic chern numbers for topological photonic
  crystals},\ }\href@noop {} {\bibfield  {journal} {\bibinfo  {journal}
  {Journal of Mathematical Physics}\ }\textbf {\bibinfo {volume} {61}},\
  \bibinfo {pages} {022901} (\bibinfo {year} {2020})}\BibitemShut {NoStop}%
\bibitem [{\citenamefont {Kohn}(1959)}]{kohn1959analytic}%
  \BibitemOpen
  \bibfield  {author} {\bibinfo {author} {\bibfnamefont {W.}~\bibnamefont
  {Kohn}},\ }\bibfield  {title} {\bibinfo {title} {Analytic {{Properties}} of
  {{Bloch Waves}} and {{Wannier Functions}}},\ }\href
  {https://doi.org/10.1103/PhysRev.115.809} {\bibfield  {journal} {\bibinfo
  {journal} {Phys. Rev.}\ }\textbf {\bibinfo {volume} {115}},\ \bibinfo {pages}
  {809} (\bibinfo {year} {1959})}\BibitemShut {NoStop}%
\bibitem [{\citenamefont {He}\ and\ \citenamefont
  {Vanderbilt}(2001)}]{he2001exponential}%
  \BibitemOpen
  \bibfield  {author} {\bibinfo {author} {\bibfnamefont {L.}~\bibnamefont
  {He}}\ and\ \bibinfo {author} {\bibfnamefont {D.}~\bibnamefont
  {Vanderbilt}},\ }\bibfield  {title} {\bibinfo {title} {Exponential decay
  properties of wannier functions and related quantities},\ }\href
  {https://doi.org/10.1103/PhysRevLett.86.5341} {\bibfield  {journal} {\bibinfo
   {journal} {Physical Review Letters}\ }\textbf {\bibinfo {volume} {86}},\
  \bibinfo {pages} {5341} (\bibinfo {year} {2001})}\BibitemShut {NoStop}%
\bibitem [{\citenamefont {Cano}\ and\ \citenamefont
  {Bradlyn}(2021)}]{cano2021band}%
  \BibitemOpen
  \bibfield  {author} {\bibinfo {author} {\bibfnamefont {J.}~\bibnamefont
  {Cano}}\ and\ \bibinfo {author} {\bibfnamefont {B.}~\bibnamefont {Bradlyn}},\
  }\bibfield  {title} {\bibinfo {title} {Band {{Representations}} and
  {{Topological Quantum Chemistry}}},\ }\href
  {https://doi.org/10.1146/annurev-conmatphys-041720-124134} {\bibfield
  {journal} {\bibinfo  {journal} {Annual Reviews of Condensed Matter Physics}\
  }\textbf {\bibinfo {volume} {12}},\ \bibinfo {pages} {225} (\bibinfo {year}
  {2021})}\BibitemShut {NoStop}%
\bibitem [{\citenamefont {Henriques}\ \emph {et~al.}(2020)\citenamefont
  {Henriques}, \citenamefont {Rappoport}, \citenamefont {Bludov}, \citenamefont
  {Vasilevskiy},\ and\ \citenamefont {Peres}}]{henriques2020topological}%
  \BibitemOpen
  \bibfield  {author} {\bibinfo {author} {\bibfnamefont {J.~C.~G.}\
  \bibnamefont {Henriques}}, \bibinfo {author} {\bibfnamefont {T.~G.}\
  \bibnamefont {Rappoport}}, \bibinfo {author} {\bibfnamefont {Y.~V.}\
  \bibnamefont {Bludov}}, \bibinfo {author} {\bibfnamefont {M.~I.}\
  \bibnamefont {Vasilevskiy}},\ and\ \bibinfo {author} {\bibfnamefont
  {N.~M.~R.}\ \bibnamefont {Peres}},\ }\bibfield  {title} {\bibinfo {title}
  {Topological photonic tamm states and the su-schrieffer-heeger model},\
  }\href {https://doi.org/10.1103/PhysRevA.101.043811} {\bibfield  {journal}
  {\bibinfo  {journal} {Physical Review A: Atomic, Molecular, and Optical
  Physics}\ }\textbf {\bibinfo {volume} {101}},\ \bibinfo {pages} {043811}
  (\bibinfo {year} {2020})}\BibitemShut {NoStop}%
\bibitem [{\citenamefont {Jiang}\ \emph {et~al.}(2021)\citenamefont {Jiang},
  \citenamefont {Liu}, \citenamefont {Xu}, \citenamefont {Gao}, \citenamefont
  {Zhu}, \citenamefont {Xie},\ and\ \citenamefont
  {Yang}}]{jiang2021topological}%
  \BibitemOpen
  \bibfield  {author} {\bibinfo {author} {\bibfnamefont {H.}~\bibnamefont
  {Jiang}}, \bibinfo {author} {\bibfnamefont {W.}~\bibnamefont {Liu}}, \bibinfo
  {author} {\bibfnamefont {J.}~\bibnamefont {Xu}}, \bibinfo {author}
  {\bibfnamefont {B.}~\bibnamefont {Gao}}, \bibinfo {author} {\bibfnamefont
  {C.}~\bibnamefont {Zhu}}, \bibinfo {author} {\bibfnamefont {S.}~\bibnamefont
  {Xie}},\ and\ \bibinfo {author} {\bibfnamefont {Y.}~\bibnamefont {Yang}},\
  }\bibfield  {title} {\bibinfo {title} {Topological edge modes in
  one-dimensional photonic crystals containing metal},\ }\href@noop {}
  {\bibfield  {journal} {\bibinfo  {journal} {OSA Continuum}\ }\textbf
  {\bibinfo {volume} {4}},\ \bibinfo {pages} {1626} (\bibinfo {year}
  {2021})}\BibitemShut {NoStop}%
\bibitem [{\citenamefont {Tan}\ \emph {et~al.}(2014)\citenamefont {Tan},
  \citenamefont {Sun}, \citenamefont {Chen},\ and\ \citenamefont
  {Shen}}]{tan2014photonic}%
  \BibitemOpen
  \bibfield  {author} {\bibinfo {author} {\bibfnamefont {W.}~\bibnamefont
  {Tan}}, \bibinfo {author} {\bibfnamefont {Y.}~\bibnamefont {Sun}}, \bibinfo
  {author} {\bibfnamefont {H.}~\bibnamefont {Chen}},\ and\ \bibinfo {author}
  {\bibfnamefont {S.-Q.}\ \bibnamefont {Shen}},\ }\bibfield  {title} {\bibinfo
  {title} {Photonic simulation of topological excitations in metamaterials},\
  }\href {https://doi.org/10.1038/srep03842} {\bibfield  {journal} {\bibinfo
  {journal} {Scientific Reports}\ }\textbf {\bibinfo {volume} {4}},\ \bibinfo
  {pages} {3842} (\bibinfo {year} {2014})}\BibitemShut {NoStop}%
\bibitem [{\citenamefont {Slobozhanyuk}\ \emph {et~al.}(2015)\citenamefont
  {Slobozhanyuk}, \citenamefont {Poddubny}, \citenamefont {Miroshnichenko},
  \citenamefont {Belov},\ and\ \citenamefont
  {Kivshar}}]{slobozhanyuk2015subwavelength}%
  \BibitemOpen
  \bibfield  {author} {\bibinfo {author} {\bibfnamefont {A.~P.}\ \bibnamefont
  {Slobozhanyuk}}, \bibinfo {author} {\bibfnamefont {A.~N.}\ \bibnamefont
  {Poddubny}}, \bibinfo {author} {\bibfnamefont {A.~E.}\ \bibnamefont
  {Miroshnichenko}}, \bibinfo {author} {\bibfnamefont {P.~A.}\ \bibnamefont
  {Belov}},\ and\ \bibinfo {author} {\bibfnamefont {Y.~S.}\ \bibnamefont
  {Kivshar}},\ }\bibfield  {title} {\bibinfo {title} {Subwavelength topological
  edge states in optically resonant dielectric structures},\ }\href@noop {}
  {\bibfield  {journal} {\bibinfo  {journal} {Physical review letters}\
  }\textbf {\bibinfo {volume} {114}},\ \bibinfo {pages} {123901} (\bibinfo
  {year} {2015})}\BibitemShut {NoStop}%
\bibitem [{\citenamefont {Niemi}\ and\ \citenamefont
  {Semenoff}(1986)}]{niemi1986fermion}%
  \BibitemOpen
  \bibfield  {author} {\bibinfo {author} {\bibfnamefont {A.~J.}\ \bibnamefont
  {Niemi}}\ and\ \bibinfo {author} {\bibfnamefont {G.~W.}\ \bibnamefont
  {Semenoff}},\ }\bibfield  {title} {\bibinfo {title} {Fermion number
  fractionization in quantum field theory},\ }\href
  {https://doi.org/10.1016/0370-1573(86)90167-5} {\bibfield  {journal}
  {\bibinfo  {journal} {Physics Reports}\ }\textbf {\bibinfo {volume} {135}},\
  \bibinfo {pages} {99} (\bibinfo {year} {1986})}\BibitemShut {NoStop}%
\bibitem [{\citenamefont {Jackiw}\ and\ \citenamefont
  {Rebbi}(1976)}]{jackiw1976solitons}%
  \BibitemOpen
  \bibfield  {author} {\bibinfo {author} {\bibfnamefont {R.}~\bibnamefont
  {Jackiw}}\ and\ \bibinfo {author} {\bibfnamefont {C.}~\bibnamefont {Rebbi}},\
  }\bibfield  {title} {\bibinfo {title} {Solitons with fermion number
  {$\frac{1}{2}$}},\ }\href {https://doi.org/10.1103/PhysRevD.13.3398}
  {\bibfield  {journal} {\bibinfo  {journal} {Phys. Rev. D}\ }\textbf {\bibinfo
  {volume} {13}},\ \bibinfo {pages} {3398} (\bibinfo {year}
  {1976})}\BibitemShut {NoStop}%
\bibitem [{\citenamefont {Thouless}(1983)}]{thouless1983quantization}%
  \BibitemOpen
  \bibfield  {author} {\bibinfo {author} {\bibfnamefont {D.~J.}\ \bibnamefont
  {Thouless}},\ }\bibfield  {title} {\bibinfo {title} {Quantization of particle
  transport},\ }\href {https://doi.org/10.1103/PhysRevB.27.6083} {\bibfield
  {journal} {\bibinfo  {journal} {Phys. Rev. B}\ }\textbf {\bibinfo {volume}
  {27}},\ \bibinfo {pages} {6083} (\bibinfo {year} {1983})}\BibitemShut
  {NoStop}%
\bibitem [{\citenamefont {Ko{\v s}ata}\ and\ \citenamefont
  {Zilberberg}(2021)}]{kovsata2021second}%
  \BibitemOpen
  \bibfield  {author} {\bibinfo {author} {\bibfnamefont {J.}~\bibnamefont
  {Ko{\v s}ata}}\ and\ \bibinfo {author} {\bibfnamefont {O.}~\bibnamefont
  {Zilberberg}},\ }\bibfield  {title} {\bibinfo {title} {Second-order
  topological modes in two-dimensional continuous media},\ }\href@noop {}
  {\bibfield  {journal} {\bibinfo  {journal} {Physical Review Research}\
  }\textbf {\bibinfo {volume} {3}},\ \bibinfo {pages} {L032029} (\bibinfo
  {year} {2021})}\BibitemShut {NoStop}%
\bibitem [{\citenamefont {Cornean}\ \emph {et~al.}(2016)\citenamefont
  {Cornean}, \citenamefont {Herbst},\ and\ \citenamefont
  {Nenciu}}]{cornean2016construction}%
  \BibitemOpen
  \bibfield  {author} {\bibinfo {author} {\bibfnamefont {H.~D.}\ \bibnamefont
  {Cornean}}, \bibinfo {author} {\bibfnamefont {I.}~\bibnamefont {Herbst}},\
  and\ \bibinfo {author} {\bibfnamefont {G.}~\bibnamefont {Nenciu}},\
  }\bibfield  {title} {\bibinfo {title} {On the construction of composite
  {{Wannier}} functions},\ }in\ \href@noop {} {\emph {\bibinfo {booktitle}
  {Annales Henri Poincar\'e}}},\ Vol.~\bibinfo {volume} {17}\ (\bibinfo
  {organization} {{Springer}},\ \bibinfo {year} {2016})\ pp.\ \bibinfo {pages}
  {3361--3398}\BibitemShut {NoStop}%
\bibitem [{\citenamefont {Griffiths}(2016)}]{griffiths2016introduction}%
  \BibitemOpen
  \bibfield  {author} {\bibinfo {author} {\bibfnamefont {D.~J.}\ \bibnamefont
  {Griffiths}},\ }\href@noop {} {\emph {\bibinfo {title} {Introduction to
  Quantum Mechanics}}}\ (\bibinfo  {publisher} {{Cambridge University Press}},\
  \bibinfo {year} {2016})\BibitemShut {NoStop}%
\end{thebibliography}%
\end{document}